\documentclass[conference]{IEEEtran}
\IEEEoverridecommandlockouts
\usepackage{cite}
\usepackage{amsmath,amssymb,amsfonts}
\usepackage{algorithm}
\usepackage{algpseudocode}
\usepackage{graphicx}
\usepackage{textcomp}
\usepackage{xcolor}

\usepackage{blkarray}
\usepackage{amsmath}
\usepackage{dsfont}
\usepackage{amsthm}
\usepackage{lipsum}
\usepackage{graphicx}

\usepackage[utf8]{inputenc}
\usepackage[english]{babel}
\usepackage{listings}
\usepackage{fancyvrb}
\usepackage{comment}

\usepackage{tabto}

\def\BibTeX{{\rm B\kern-.05em{\sc i\kern-.025em b}\kern-.08em
    T\kern-.1667em\lower.7ex\hbox{E}\kern-.125emX}}
\begin{document}

\title{Node and Edge Centrality based Failures in Multi-layer Complex Networks\\
\thanks{This research is funded by \emph{Tejas Networks}, Bangalore, India}
}

\author{\IEEEauthorblockN{Dibakar Das}
\IEEEauthorblockA{
\textit{IIIT Bangalore}\\
Bangalore, India \\
dibakard@ieee.org}
\and
\IEEEauthorblockN{Jyotsna Bapat}
\IEEEauthorblockA{
\textit{IIIT Bangalore}\\
Bangalore, India \\
jbapat@iiitb.ac.in}
\and
\IEEEauthorblockN{Debabrata Das}
\IEEEauthorblockA{
\textit{IIIT Bangalore}\\
Bangalore, India\\
ddas@iiitb.ac.in}
}

\maketitle

\begin{abstract}
Multi-layer complex networks (MLCN) appears in various domains, such as, transportation, supply chains, etc. Failures in MLCN can lead to major disruptions in systems. Several research have focussed on different kinds of failures, such as,  cascades, their reasons and ways to avoid them. This paper considers failures in a specific type of MLCN where the lower layer provides services to the higher layer without cross layer interaction, typical of a computer network. A three layer MLCN is constructed with the same set of nodes where each layer has different characteristics, the bottom most layer is Erdos-Renyi (ER) random graph with shortest path hop count among the nodes as gaussian, the middle layer is ER graph with higher number of edges from the previous, and the top most layer is scale free graph with even higher number of edges. Both edge and node failures are considered. Failures happen with decreasing order of centralities of edges and nodes in static batch mode and when the centralities change dynamically with progressive failures. Emergent pattern of three key parameters, namely, average shortest path length (ASPL), total shortest path count (TSPC) and total number of edges (TNE) for all the three layers after node or edge failures are studied. Extensive simulations show that all but one parameters show definite degrading patterns. Surprising, ASPL for the middle layer starts showing a chaotic behavior beyond a certain point for all types of failures.
\end{abstract}

\begin{IEEEkeywords}
complex networks, multi-layer, failure, centrality, chaotic
\end{IEEEkeywords}

\section{Introduction}
Complex networks are of great research interest over the years \cite{cite_book_complex_networks_principles_methods_apps}. They occur in different types of systems, such as, transportation, biological networks, internet, social networks, etc., \cite{cite_complex_networks_apps_survey}. Multi-layer networks are a type of complex network where each node has different channels of interaction with different neighbours \cite{cite_multilayer_complex_networks_structure_and_dynamics}. For example, if cities are nodes then air routes, roadways, rail routes can form a multi-layer network.

Failures in multi-layer networks are of special interest since they can have severe consequences in the function of systems. Several research works have focussed on cascading failures in complex networks \cite{cite_cn_cascade_failures_detailed_study}, such as, due to cyber attacks \cite{cite_cn_cascade_attacks}, effect on load distribution after those events \cite{cite_cn_cascade_failures_in_er_and_scale_free_nets_load_dist}, prediction models \cite{cite_cn_dynamic_topological_models_to_predict_system_failure_from_cascade}, impact of coupling \cite{cite_cn_multiplex_network_cascade_failures_emergent_coupling}.
A major body of literature has focussed on propagation \cite{cite_cn_scale_free_interdependen_network_cascade_prevent_with_different_properties}, mitigation \cite{cite_cn_mitigate_cascade_failures_protect_critical_nodes_and_local_structure}, avoidance \cite{cite_cn_cascade_failures_avoided_with_reinforced_nodes} and restoration \cite{cite_cn_self_heal_reconstruct_after_failure_using_local_info} of the network after cascading failures.

This paper analyses node and edge failures in MLCN under settings more akin to a computer network. Firstly, a three layer MLCN ($L_1$, $L_2$ and $L_3$) is considered where a node in any layer can interact with another node (have a connection) if there is a path in the underlying layer (obviously this does not apply to the bottommost layer). Each layer has different properties. $L_1$ is an Erdos-Renyi (ER) random graph whose shortest path hop count has a gaussian distribution. $L_2$ is also and ER graph but with shortest path hop count to be uniformly distributed. $L_3$ is scale free in nature. Number of edges increases at a faster rate from bottom to top layer. Both edge and node failures are considered with the following properties.
Two types on node centralities are considered, static and dynamic. Static node betweenness centrality (SNBC) means the betweenness centralities of the nodes in each layer decided a the time of construction of the MLCN. Dynamic node betweenness centrality (DNBC) means re-evaluated betweenness centralities of the nodes in each layer when nodes fail progressively at runtime. In the same way, static edge betweenness centralities (SEBC) and dynamic edge betweenness centralities (DEBC) exists for edges in each layer. Nodes and edges fail decreasing order of centralities. In such a setting, the paper studies the emergent behaviour of three key parameters of the three layer MLCN, namely, average shortest path length (ASPL), total shortest path count (TSPC) and total number of edges (TNE) in each layer as nodes and edges fail. One of the key result shows that ASPL of $L_2$ starts showing a chaotic behaviour after certain amount of edges or nodes failure whereas rest of the parameters show a definite declining trend. To the best of the knowledge of the authors, none of the previous work considers these important scenarios and findings.

The paper is organized as follows. Section \ref{section_literature_survey} does a detailed survey of the relevant works. Section \ref{section_model} describes the properties of the MLCN and the type of failures considered in this work along with algorithms used to study the failures. Extensive simulation results are presented in section \ref{section_results}. Section \ref{section_conclusion} concludes this work.
\section{Literature Survey}\label{section_literature_survey}
Several research works have focussed on cascading failures in complex networks.
In one of the early works, authors show how an attack on a single node can bring large scale cascading failures in complex networks \cite{cite_cn_cascade_attacks}.
An early detailed analysis of cascade failures under their different propagation scenarios, such as, connectivity, is described in \cite{cite_cn_cascade_failures_detailed_study}. \cite{cite_cn_percolulation_of_failures_in_ER_scale_free_random_regular_nets} studies robustness of ER, scale-free and random-regular networks against attacks using a percolation framework.
In a recent interesting work, \cite{cite_cn_knowledge_from_random_attacks_lead_to_targetted_attack} shows how knowledge from random attacks in a complex network can lead to targeted ones leading to major disruptions.
Using node load redistribution in ER and scale-free graphs, \cite{cite_cn_cascade_failures_in_er_and_scale_free_nets_load_dist} studies robustness against different attack and connection strategies.
\cite{cite_cn_assortativity_decreases_robustness} shows how assortativity of a single network can decrease the robustness of the entire interdependent networks. \cite{cite_cn_dynamic_topological_models_to_predict_system_failure_from_cascade} describes dynamical models and topological network structures to quantitatively predict system failures due to cascades.
In \cite{cite_cn_assymmetry_lead_to_perculation_transition_and_node_and_its_peer_position}, authors study percolation dynamics in two layer systems and finds that damage due to a node failure in any layer depends on both on its position in the layer and its neighbour in their respective counterparts in other layers whereas percolation transition depends on degree of asymmetry. \cite{cite_cn_multiplex_network_cascade_failures_emergent_coupling} develops a model of cascade failures in multiplex networks of layers with weight heterogeneity and find that catastrophic failures are due to collective effect of coupling of layers especially the weak ones.

A mechanism to mitigate cascading failures in complex networks by protecting some of the critical nodes have been proposed in \cite{cite_cn_mitigate_cascade_failures_protect_critical_nodes_and_local_structure}.
\cite{cite_cn_cascade_failures_avoided_with_reinforced_nodes} shows how reinforcement of nodes can avoid cascading failures in interdependent networks.
An evolutionary optimization approach in dynamic complex network for resilience against cascading failure has been proposed in \cite{cite_cn_evolutionary_algo_for_cascade_failure_resilience}.
A self healing mechanism using local information at the failed nodes to restore the overall network has been described in \cite{cite_cn_self_heal_reconstruct_after_failure_using_local_info}.
In \cite{cite_cn_triple_points_help_restore_failures}, authors find that triple points play dominant role in restoring failures in complex networks.
Impact of sparse coupling and enhanced coupling probability to improve robustness against intentional attack on data network has been suggested in \cite{cite_cn_sparse_coupling_enhanced_coupling_prob_for_robustness_against_cascade}.
Dealing with cascade failures by combining and splitting layers in multiplex network is proposed in \cite{cite_cn_cascade_failures_avoid_with_split_and_combine_layers}.
\cite{cite_cn_node_cascade_node_community_cascade} studies robustness of multiplex networks in the event of node cascade failures and shows the impact on the network when random node fails during node community cascades.
\cite{cite_cn_recovery_coupling_from_recoveries_of_power_grid_failures} studies recovery coupling using data of restoration from failures of numerous power grids. Optimal interaction among subnetworks to enhance resilience to failures has been explained in \cite{cite_cn_optimal_subnetwork_interaction_for_resilience}.
Cascade failure propagation using different network topologies of scale-free networks with heterogeneous degree distributions along with intra and inter layer degree correlations, \cite{cite_cn_scale_free_interdependen_network_cascade_prevent_with_different_properties} shows that scale-free degree distribution, internal network assortativity and cross-network hub-to-hub connections are necessary to reduce large cascades in Bak-Tang-Wiesenfeld sandpile model. Considering two interdependent network, \cite{cite_cn_degree_heterogeneity_increase_chances_of_failures} shows that degree heterogeneity increases the vulnerability to failures and enhanced coupling can make the network fragile towards targeted attacks.

From the above survey, it is evident that major body of research has concentration on cascading failures in MLCNs, targeted attacks, cyberattacks, etc. Several works have tried to understand the reason behind the failures and their consequences, how failures propagate and prediction models. Some of the works have emphasized on load balancing when failures occur and ways to mitigate, restore and reduce the impact of such events. The paper considers a MLCN where each layer is dependent on underlying one to connect to a node and number of edges increase exponentially from lower layer to the uppermost and each layer having different properties. It studies the emergent behaviour ASPL, TSPC and TNE when nodes and edges fail with decreasing order of their static and dynamic betweenness centralities. To the best of the knowledge of the authors, none of the previous works have considered this scenario.
\section{System Model}\label{section_model}
To study the impact of failures of high centrality nodes and edges in lower layers on higher ones, a MLCN model needs to be developed which is explained below. An MLCN of $N$ nodes consisting of $L_1$, $L_2$ and $L_3$ layers each having different proper characteristics is constructed.
\subsection{Construction of $L_1$}
$L_1$ layer is modeled as  a sparse Erdos-Renyi (ER) random graph. Each of the $N$ nodes in this layer has at least one edge. The graph is created in such as way that the shortest path hop count among the nodes is gaussian in nature. The edges in this layer can be imagined as physical connections which provides logical services to the higher layers. Edges in this layer and the nodes may fail which will impact the layers above, i.e., $L_2$ and $L_3$. Thus, $L_1$ has $N$ vertices and $E_{L_1}$ edges.
\subsection{Construction of $L_2$}
$L_2$ layer uses the services of $L_1$ to set up a logical connection between any two of the $N$ nodes. An $L_2$ exists if there is at least one underlying $L_1$ path. It is assumed that $L_2$ uses the shortest $L_1$ to set up the connection between two nodes. $L_2$ is also modeled as ER random graph using the same set of $N$ nodes with higher number of edges compared to $L_1$ but much lower than those of $L_3$.
Thus, $L_2$ has $N$ vertices and $E_{L_2}$ edges.
\subsection{Construction of $L_3$}
$L_3$ is constructed as a scale free graph with the same set of nodes. This essentially means that a few nodes have very high degree centrality and there are many more nodes which have much less of the same. $L_3$ network is used to set up direct end-to-end logical connection between two (remote) nodes in the network.
$L_3$ connects two end nodes in the network if there is $L_2$ (shortest) path between them. There are many more edges in $L_3$ than $L_2$. Thus, $L_3$ has $N$ vertices and $E_{L_3}$ edges.

\subsection{Edge and node failures}
Two types of failures are considered here, both nodes and edges can fail. A node failure leads to disruption in all the three layers with their respective edges being removed from the MLCN. If there are $E^{(i)}_{L_1}$, $E^{(i)}_{L_2}$ and $E^{(i)}_{L_3}$ number of edges from $L_1$, $L_2$ and $L_3$ layers attached to a node $V_i$, then failure of the node will lead to removal of $E^{(i)}_{L_1} + E^{(i)}_{L_2}$ + $E^{(i)}_{L_3}$ edges from the MLCN. For edge failure, only $L_1$ links can fail. This will obviously lead to redirection of $L_2$ services over a different (shortest) $L_1$ path. This redirection at $L_2$ may in turn cause a change in the (shortest) $L_2$ path for an $L_3$ edge. An $L_2$ or an $L_3$ connection (or edge) can fail if there is no path in the underlying $L_1$ and $L_2$ layers respectively. Thus, node failure can have a much more adverse impact on the functioning of the MLCN compared to a failed edge.

\subsection{Centralities}
Centralities of nodes and edges are critical attributes of MLCNs. Node betweenness centrality (NBC) and edge betweenness centrality (EBC) are both considered in this work.
\subsubsection{Node betweenness centrality}
NBC of a node is the sum of fraction of all pairs of shortest path between two other vertices that pass through it \cite{cite_centrality_definitions}. For layer ${L_k}, k = 1,2,3$, the NBC of $v_i \in V$,
\begin{equation}
NBC_{L_k}(v_i) = \sum_{x,y \in V}\frac{N^{(V)}_{L_k}(x,y|z)}{N^{(V)}_{L_k}(x,y)}
\end{equation}
where ${N^{(V)}_{L_k}(x,y)}$ is number of shortest paths between $x$ and $y$ and ${N^{(V)}_{L_k}(x,y|v_i)}$ is the number paths through node $v_i$ where $v_i \neq x$ and  $v_i \neq y$ .

Bigger the NBC of a node the higher is its importance in the network. For a MLCN, a node has its NBC specific to each layer, since there is no cross layer interaction in this model though there is dependency on underlying layer in terms of availing services. Thus, node $v_i$ will have NBCs of $d^{(i)}_{L_1}$, $d^{(i)}_{L_2}$ and $d^{(i)}_{L_3}$ for layers $L_1$, $L_2$ and $L_3$ respectively. Generally, if an edge fails then the NBCs of connecting nodes change. For the type of MLCN considered here, an edge failure in any layer changes the NBCs of the nodes only for that specific layer.
\subsubsection{Edge betweenness centrality}
EBC of an edge defines the sum of the fraction of all pairs of shortest paths that pass through it \cite{cite_centrality_definitions}. The higher the value of EBC for an edge the larger the number of geodesics that pass through it and more critical is link for the network. For layer ${L_k}$, the EBC of $e^{(j)}_{L_k}  \in E_{L_k}$,
\begin{equation}
EBC(e^{(j)}_{L_k}) =\sum_{x,y \in V}\frac{N^{(E)}_{L_k}(x, y|e^{(j)}_{L_k})}{N^{(E)}_{L_k}(x, y)}
\end{equation}
where ${N^{(E)}_{L_k}(x,y)}$ is number of shortest paths between $x$ and $y$ and ${N^{(E)}_{L_k}(x,y|e^{(j)}_{L_k})}$ is the number paths through edge $e^{(j)}_{L_k}$ .

For MLCN considered here, an edge of any layer will have its own EBC. Generally, if an edge or a node fails the geodesics changes. Since, there is no direct cross layer interaction, an edge failure only affects the EBC for that particular layer. If an edge $e^{(j)}_{L_2}$ fails then only the EBCs $e^{(l)}_{L_2} \in E_{L_2}\setminus \{e^{(j)}_{L_2}\}$ change and those of $L_1$ and $L_3$ remain the same. However, if a node $v_i$ fails, then $e^{(l)}_{L_k} \in E_{L_k}\setminus \{e^{(j)}_{L_k}\}$, $k = 1,2,3$ of all three layers change.
\subsubsection{Static and dynamic node betweenness centralities}
This paper considers nodes failures and edges failures only in $L_1$. As already mentioned, node failures affect edges from all the layers which are incident on them. In the context of this paper, SNBC means that nodes have their fixed values (based on the initial construction of the MLCN), arranged in descending order and then a certain number of nodes are assumed to fail at the same time. DNBC in the current context means that nodes have their NBC values (initialized at the time of construction) and then the node with  highest value fails at a time which leads to change in the NBCs in all layers. Then, in the next iteration another node with the current highest value fails which triggers further changes in NBC values. This process leads to dynamic changes in NBC values across all layers.
\subsubsection{Static and dynamic edge betweenness centralities}
Similar to two types of NBCs described above, EBCs can be static and dynamic. For SEBC, all the selected edges with decreasing centralities fail at the same time whereas for the DEBC one edge with the highest EBC currently fail at a time.
\subsection{Evaluation parameters}
As nodes and edges fail, the graph topology changes in each layer of the MLCN. The emergent behaviour of following three parameters are studied.
\subsubsection{Average shortest path length}
Average shortest path length (ASPL) is the mean of all the shortest paths between any two nodes. This is calculated for each layer separately as the graph changes due to a node or an edge failure.
\subsubsection{Total shortest path count}
Total shortest path count (TSPC) is the number of all the shortest paths between any two nodes. This is also calculated for each layer separately with the change in graph due to a node or an edge failure.
\subsubsection{Total number of edges}
Total number of edges (TNE) is the number of edges in each layer. This is calculated for each layer separately after a node or an edge failure.
\subsection{Dynamics of the MLCN}
The dynamic behaviour of the MLCN and evaluation of the emergent behaviour of ASPL, TSPC and TNEs in terms of failures of edges and nodes based on the four types of centralities is described with the following algorithms.
\subsubsection{Static edge betweennness centrality}
\textbf{Algorithm \ref{algo_static_sebc}} explains the dynamics of failures of edges based on SEBC. Lines 3-5 the $L_1$ is created as an ER graph with $N$ ensuring no node has zero edges. Then, the $L_2$ ER graph is created on the same set of $N$ nodes in lines 7-9 with higher number of edges compared to $L_1$. Scale free $L_3$ graph is created with much higher number of edges compared to $L_2$ in lines 11-13. Lines 15-18 and 20-23 calculates the ASPL, TSPC and TNE of $L_1$ and $L_2$ respectively. Line 25-26 calculates the TNE of $L_3$. Lines 28-29 and 31-32 calculates the EBCs of $L_1$ and then sorts them in descending order respectively. Lines 38-44 deletes the required number of edges with top EBCs. After deletion, the evaluation parameters of the three layers change and the new values of are again calculated in lines 46-57.
\begin{algorithm}
\scriptsize
\caption{Static edge betweennness centrality}\label{algo_static_sebc}
\begin{algorithmic}[1]
\Procedure{EvaluateStaticEBC()}{}
\\
\\\hspace{5mm}/* Create ER $L_1$ graph, ensure no node has 0 edges */
\\\hspace{5mm}$GL1 = L1CreateGraph(NumberOfVertices, $
\\\hspace{30mm}$L1EdgeProbability)$
\\
\\\hspace{5mm}/* Create ER $L_2$ graph */
\\\hspace{5mm}$GL2 = L2CreateGraph(NumberOfVertices, $
\\\hspace{30mm}$L3EdgeProbability)$
\\
\\\hspace{5mm}/* Create scale free $L_3$ graph */
\\\hspace{5mm}$GL3 = L3CreateGraph(NumberOfVertices, $
\\\hspace{30mm}$L4EdgeProbability)$
\\
\\\hspace{5mm}/* Calculate ASPL, TSPC and TNE of $L_1$ */
\\\hspace{5mm}$CalculateASPL(GL1)$
\\\hspace{5mm}$CalculateTSPC(GL1)$
\\\hspace{5mm}$CalculateTNE(GL1)$
\\
\\\hspace{5mm}/* Calculate ASPL, TSPC and TNE of $L_2$ */
\\\hspace{5mm}$CalculateASPL(GL2)$
\\\hspace{5mm}$CalculateTSPC(GL2)$
\\\hspace{5mm}$CalculateTNE(GL2)$
\\
\\\hspace{5mm}/* Calculate TNE of $L_3$*/
\\\hspace{5mm}$CalculateTNE(GL3)$
\\
\\\hspace{5mm}/* Calculate EBCs of $L_1$ edges */
\\\hspace{5mm}$EbcL1 = CalculateEBC(GL1)$
\\
\\\hspace{5mm}/* Sort EBCs of $L_1$ edges in descending order*/
\\\hspace{5mm}$EbcL1 = SortEBCDescending(GL1)$
\\
\\\hspace{5mm}/*
\\\hspace{5mm} * Remove top $EdgesToBeRemoved$
\\\hspace{5mm} * EBC $L_1$ edges
\\\hspace{5mm} */
\\\hspace{5mm}$for\ count\ =\ 1\ to\ EdgesToBeRemoved$
\\\hspace{5mm}$begin$
\\
\\\hspace{10mm}/* Delete edge in $L_1$ */
\\\hspace{10mm}$DeleteHighestEBCEdge(GL1, EbcL1)$
\\
\\\hspace{5mm}$endfor$
\\
\\\hspace{5mm}/* Calculate new ASPL, TSPC and TNE of $L_1$ */
\\\hspace{5mm}$CalculateASPL(GL1)$
\\\hspace{5mm}$CalculateTSPC(GL1)$
\\\hspace{5mm}$CalculateTNE(GL1)$
\\
\\\hspace{5mm}/* Calculate new ASPL, TSPC and TNE of $L_2$ */
\\\hspace{5mm}$CalculateASPL(GL2)$
\\\hspace{5mm}$CalculateTSPC(GL2)$
\\\hspace{5mm}$CalculateTNE(GL2)$
\\
\\\hspace{5mm}/* Calculate new TNE of $L_3$*/
\\\hspace{5mm}$CalculateTNE(GL3)$
\\
\EndProcedure
\end{algorithmic}
\end{algorithm}
\subsubsection{Dynamic edge betweennness centrality}
\textbf{Algorithm \ref{algo_dynamic_debc}} explains the dynamics of failures of edges based on DEBC. In this case, $L_1$ graph is created only once and progressively edges fail. Lines 3-5, 7-9 and 11-13 create the $L_1$, $L_2$ and $L_3$ respectively similar to SEBC. Lines 15-60 following things happen in a loop. Firstly, lines 19-30 calculate ASPLs, TSPCs of $L_1$ and $L_2$, and TNEs of all three layers. Lines 32-33 calculate the EBC of $L_1$ and then the edge with highest EBC is deleted in lines 35-36. Again, the parameters are again calculated for the changed graph in lines 38-49.  Lines 49-57 reconstruct new $L_2$ and $L_3$ graph for the next iteration.
\begin{algorithm}
\scriptsize
\caption{Dynamic edge betweennness centrality}\label{algo_dynamic_debc}
\begin{algorithmic}[1]
\Procedure{EvaluateDynamicEBC()}{}
\\
\\\hspace{5mm}/* Create ER $L_1$ graph, ensure no node has 0 edges */
\\\hspace{5mm}$GL1 = L1CreateGraph(NumberOfVertices, $
\\\hspace{30mm}$L1EdgeProbability)$
\\
\\\hspace{5mm}/* Create ER $L_2$ graph */
\\\hspace{5mm}$GL2 = L2CreateGraph(NumberOfVertices, $
\\\hspace{30mm}$L3EdgeProbability)$
\\
\\\hspace{5mm}/* Create scale free $L_3$ graph */
\\\hspace{5mm}$GL3 = L3CreateGraph(NumberOfVertices, $
\\\hspace{30mm}$L4EdgeProbability)$
\\
\\\hspace{5mm}/* Remove each $L_1$ edge one at a time */
\\\hspace{5mm}$for\ count\ =\ 1\ to\ EdgesToBeRemoved$
\\\hspace{5mm}$begin$
\\
\\\hspace{10mm}/* Calculate ASPL, TSPC and TNE of $L_1$ */
\\\hspace{10mm}$CalculateASPL(GL1)$
\\\hspace{10mm}$CalculateTSPC(GL1)$
\\\hspace{10mm}$CalculateTNE(GL1)$
\\
\\\hspace{10mm}/* Calculate ASPL, TSPC and TNE of $L_2$ */
\\\hspace{10mm}$CalculateASPL(GL2)$
\\\hspace{10mm}$CalculateTSPC(GL2)$
\\\hspace{10mm}$CalculateTNE(GL2)$
\\
\\\hspace{10mm}/* Calculate TNE of $L_3$*/
\\\hspace{10mm}$CalculateTNE(GL3)$
\\
\\\hspace{10mm}/* Calculate EBCs of $L_1$ edges */
\\\hspace{10mm}$EbcL1 = CalculateEBC(GL1)$
\\
\\\hspace{10mm}/* Delete edge with highest EBC in $L_1$ */
\\\hspace{10mm}$DeleteHighestEBCEdge(GL1, EbcL1)$
\\
\\\hspace{10mm}/* Calculate ASPL, TSPC and TNE of $L_1$ */
\\\hspace{10mm}$CalculateASPL(GL1)$
\\\hspace{10mm}$CalculateTSPC(GL1)$
\\\hspace{10mm}$CalculateTNE(GL1)$
\\
\\\hspace{10mm}/* Calculate ASPL, TSPC and TNE of $L_2$ */
\\\hspace{10mm}$CalculateASPL(GL2)$
\\\hspace{10mm}$CalculateTSPC(GL2)$
\\\hspace{10mm}$CalculateTNE(GL2)$
\\
\\\hspace{10mm}/* Calculate TNE of $L_3$*/
\\\hspace{10mm}$CalculateTNE(GL3)$
\\
\\\hspace{10mm}/* Create ER $L_2$ graph */
\\\hspace{10mm}$GL2 = L2CreateGraph(NumberOfVertices, $
\\\hspace{35mm}$L2EdgeProbability)$
\\
\\\hspace{10mm}/* Create scale free $L_3$ graph */
\\\hspace{10mm}$GL3 = L3CreateGraph(NumberOfVertices, $
\\\hspace{35mm}$L3EdgeProbability)$
\\
\\
\\\hspace{5mm}$endfor$
\EndProcedure
\end{algorithmic}
\end{algorithm}
\subsubsection{Static node betweennness centrality}
\textbf{Algorithm \ref{algo_static_snbc}} explains the dynamics of failures of nodes based on SNBC. This algorithm works same as \textbf{Algorithm \ref{algo_static_sebc}} from 3-26. Lines 28-29 calculate the NBCs of the $N$ nodes. These are sorted in descending order in lines 31-32. The required number of nodes and their associated $L_1$ are deleted in lines 38-47. If the nodes have any $L_2$ and $L_3$ edges they are also deleted.
\begin{algorithm}
\scriptsize
\caption{Static node betweennness centrality}\label{algo_static_snbc}
\begin{algorithmic}[1]
\Procedure{EvaluateStaticNBC()}{}
\\
\\\hspace{5mm}/* Create ER $L_1$ graph, ensure no node has 0 edges */
\\\hspace{5mm}$GL1 = L1CreateGraph(NumberOfVertices, $
\\\hspace{30mm}$L1EdgeProbability)$
\\
\\\hspace{5mm}/* Create ER $L_2$ graph */
\\\hspace{5mm}$GL2 = L2CreateGraph(NumberOfVertices, $
\\\hspace{30mm}$L3EdgeProbability)$
\\
\\\hspace{5mm}/* Create scale free $L_3$ graph */
\\\hspace{5mm}$GL3 = L3CreateGraph(NumberOfVertices, $
\\\hspace{30mm}$L4EdgeProbability)$
\\
\\\hspace{5mm}/* Calculate ASPL, TSPC and TNE of $L_1$ */
\\\hspace{5mm}$CalculateASPL(GL1)$
\\\hspace{5mm}$CalculateTSPC(GL1)$
\\\hspace{5mm}$CalculateTNE(GL1)$
\\
\\\hspace{5mm}/* Calculate ASPL, TSPC and TNE of $L_2$ */
\\\hspace{5mm}$CalculateASPL(GL2)$
\\\hspace{5mm}$CalculateTSPC(GL2)$
\\\hspace{5mm}$CalculateTNE(GL2)$
\\
\\\hspace{5mm}/* Calculate TNE of $L_3$*/
\\\hspace{5mm}$CalculateTNE(GL3)$
\\
\\\hspace{5mm}/* Calculate NBCs of $L_1$ edges */
\\\hspace{5mm}$NbcL1 = CalculateNBC(GL1)$
\\
\\\hspace{5mm}/* Sort NBCs of $L_1$ edges in descending order*/
\\\hspace{5mm}$NbcL1= SortNBCDescending(NbcL1)$
\\
\\\hspace{5mm}/*
\\\hspace{5mm} * Remove top $NodesToBeRemoved$
\\\hspace{5mm} * NBC $L_1$ nodes
\\\hspace{5mm} */
\\\hspace{5mm}$for\ count\ =\ 1\ to\ NodesToBeRemoved$
\\\hspace{5mm}$begin$
\\
\\\hspace{10mm}/* Delete node and its edges in $L_1$, $L_2$ and $L_3$ */
\\\hspace{10mm}$Node = DeleteHighestNBCNode(GL1, NbcL1)$
\\\hspace{10mm}$DeleteEdgesOfNode(GL1, Node)$
\\\hspace{10mm}$DeleteEdgesOfNode(GL2, Node)$ /* if any */
\\\hspace{10mm}$DeleteEdgesOfNode(GL3, Node)$ /* if any */
\\
\\\hspace{5mm}$endfor$
\\
\\\hspace{5mm}/* Calculate new ASPL, TSPC and TNE of $L_1$ */
\\\hspace{5mm}$CalculateASPL(GL1)$
\\\hspace{5mm}$CalculateTSPC(GL1)$
\\\hspace{5mm}$CalculateTNE(GL1)$
\\
\\\hspace{5mm}/* Calculate new ASPL, TSPC and TNE of $L_2$ */
\\\hspace{5mm}$CalculateASPL(GL2)$
\\\hspace{5mm}$CalculateTSPC(GL2)$
\\\hspace{5mm}$CalculateTNE(GL2)$
\\
\\\hspace{5mm}/* Calculate new TNE of $L_3$*/
\\\hspace{5mm}$CalculateTNE(GL3)$
\\
\EndProcedure
\end{algorithmic}
\end{algorithm}
\subsubsection{Dynamic node betweennness centrality}
\textbf{Algorithm \ref{algo_dynamic_dnbc}} explains the dynamics of failures of nodes based on DNBC. This algorithm works same as \textbf{Algorithm \ref{algo_dynamic_debc}}. Firstly, the three layers of graphs are created in lines 3-13. Then in a loop, the three parameters of ASPL, TSPC and TNE are calculated in lines 19-33, highest NBC is calculated in in line 33, the highest NBC node is deleted along with the $L_1$ edges in lines 39-40. If there are any $L_2$ and $L_3$ edges with the node then they are also deleted in lines 40-41. Again, the three parameters are calculate to know their change in lines 44-55. Then, new $L_2$ and $L_3$ graph are created for the next iteration in lines 57-63.
\begin{algorithm}
\scriptsize
\caption{Dynamic node betweennness centrality}\label{algo_dynamic_dnbc}
\begin{algorithmic}[1]
\Procedure{EvaluateDynamicEBC()}{}
\\
\\\hspace{5mm}/* Create ER $L_1$ graph, ensure no node has 0 edges */
\\\hspace{5mm}$GL1 = L1CreateGraph(NumberOfVertices, $
\\\hspace{30mm}$L1EdgeProbability)$
\\
\\\hspace{5mm}/* Create ER $L_2$ graph */
\\\hspace{5mm}$GL2 = L2CreateGraph(NumberOfVertices, $
\\\hspace{30mm}$L3EdgeProbability)$
\\
\\\hspace{5mm}/* Create scale free $L_3$ graph */
\\\hspace{5mm}$GL3 = L3CreateGraph(NumberOfVertices, $
\\\hspace{30mm}$L4EdgeProbability)$
\\
\\\hspace{5mm}/* Remove each $L_1$ node one at a time */
\\\hspace{5mm}$for\ count\ =\ 1\ to\ NodesToBeRemoved$
\\\hspace{5mm}$begin$
\\
\\\hspace{10mm}/* Calculate ASPL, TSPC and TNE of $L_1$ */
\\\hspace{10mm}$CalculateASPL(GL1)$
\\\hspace{10mm}$CalculateTSPC(GL1)$
\\\hspace{10mm}$CalculateTNE(GL1)$
\\
\\\hspace{10mm}/* Calculate ASPL, TSPC and TNE of $L_2$ */
\\\hspace{10mm}$CalculateASPL(GL2)$
\\\hspace{10mm}$CalculateTSPC(GL2)$
\\\hspace{10mm}$CalculateTNE(GL2)$
\\
\\\hspace{10mm}/* Calculate TNE of $L_3$*/
\\\hspace{10mm}$CalculateTNE(GL3)$
\\
\\\hspace{10mm}/* Calculate NBCs of $L_1$ edges */
\\\hspace{10mm}$NbcL1 = CalculateNBC(GL1)$
\\
\\\hspace{10mm}/*
\\\hspace{10mm} * Delete node and its edges
\\\hspace{10mm} * with highest NBC in $L_1$
\\\hspace{10mm} */
\\\hspace{10mm}$Node = DeleteHighestNBCNode(GL1, NbcL1)$
\\\hspace{10mm}$DeleteEdgesOfNode(GL1, Node)$
\\\hspace{10mm}$DeleteEdgesOfNode(GL2, Node)$ /* if any */
\\\hspace{10mm}$DeleteEdgesOfNode(GL3, Node)$ /* if any */
\\
\\\hspace{10mm}/* Calculate ASPL, TSPC and TNE of $L_1$ */
\\\hspace{10mm}$CalculateASPL(GL1)$
\\\hspace{10mm}$CalculateTSPC(GL1)$
\\\hspace{10mm}$CalculateTNE(GL1)$
\\
\\\hspace{10mm}/* Calculate ASPL, TSPC and TNE of $L_2$ */
\\\hspace{10mm}$CalculateASPL(GL2)$
\\\hspace{10mm}$CalculateTSPC(GL2)$
\\\hspace{10mm}$CalculateTNE(GL2)$
\\
\\\hspace{10mm}/* Calculate TNE of $L_3$*/
\\\hspace{10mm}$CalculateTNE(GL3)$
\\
\\\hspace{10mm}/* Create ER $L_2$ graph */
\\\hspace{10mm}$GL2 = L2CreateGraph(NumberOfVertices, $
\\\hspace{35mm}$L3EdgeProbability)$
\\
\\\hspace{10mm}/* Create scale free $L_3$ graph */
\\\hspace{10mm}$GL3 = L3CreateGraph(NumberOfVertices, $
\\\hspace{35mm}$L4EdgeProbability)$
\\
\\\hspace{5mm}$endfor$
\EndProcedure
\end{algorithmic}
\end{algorithm}
\section{Results}\label{section_results}
This section presents the results of emergent patterns of the three parameters ASPL, TSPC and TNE due to edge and node failures based on their centralities in the MLCN model. The simulation model is implemented using the \verb|python networkx| library.
Number of nodes $N$ is fixed at 100.

\emph{All the parameters in the plots are normalized between 0 and 1. Thus, a 0 value along y-axis does not mean 0 in absolute terms. This is necessary because the range of parameters are very different and the objective is to compare their behaviours rather than their absolute values.}

\subsection{Construction of $L_1$ graph}
\begin{figure}[ht]
\centering
\includegraphics[width=\columnwidth]{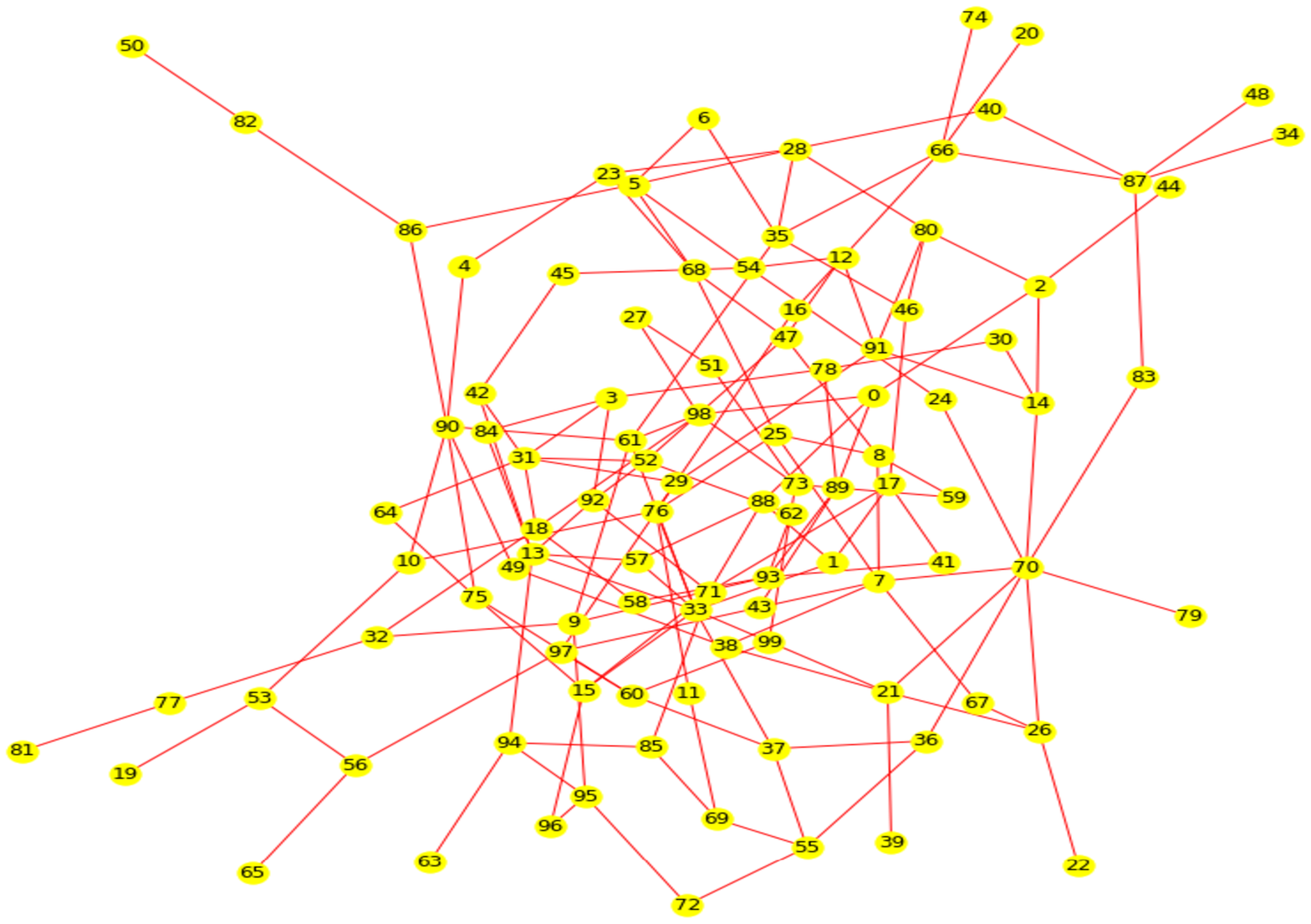}
\caption{Example $L_1$ graph}
\label{fig_plot_l1_graph_ink}
\end{figure}

\begin{figure}[ht]
\centering
\includegraphics[width=\columnwidth]{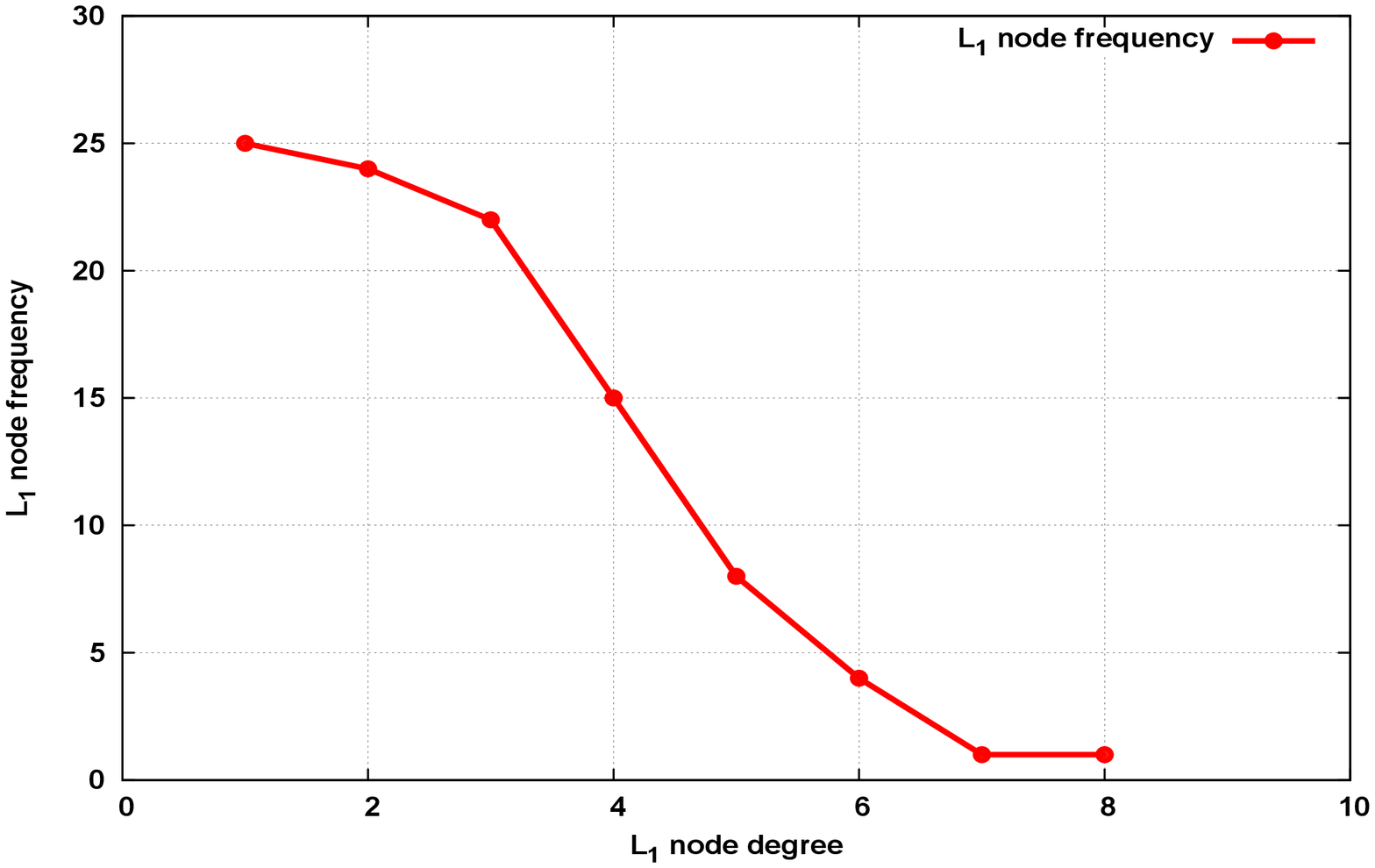}
\caption{$L_1$ node degree distribution}
\label{fig_plot_l1_node_freq_dist_ink}
\end{figure}

\begin{figure}[ht]
\centering
\includegraphics[width=\columnwidth]{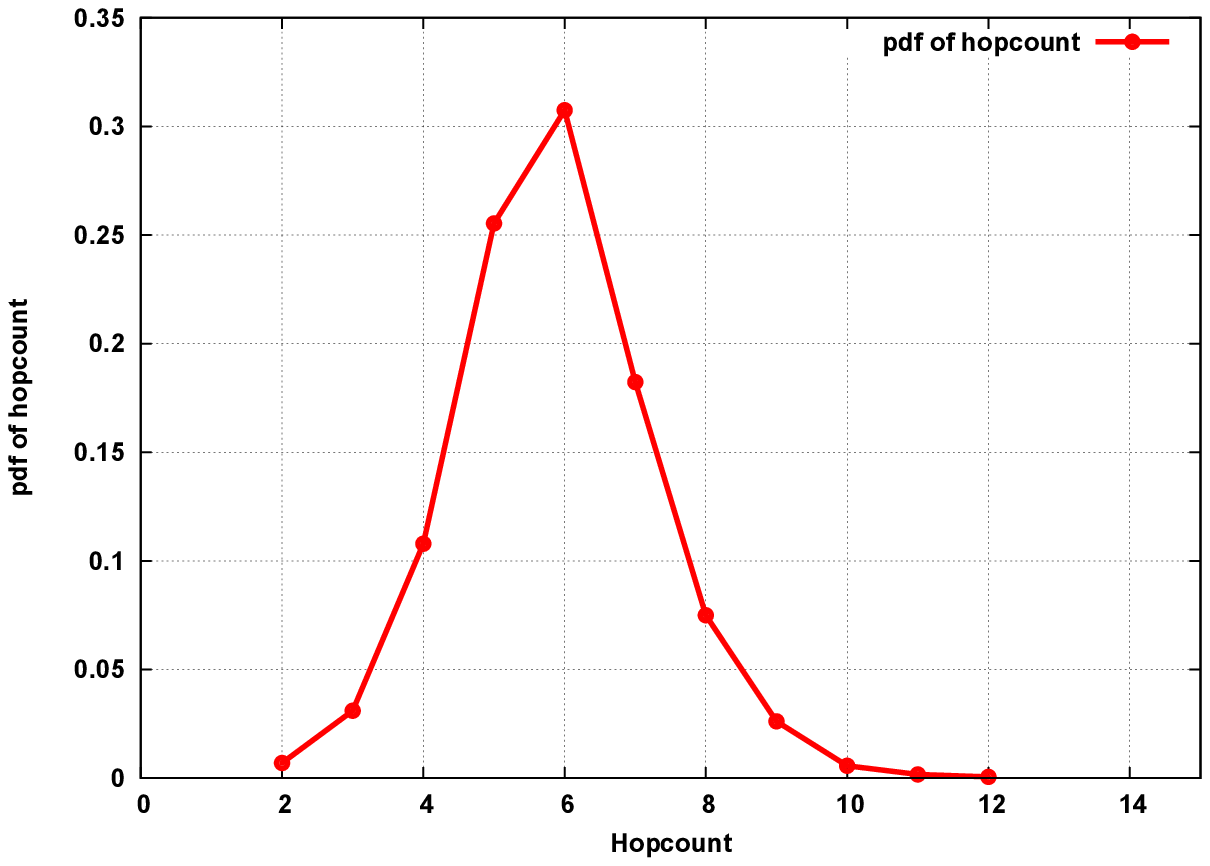}
\caption{Frequency distribution of hopcount of $L_1$}
\label{fig_plot_l1_hopcount_prob_ink}
\end{figure}

The key thing to ensure during the construction of $L_1$ graph is that the hopcount follows a gaussian distribution. For each $L_2$ edge, the shortest path in $L_1$ graph is found out first. Number of nodes in this shortest path is the hopcount. Fig. \ref{fig_plot_l1_graph_ink} shows such a $L_1$ graph. The degree distribution of the nodes is shown in Fig. \ref{fig_plot_l1_node_freq_dist_ink}.
Fig. \ref{fig_plot_l1_hopcount_prob_ink} shows the distribution of the hopcount. 
\subsection{Construction of $L_2$ graph}
\begin{figure}[ht]
\centering
\includegraphics[width=\columnwidth]{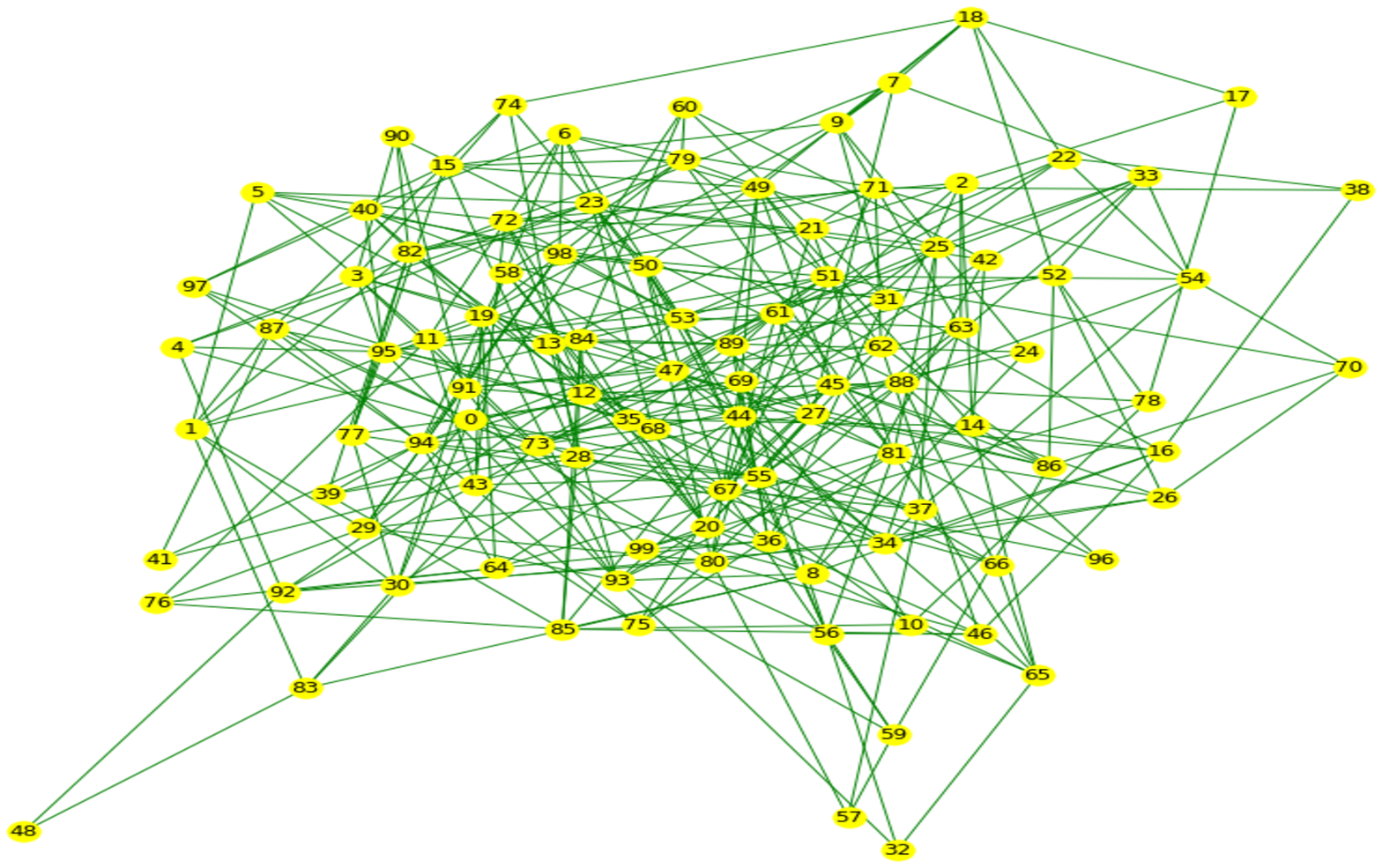}
\caption{$L_2$ graph}
\label{fig_plot_l2_graph_ink}
\end{figure}

\begin{figure}[ht]
\centering
\includegraphics[width=\columnwidth]{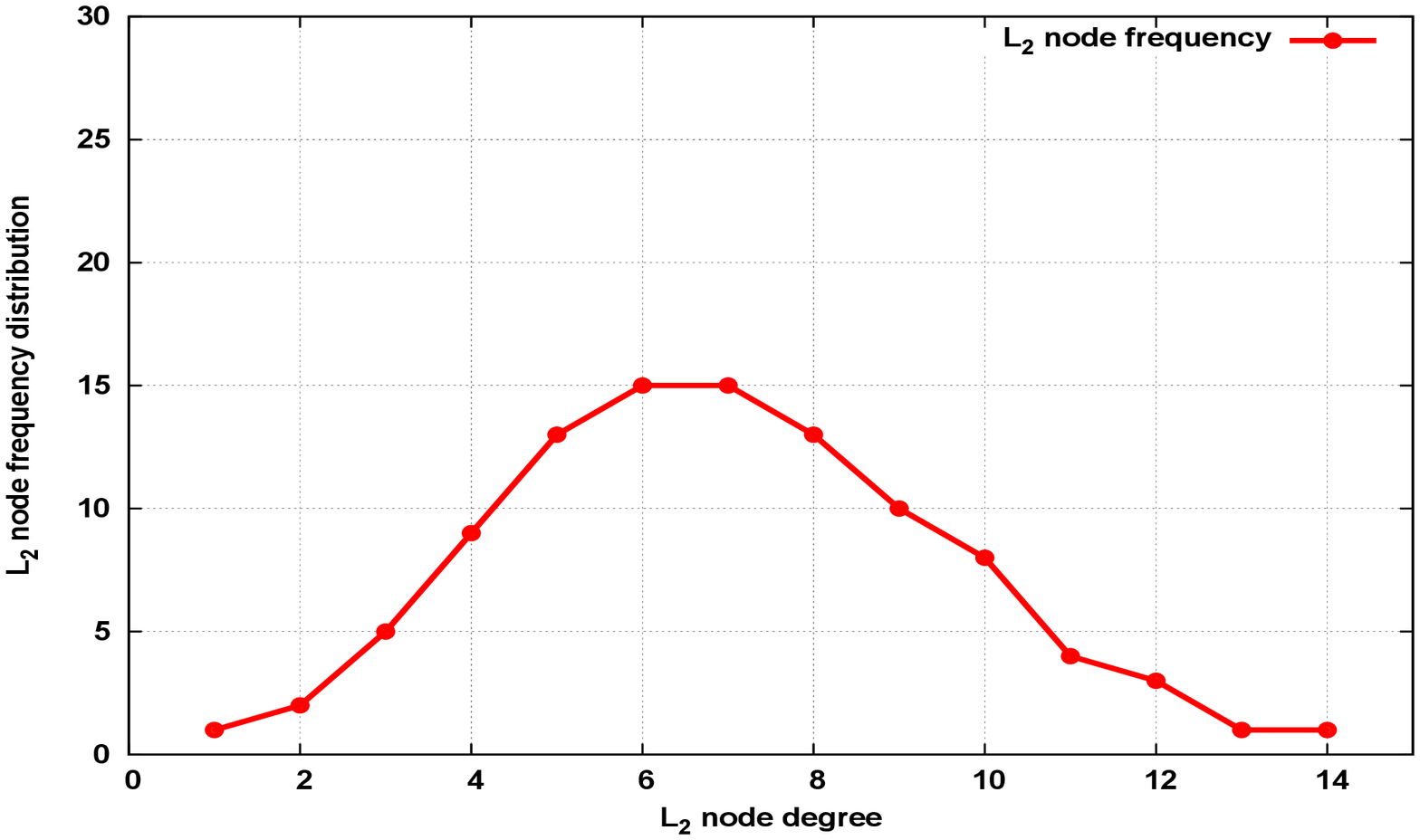}
\caption{$L_2$ degree distribution}
\label{fig_plot_l2_node_freq_dist_ink}
\end{figure}

$L_2$ is a ER random graph generated with addition condition that the number of edges should be substantially lower than $L_2$ (Fig. \ref{fig_plot_l2_graph_ink}). Fig. \ref{fig_plot_l2_node_freq_dist_ink} shows the degree distribution of $L_2$ which follows almost a gaussian function.
\subsection{Construction $L_3$ graph}
As mentioned above, the $L_3$ edges are end-to-end connections and the graph follows a scale-free distribution with some nodes having very high degrees and many more having lower degrees. 

\begin{figure}[ht]
\centering
\includegraphics[width=\columnwidth]{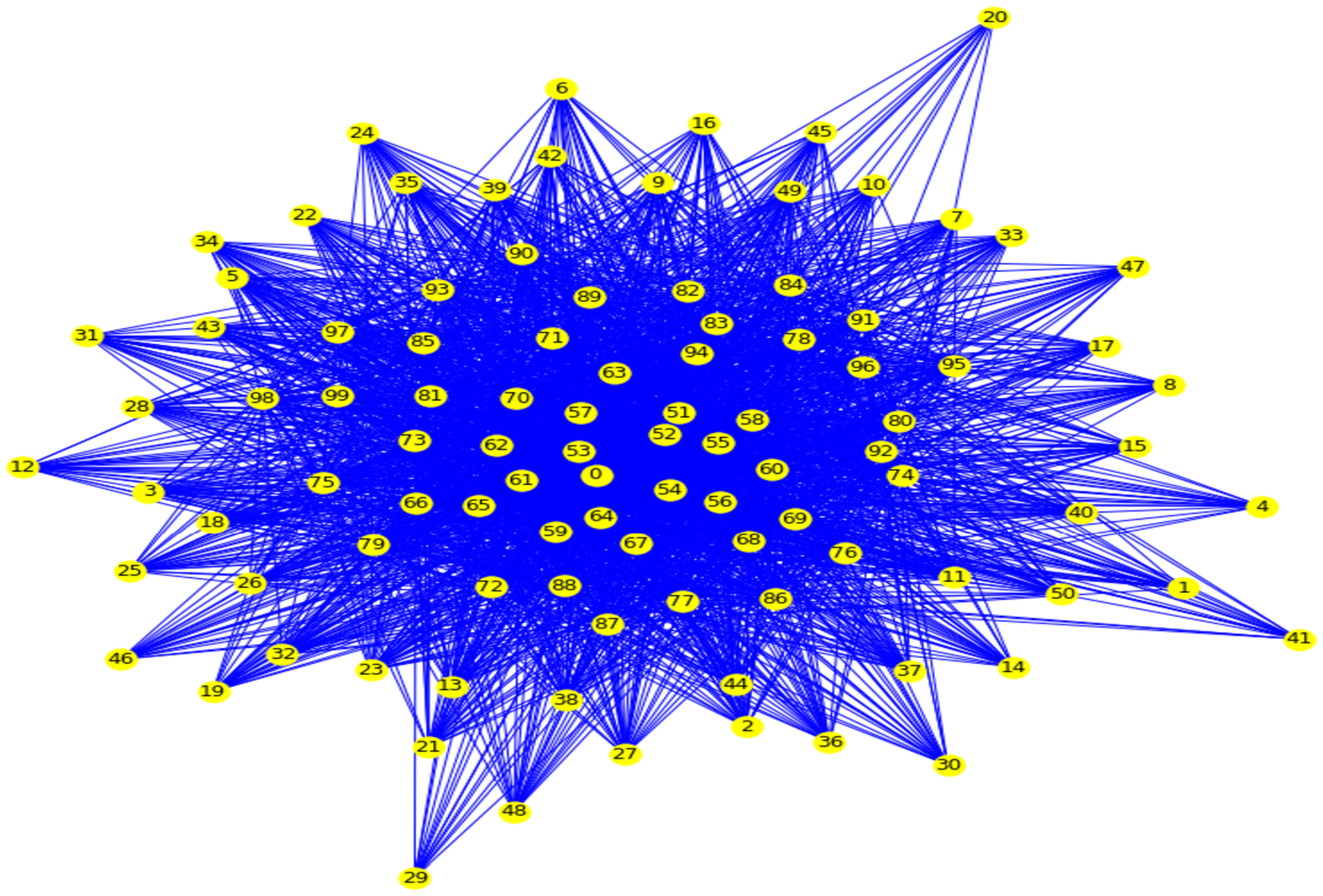}
\caption{$L_3$ graph}
\label{fig_plot_l3_graph_ink}
\end{figure}

\begin{figure}[ht]
\centering
\includegraphics[width=\columnwidth]{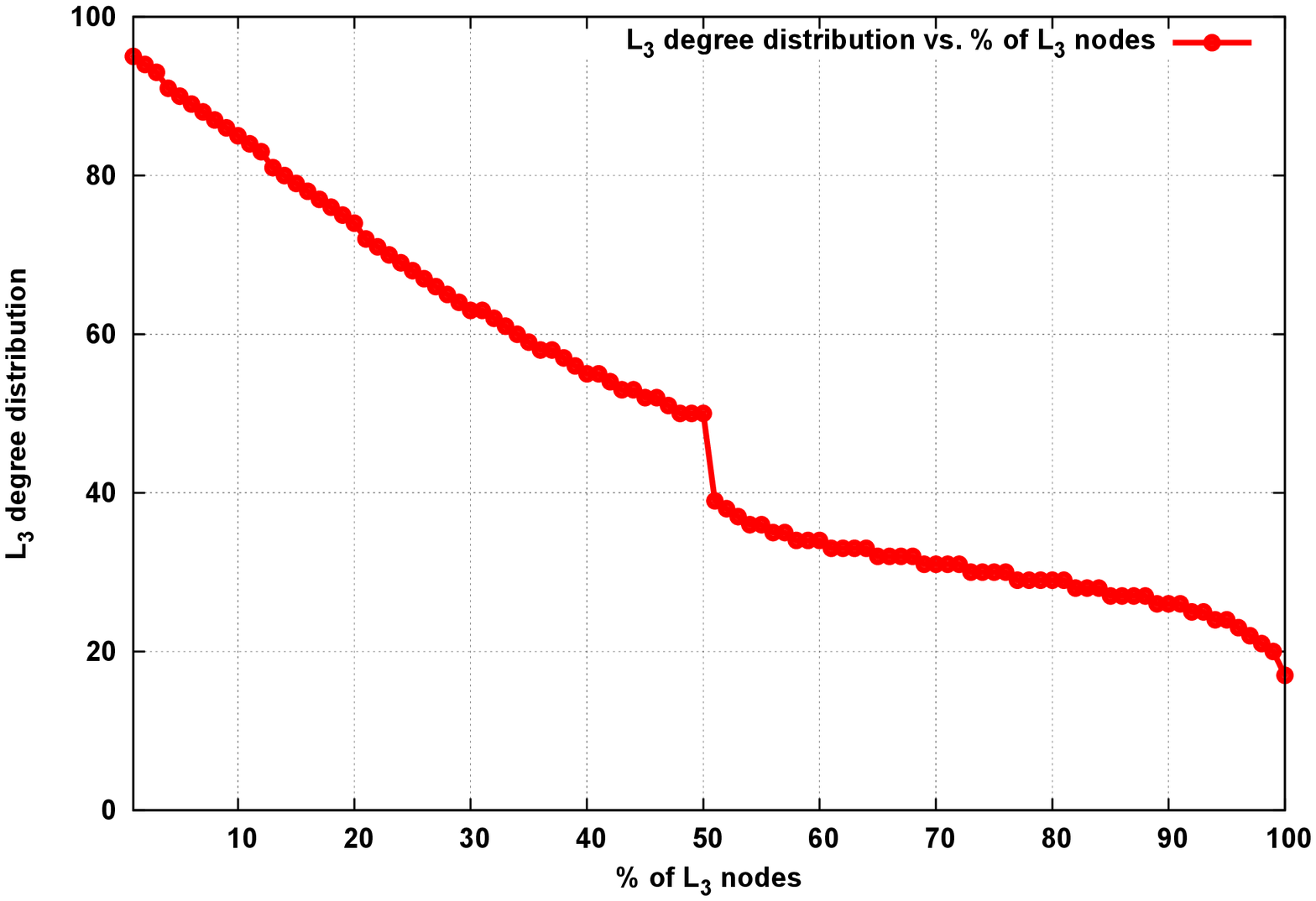}
\caption{Degree distribution of $L_3$ graph}
\label{fig_plot_l3_degree_dist_ink}
\end{figure}

A constructed graph is shown in \ref{fig_plot_l3_graph_ink}. The degrees distribution of the graph are shown in Fig. \ref{fig_plot_l3_degree_dist_ink} where \emph{x}-axis shows the percentage of nodes versus degrees along \emph{y}-axis. Around 15\% of the nodes have degrees higher than 80 and roughly 50\% of the nodes have below degrees below 40.

\begin{figure}[ht]
\centering
\includegraphics[width=\columnwidth]{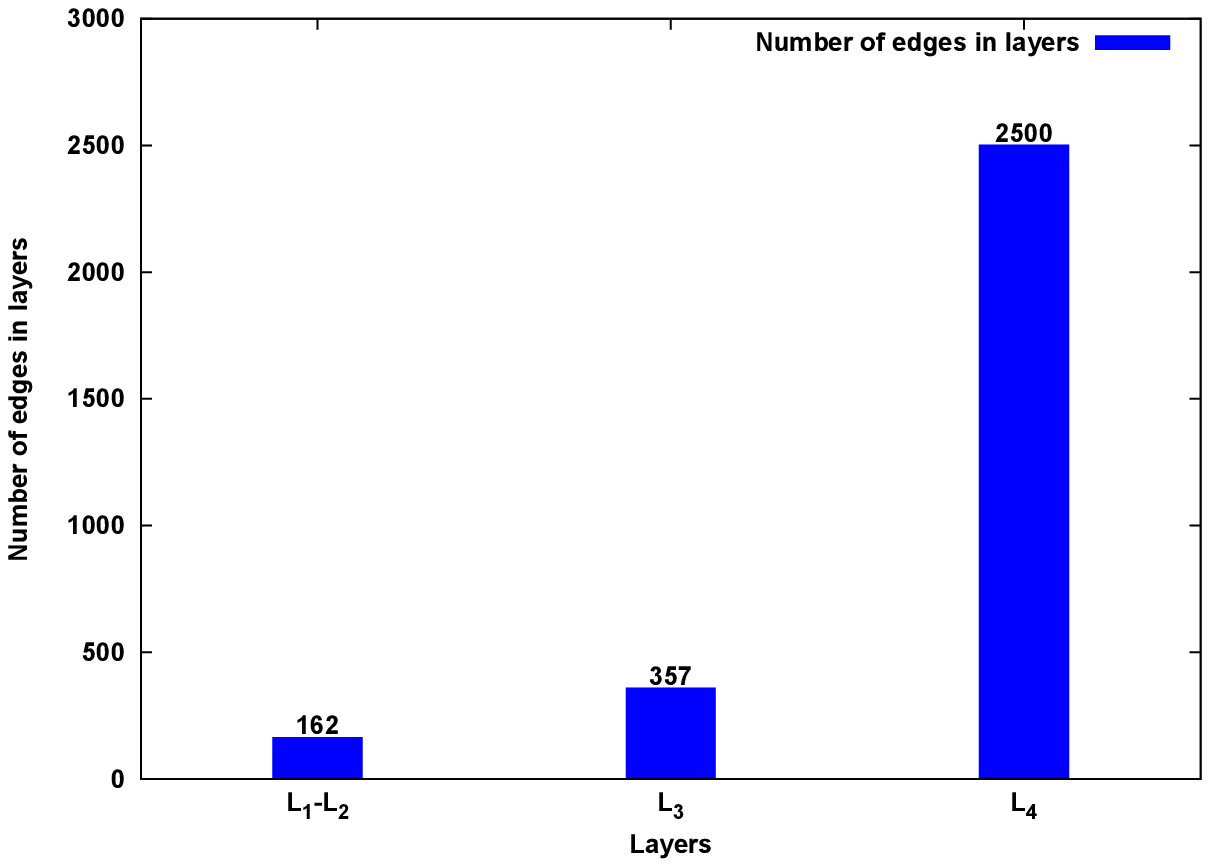}
\caption{Number of edges in all layers}
\label{fig_plot_number_of_edges_all_layers_ink}
\end{figure}

Thus, the MLCN have sufficiently increasing number of edges for the same number of nodes going from lower to higher layers. The number edges in all the layers is shown in Fig. \ref{fig_plot_number_of_edges_all_layers_ink}.
\subsection{SEBC behaviour}
In this scenario, each iteration new $L_1$, $L_2$ and $L_3$ graphs are constructed and then the required number of edges with highest centrality are failed in one go using \textbf{Algorithm \ref{algo_static_sebc}}.
For all the figures in this section, \emph{y}-axis shows the relevant parameters and \emph{x}-axis is always the number of failed edges.
\subsubsection{Behaviour of $L_1$ parameters}
\begin{figure}[ht]
\centering
\includegraphics[width=\columnwidth]{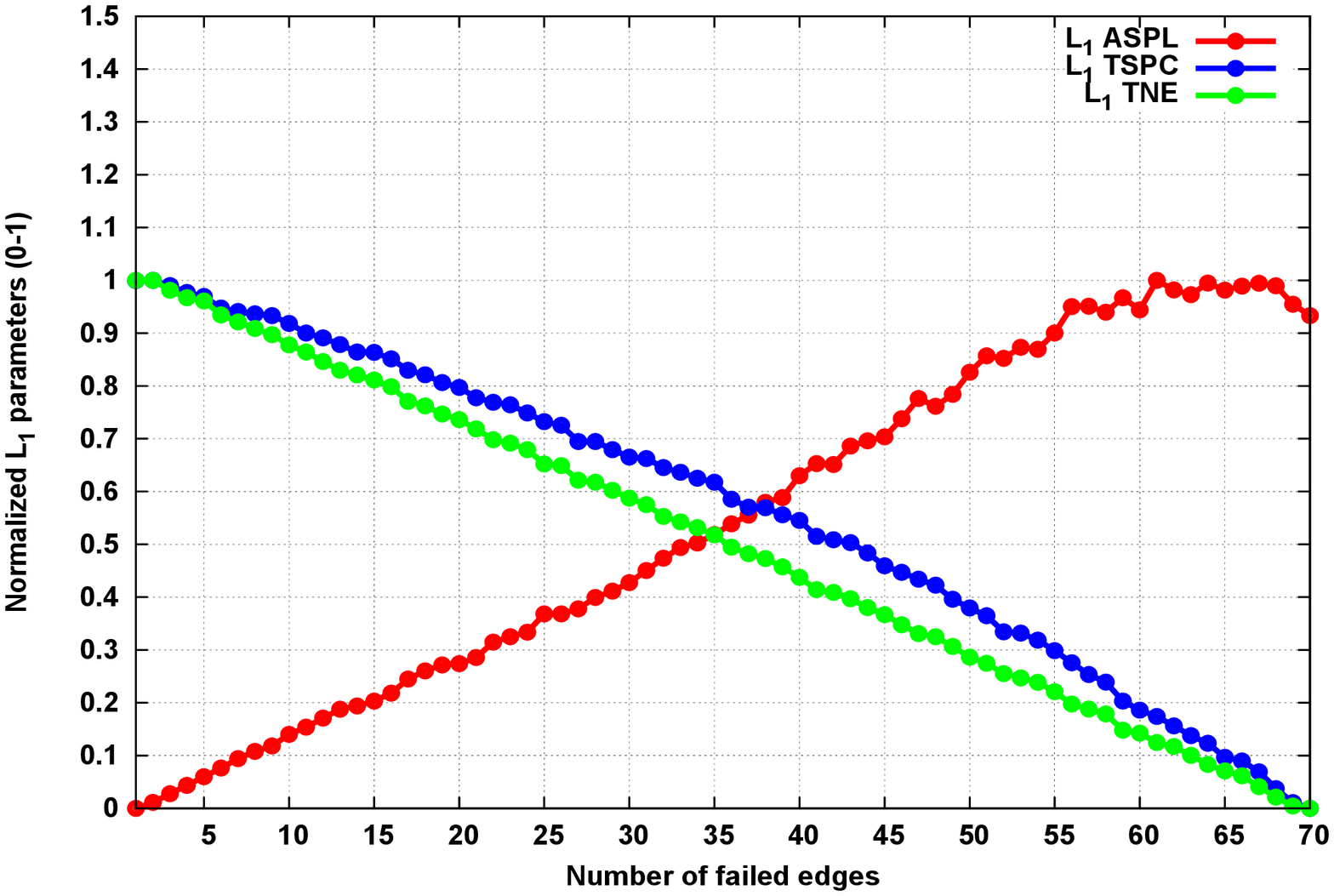}
\caption{SEBC - New $L_1$, $L_2$ and $L_3$ created. Behaviour of $L_1$ ASPL,
TSPC, and TNE vs. number of failed edges}
\label{fig_plot_edge_all_l1_l2_params_scaled_down_l1_l2_change_ink}
\end{figure}

Fig. \ref{fig_plot_edge_all_l1_l2_params_scaled_down_l1_l2_change_ink} shows the behaviour of $L_1$  ASPL, TSPC, TNE vs. the number of failed edges. $L_1$ TNE decrease linearly as expected. $L_1$ ASPL increases gradually almost linearly till it saturates and subsequently starts dropping with decrease in number of edges in the graph. $L_1$  TSPC decreases slowly and non-linearly as compared to $L_1$ TNE.
\subsubsection{Behaviour of $L_2$ parameters}
\begin{figure}[ht]
\centering
\includegraphics[width=\columnwidth]{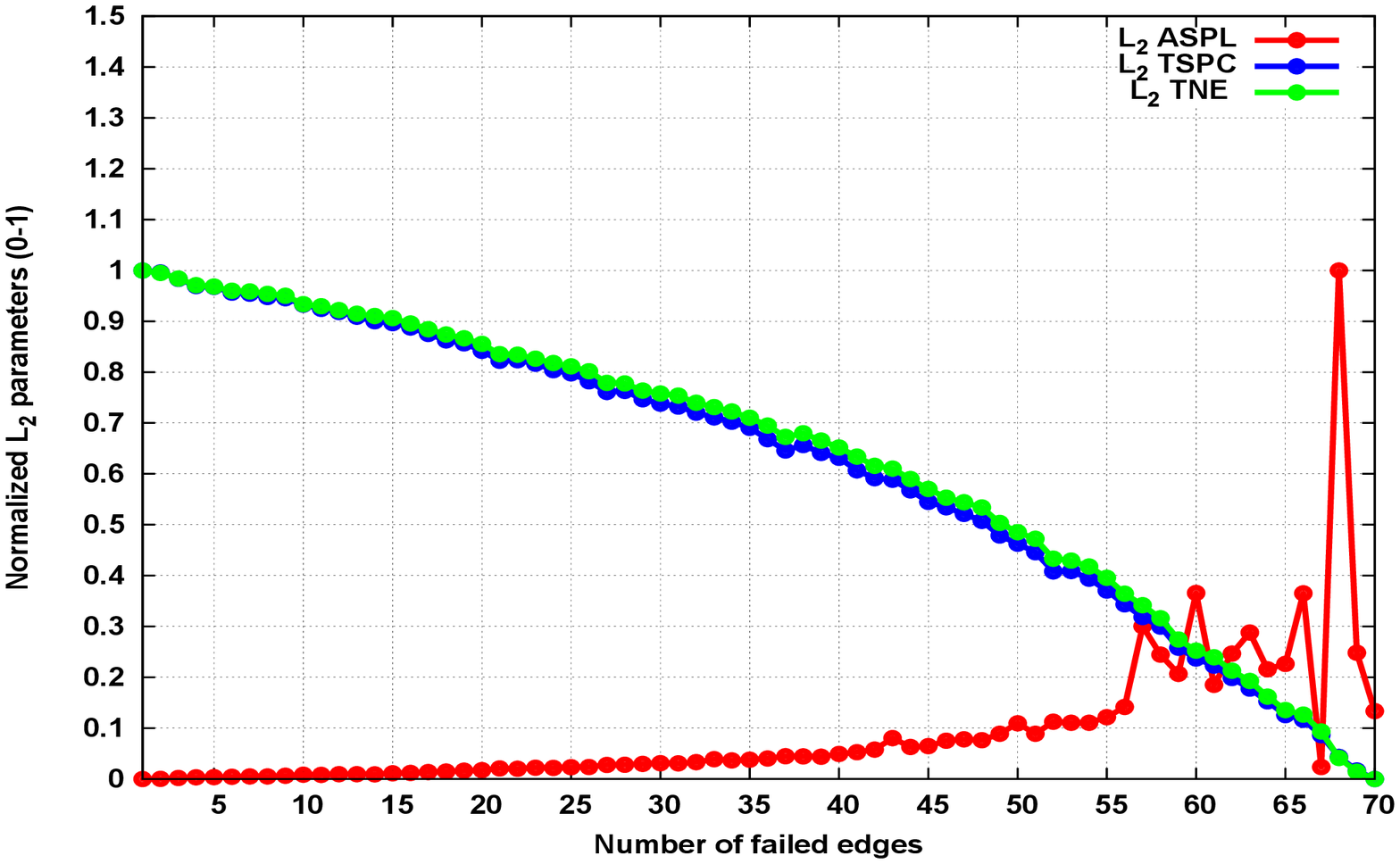}
\caption{SEBC - New $L_1$, $L_2$ and $L_3$ created. Behaviour of $L_2$ ASPL,
TSPC, and TNE vs. number of failed edges}
\label{fig_plot_edge_all_l3_params_scaled_down_l1_l2_change_ink}
\end{figure}

Fig. \ref{fig_plot_edge_all_l3_params_scaled_down_l1_l2_change_ink} shows the behaviour of $L_2$ ASPL, TSPC, TNE vs. the number of failed edges. $L_2$ ASPL increases progressively till 56 deleted edges and then shows chaotic behaviour. The chaotic behaviour is due to the fact that failure of certain edges cause much serious impact on the ASPL compared to others. $L_2$ TNE and $L_2$ TSPC show continuous decreasing trend.
\subsubsection{Behaviour of $L_3$ parameters}
\begin{figure}[ht]
\centering
\includegraphics[width=\columnwidth]{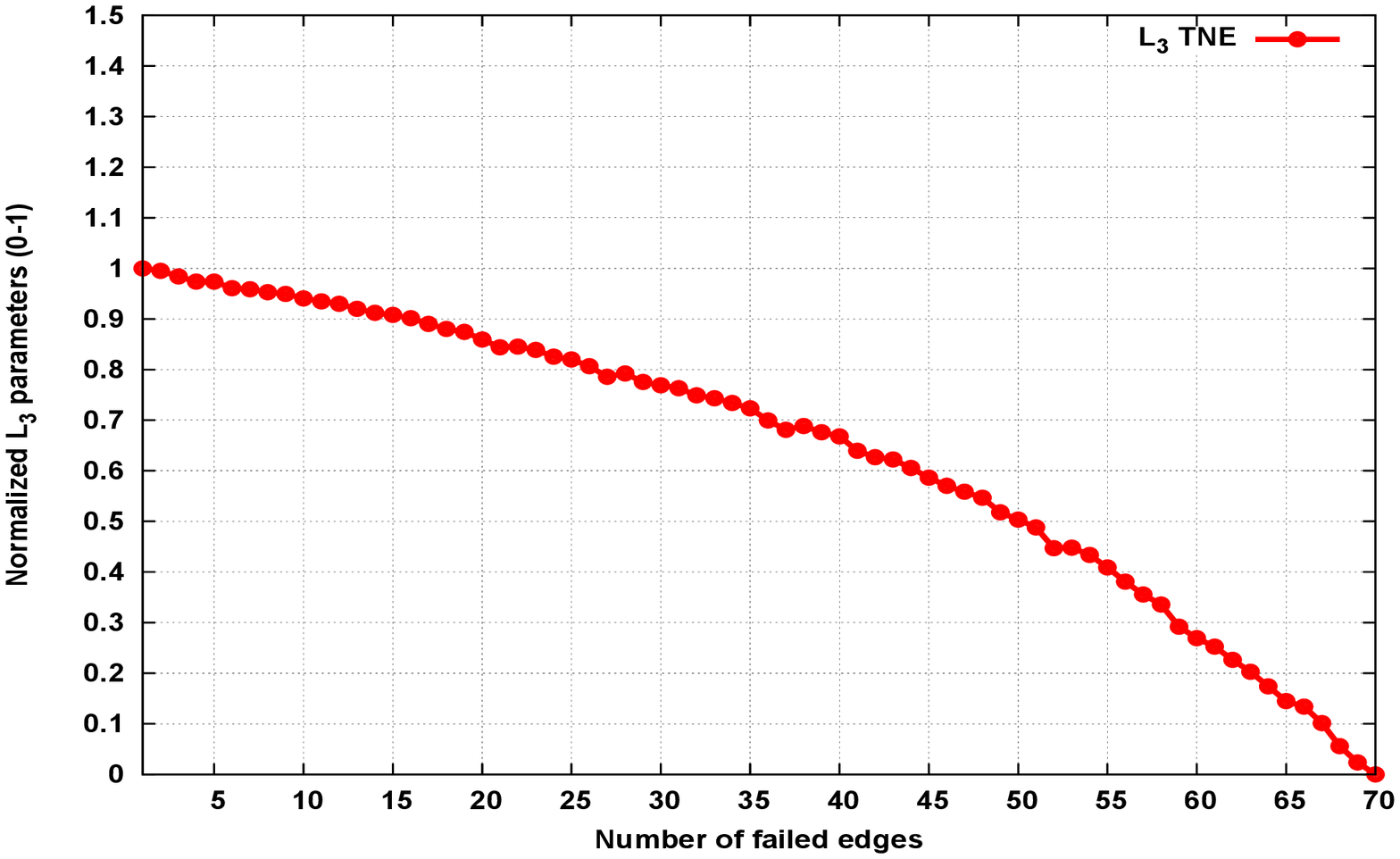}
\caption{SEBC - New $L_1$, $L_2$ and $L_3$ created. Behaviour of $L_3$ TNE vs. number of failed edges}
\label{fig_plot_edge_all_l4_params_scaled_down_l1_l2_change_ink}
\end{figure}

Fig. \ref{fig_plot_edge_all_l4_params_scaled_down_l1_l2_change_ink} shows the $L_3$ TNE which shows a decreasing curve.
\subsubsection{Comparison of ASPLs of $L_1$ and $L_2$}
\begin{figure}[ht]
\centering
\includegraphics[width=\columnwidth]{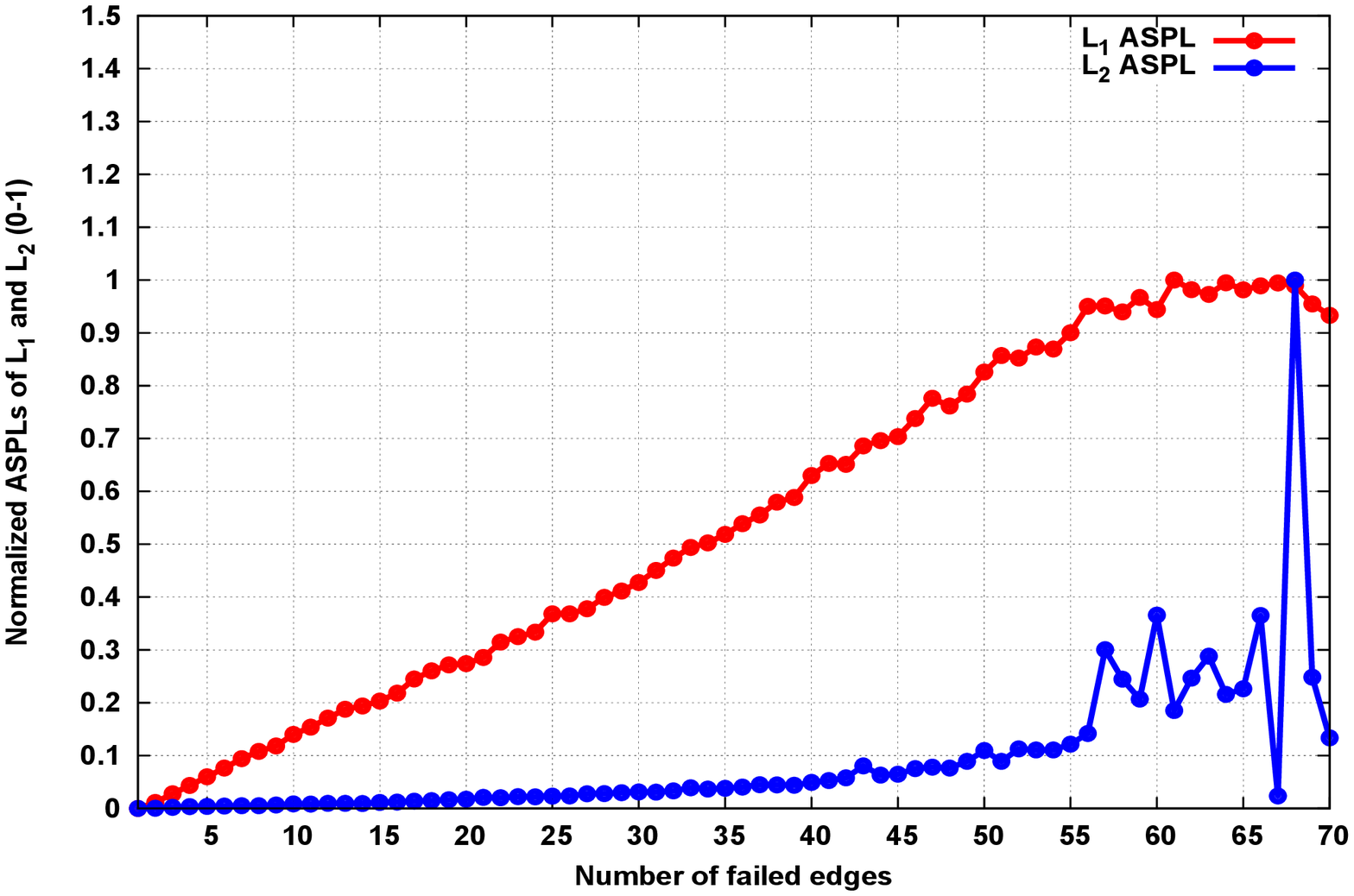}
\caption{SEBC - New $L_1$, $L_2$ and $L_3$ created. Comparison of ASPLs of $L_1$ and $L_2$ vs. number of failed edges}
\label{fig_plot_edge_avg_shortest_path_length_all_layers_scaled_down_l1_l2_change_ink}
\end{figure}

The behaviours of ASPL of $L_1$ and $L_2$ are shown in Fig. \ref{fig_plot_edge_avg_shortest_path_length_all_layers_scaled_down_l1_l2_change_ink}. $L_1$ ASPL increases much faster than that of $L_2$. For failure of 20 $L_1$ edges there is hardly any impact on $L_2$ ASPL. However, $L_2$ ASPL shows chaotic behaviour beyond 56 failed edges.
\subsubsection{Comparison of TSPCs of $L_1$ and $L_2$}
\begin{figure}[ht]
\centering
\includegraphics[width=\columnwidth]{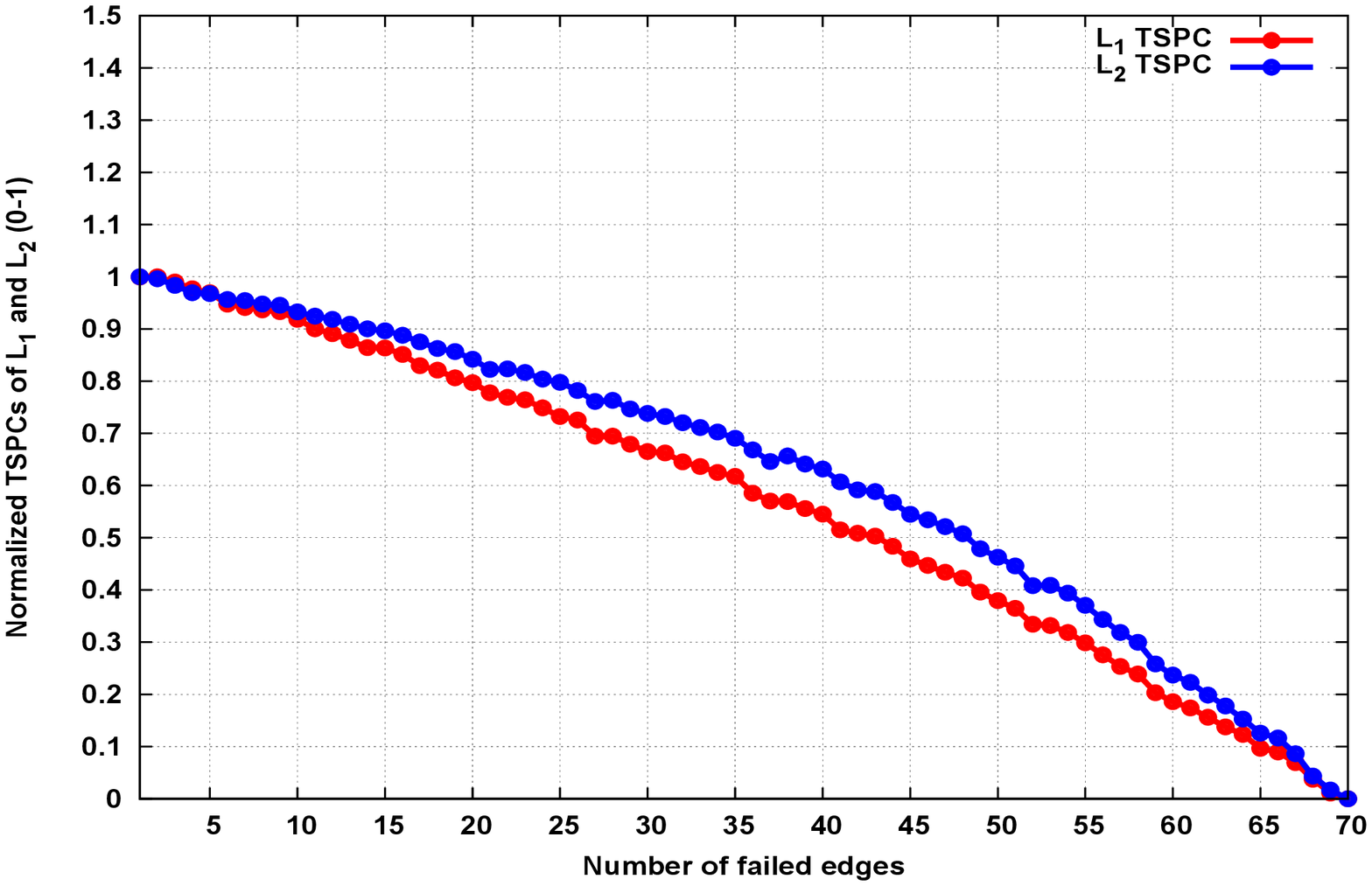}
\caption{SEBC - New $L_1$, $L_2$ and $L_3$ created. Comparison of TSPCs of $L_1$ and $L_2$ vs. number of failed edges}
\label{fig_plot_edge_total_shortest_paths_all_layers_scaled_down_l1_l2_change_ink}
\end{figure}

TSPCs of $L_1$ and $L_2$ shows almost similar non-linear behaviour with the former decreasing at a faster rate in the middle regions than the latter (Fig. \ref{fig_plot_edge_total_shortest_paths_all_layers_scaled_down_l1_l2_change_ink}). 
\subsubsection{Comparison of TNEs of $L_1$, $L_2$ and $L_3$}
\begin{figure}[ht]
\centering
\includegraphics[width=\columnwidth]{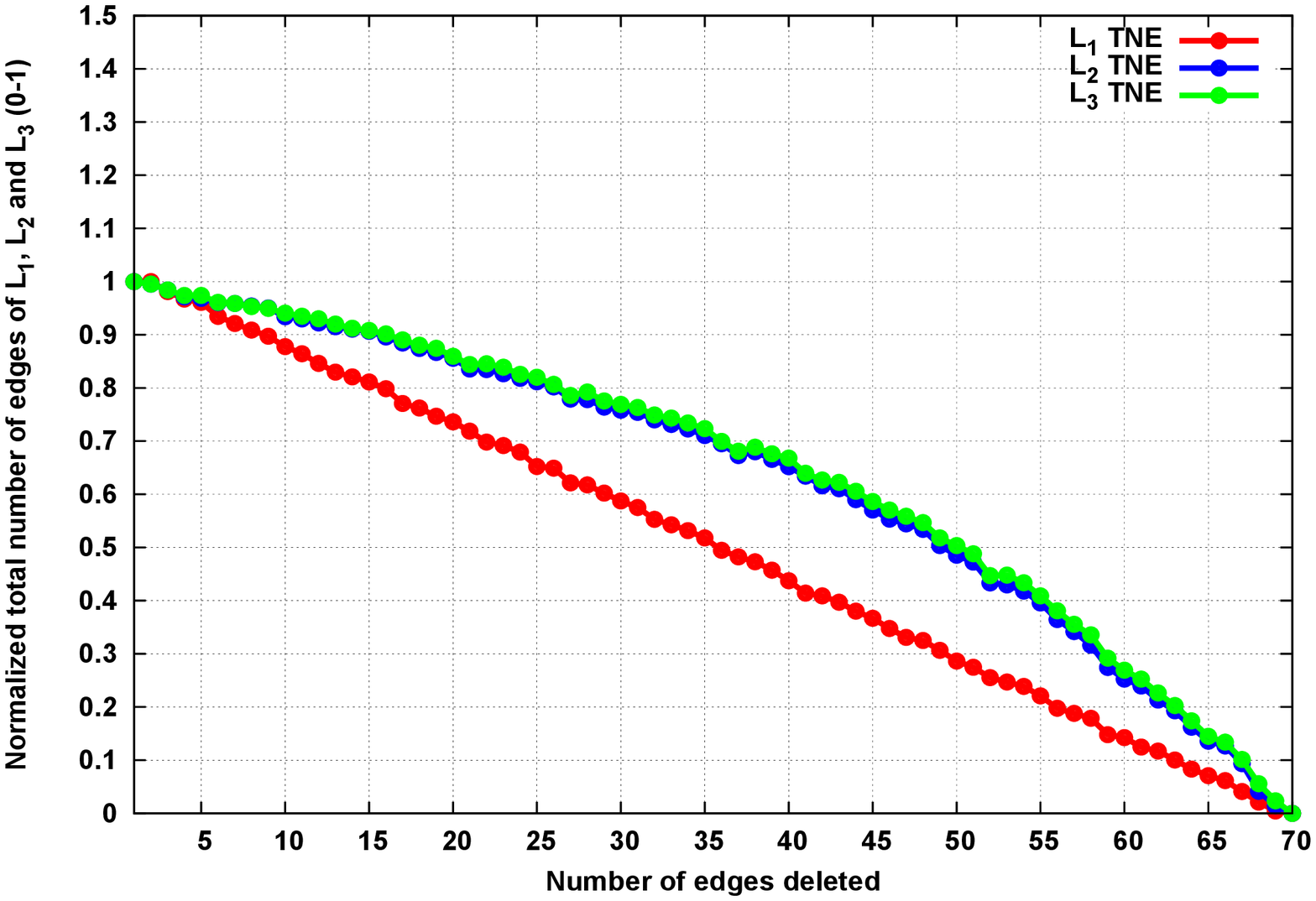}
\caption{SEBC - New $L_1$, $L_2$ and $L_3$ created. Comparison of TNEs of $L_1$, $L_2$ and $L_3$ vs. number of failed edges}
\label{fig_plot_edge_total_edges_all_layers_scaled_down_l1_l2_change_ink}
\end{figure}

TNEs of $L_1$, $L_2$ and $L_3$ are shown in Fig. \ref{fig_plot_edge_total_edges_all_layers_scaled_down_l1_l2_change_ink}. $L_1$ TNE decreases linearly as expected. However, TNEs of $L_2$ and $L_3$ shows similar but non-linear rate of change.
\subsection{DEBC behaviour}
This section describes the emergent behaviour of different parameters when the $L_1$ graph topology is not regenerated but $L_2$ and $L_3$ graph are reconstructed new in each iteration and only the edge with highest centrality fails progressively as explained in \textbf{Algorithm \ref{algo_dynamic_debc}}. The following sections presents the impact of these failures on various parameters. The 0-1 normalized values of parameters are shown along \emph{y}-axis and \emph{x}-axis shows the number of failed edges for all the plots in the section.
\subsubsection{Behaviour of $L_1$ parameters}
\begin{figure}[ht]
\centering
\includegraphics[width=\columnwidth]{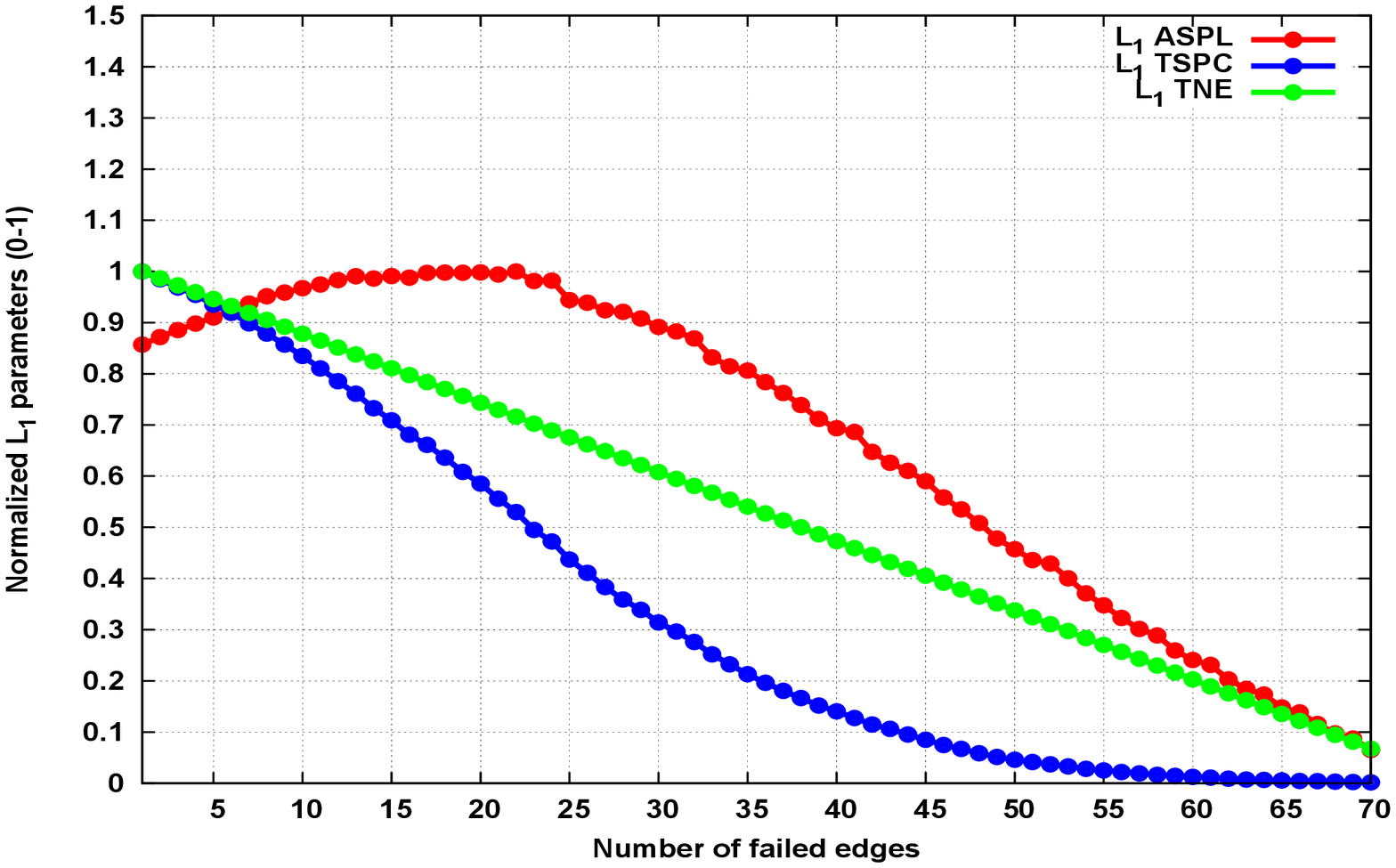}
\caption{DEBC - $L_1$ graph created only once. Behaviour of $L_1$ ASPL,
TSPC, and TNE vs. number of failed edges}
\label{fig_plot_edge_all_1l_l2_params_scaled_down_l1_l2_fixed_ink}
\end{figure}

The behaviour of $L_1$ parameters, such as, ASPL, TSPC and TNE is shown in Fig. \ref{fig_plot_edge_all_1l_l2_params_scaled_down_l1_l2_fixed_ink}. $L_1$ TNE is linearly decreasing (as expected since one edge fail each time), $L_1$ ASPL and TSPC both shows truncated gaussian decreasing emergent pattern. $L_1$ ASPL first increases till 22 edges have failed due to increase in shortest path length but then starts decreasing following a gaussian tail with the shrinking of the graph.
\subsubsection{Behaviour of $L_2$ parameters}
\begin{figure}[ht]
\centering
\includegraphics[width=\columnwidth]{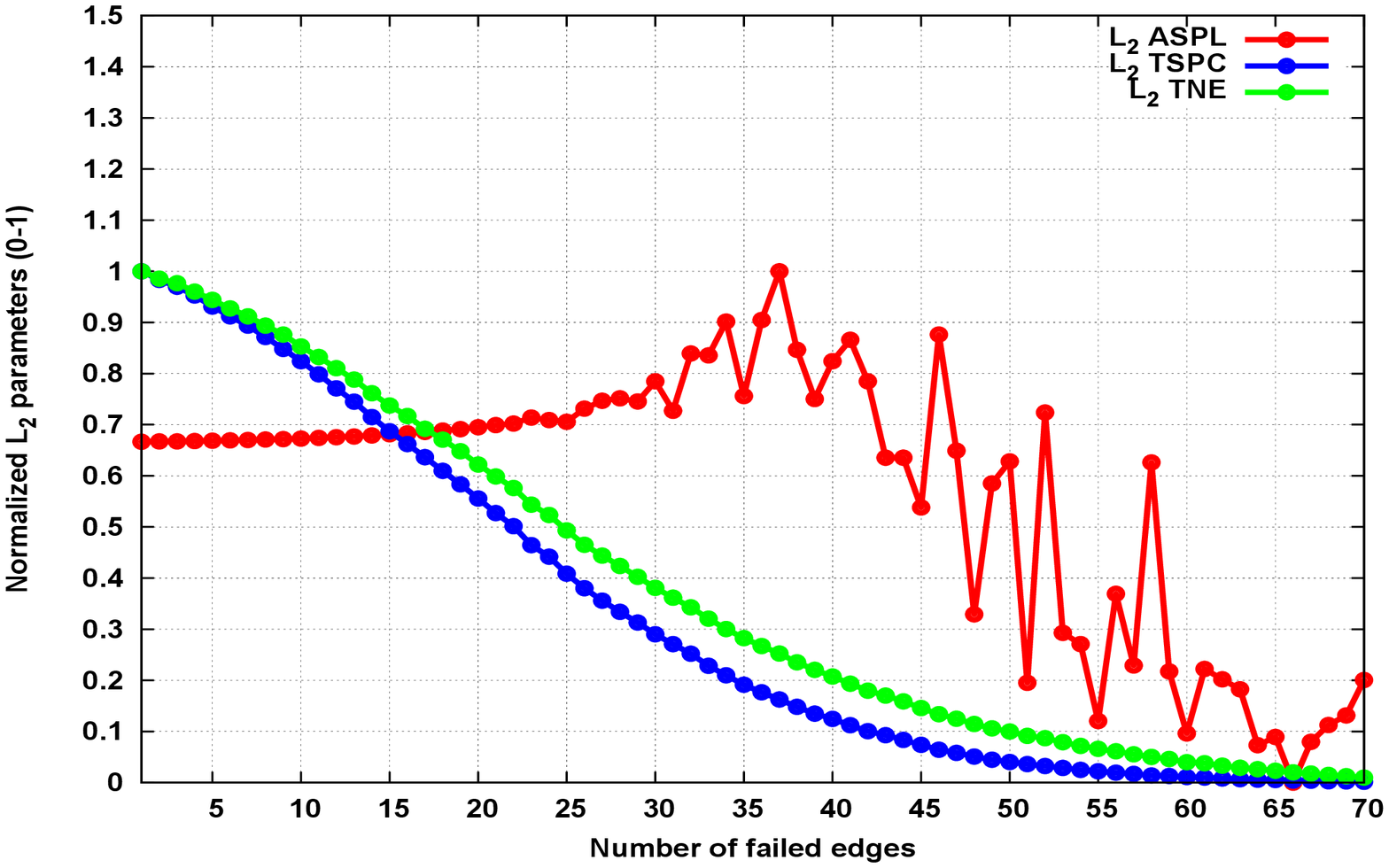}
\caption{DEBC - $L_1$ graph created only once. Behaviour of $L_2$ ASPL,
TSPC, and TNE vs. number of failed edges}
\label{fig_plot_edge_all_l3_params_scaled_down_l1_l2_fixed_ink}
\end{figure}

The behaviour of $L_2$ parameters, such as, ASPL, TSPC and TNE is depicted in Fig. \ref{fig_plot_edge_all_l3_params_scaled_down_l1_l2_fixed_ink}. While the $L_2$ TSPC and $L_2$ TNE shows a decreasing truncated guassian characteristics, interestingly, $L_2$ ASPL starts showing  a chaotic behaviour with a decreasing trend after $\approx$ 22 edges have failed due to the same reason mentioned above. Till 22 edges have failed $L_2$ ASPL increases progressively.
\subsubsection{Behaviour of $L_3$ parameters}
\begin{figure}[ht]
\centering
\includegraphics[width=\columnwidth]{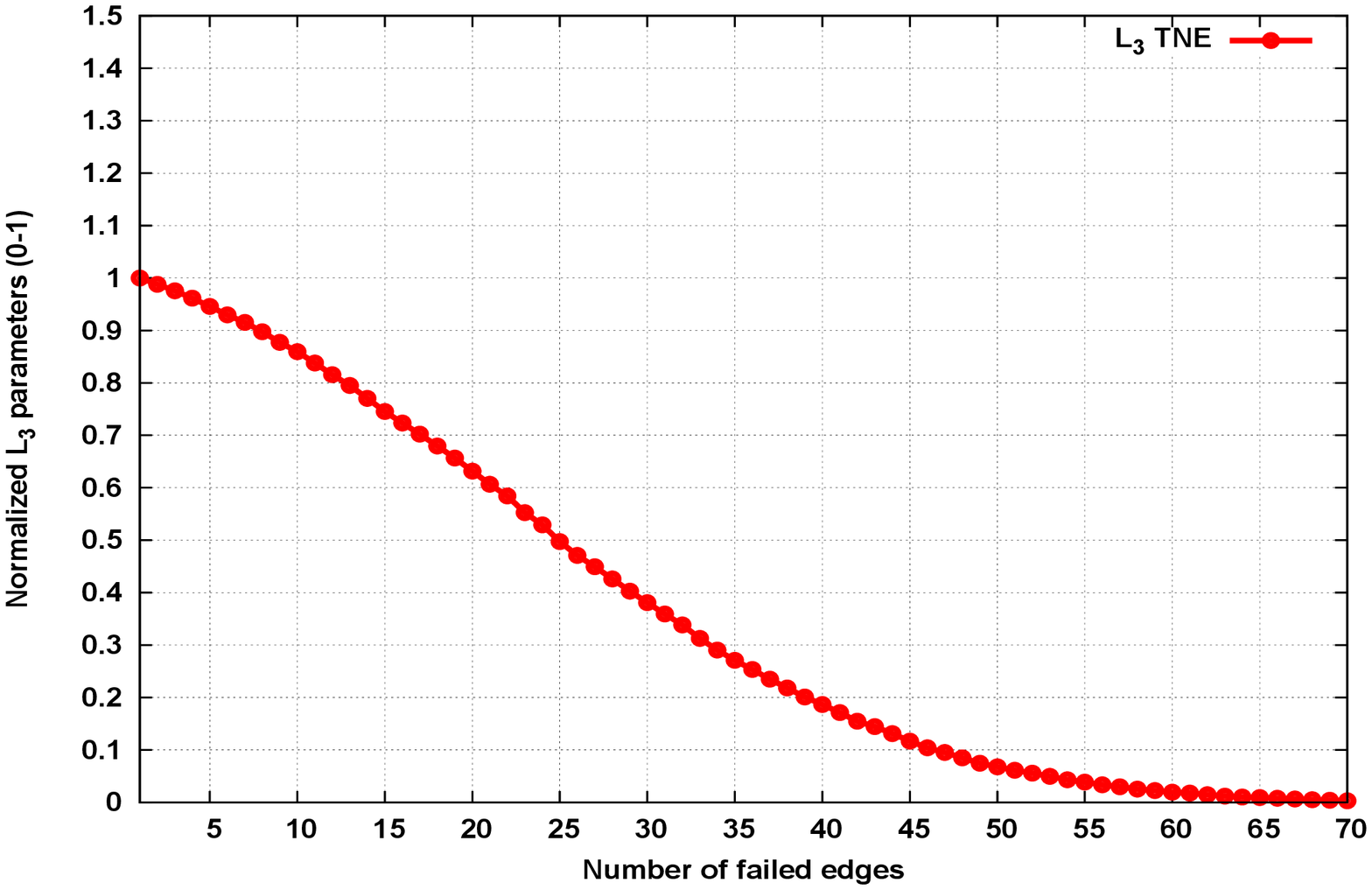}
\caption{DEBC - $L_1$ graph created only once. Behaviour of $L_3$ TNE vs. number of failed edges}
\label{fig_plot_edge_all_l4_params_scaled_down_l1_l2_fixed_ink}
\end{figure}

The behaviour of only $L_3$ TNE is provided in Fig. \ref{fig_plot_edge_all_l4_params_scaled_down_l1_l2_fixed_ink}. $L_3$ TNE also shows truncated gaussian behaviour.
\subsubsection{Comparison of ASPLs of $L_1$ and $L_2$}
\begin{figure}[ht]
\centering
\includegraphics[width=\columnwidth]{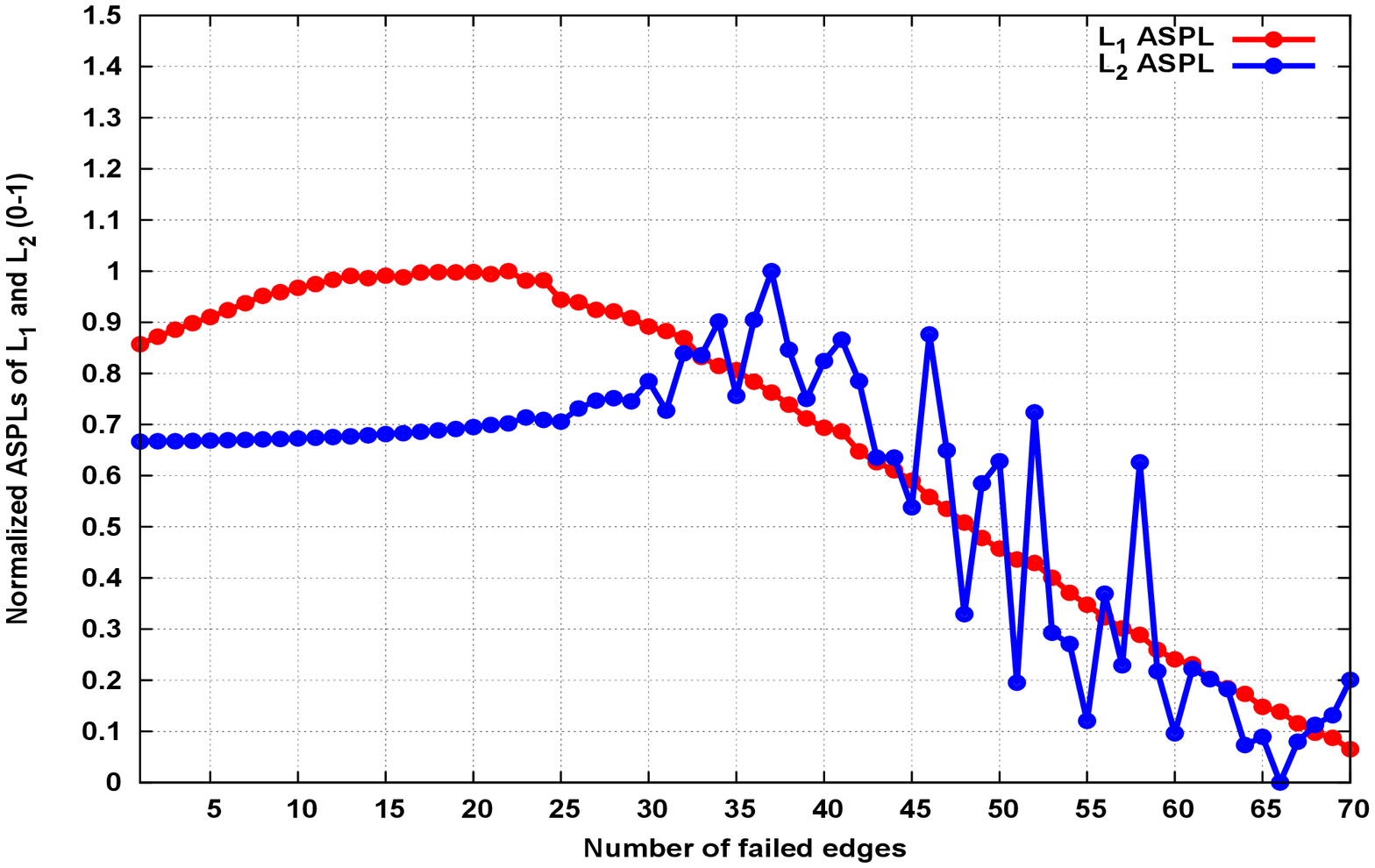}
\caption{DEBC - $L_1$ graph created only once. Comparison of ASPLs of $L_1$ and $L_2$ vs. number of failed edges}
\label{fig_plot_edge_avg_shortest_path_length_all_layers_scaled_down_l1_l2_fixed_ink}
\end{figure}

Fig. \ref{fig_plot_edge_avg_shortest_path_length_all_layers_scaled_down_l1_l2_fixed_ink} compares the ASPL of $L_1$  and $L_2$. Note that $L_3$ with end to end connections does not have ASPL. Till 22 $L_1$  edges have failed, $L_1$ ASPL increases as expected. Also, $L_2$ ASPL shows nominal increase. Thereafter, $L_2$ ASPL starts fluctuating and shows a chaotic behaviour later with a decreasing trend. After saturation for some time at the maximum value, $L_1$ ASPL shows a decrease. Thus, some amount of $L_1$ edge failures is withstood by $L_2$ without major increase in $L_2$ ASPL.
\subsubsection{Comparison of TSPCs of $L_1$ and $L_2$}
\begin{figure}[ht]
\centering
\includegraphics[width=\columnwidth]{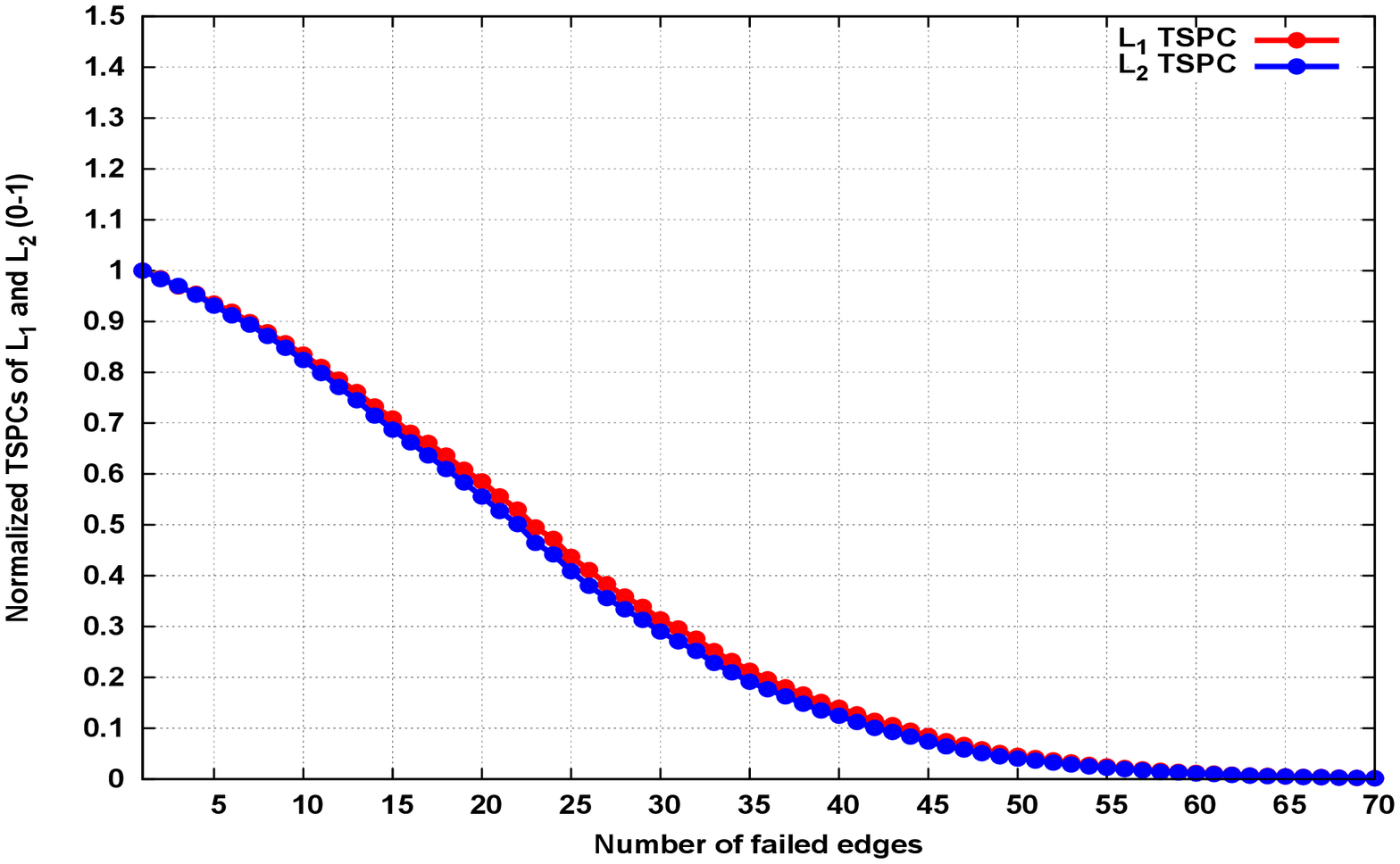}
\caption{DEBC - $L_1$ graph created only once. Comparison of TSPCs of $L_1$ and $L_2$ vs. number of failed edges}
\label{fig_plot_edge_total_shortest_paths_all_layers_scaled_down_l1_l2_fixed_ink}
\end{figure}

Fig. \ref{fig_plot_edge_total_shortest_paths_all_layers_scaled_down_l1_l2_fixed_ink} shows the behaviour of TSPC of $L_1$ and $L_2$. $L_2$ TSPC decreases slightly faster than $L_1$ TSPC in some middle stretches and both show truncated gaussian behaviour.
\subsubsection{Comparison of TNEs of $L_1$, $L_2$ and $L_3$}
\begin{figure}[ht]
\centering
\includegraphics[width=\columnwidth]{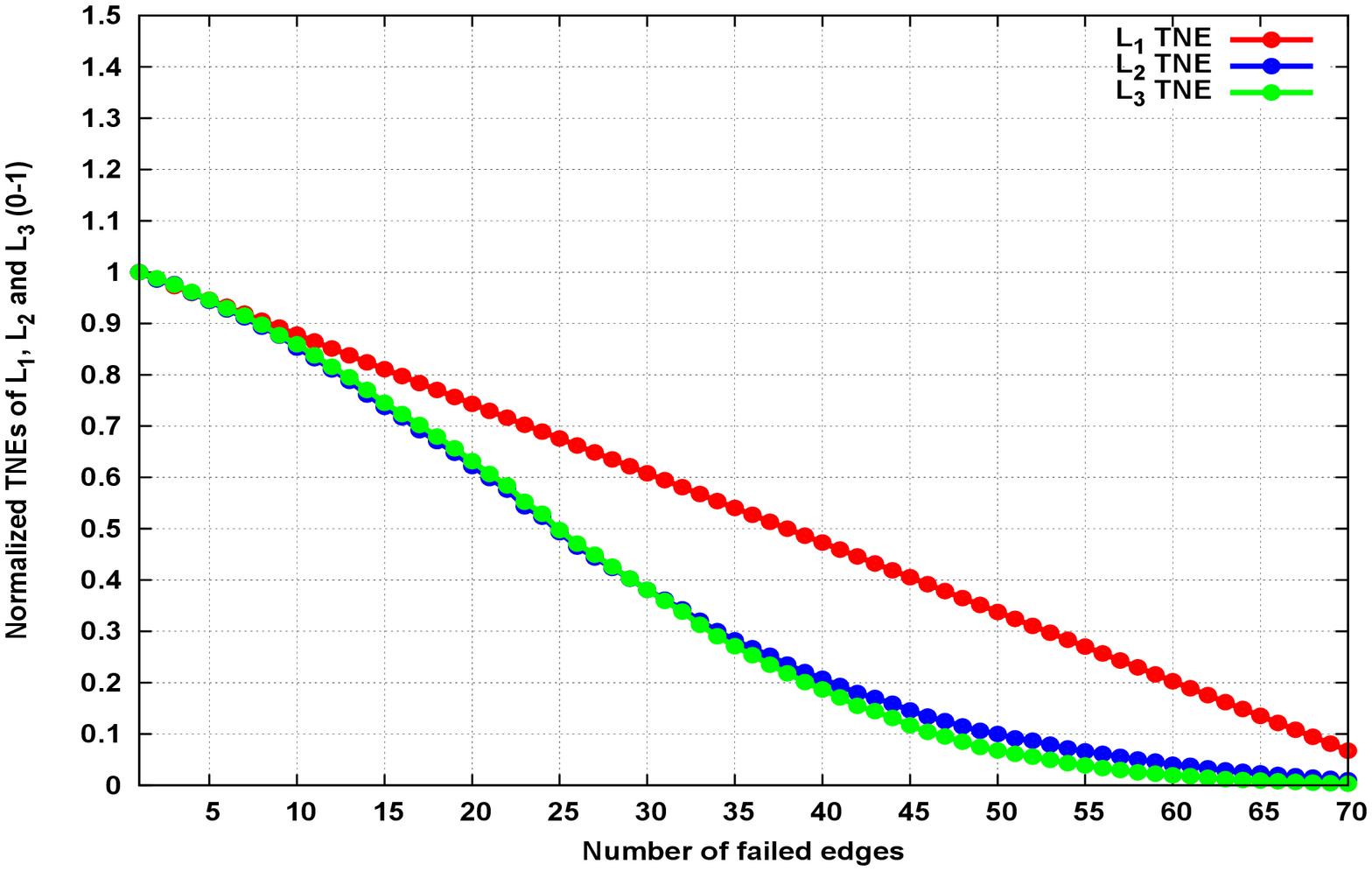}
\caption{DEBC - $L_1$ graph created only once. Comparison of TNEs of $L_1$, $L_2$ and $L_3$ vs. number of failed edges}
\label{fig_plot_edge_total_edges_all_layers_scaled_down_l1_l2_fixed_ink}
\end{figure}

TNEs of $L_1$, $L_2$ and $L_3$ are plotted in Fig. \ref{fig_plot_edge_total_edges_all_layers_scaled_down_l1_l2_fixed_ink}. $L_1$ decreases linearly as expected. Both $L_2$ and $L_3$ TNEs decrease much faster compared to that of $L_1$ which exactly opposite of SEBC. $L_2$ and $L_3$ TNEs decrease at almost same rate for the first half but for the latter half $L_3$ TNE decreases slightly faster than that of $L_2$.
\subsection{Comparison of SEBC and DEBC }
This section compares the parameters of the two scenarios against number of edges failed.
\subsubsection{Behaviour of $L_1$ ASPL}
\begin{figure}[ht]
\centering
\includegraphics[width=\columnwidth]{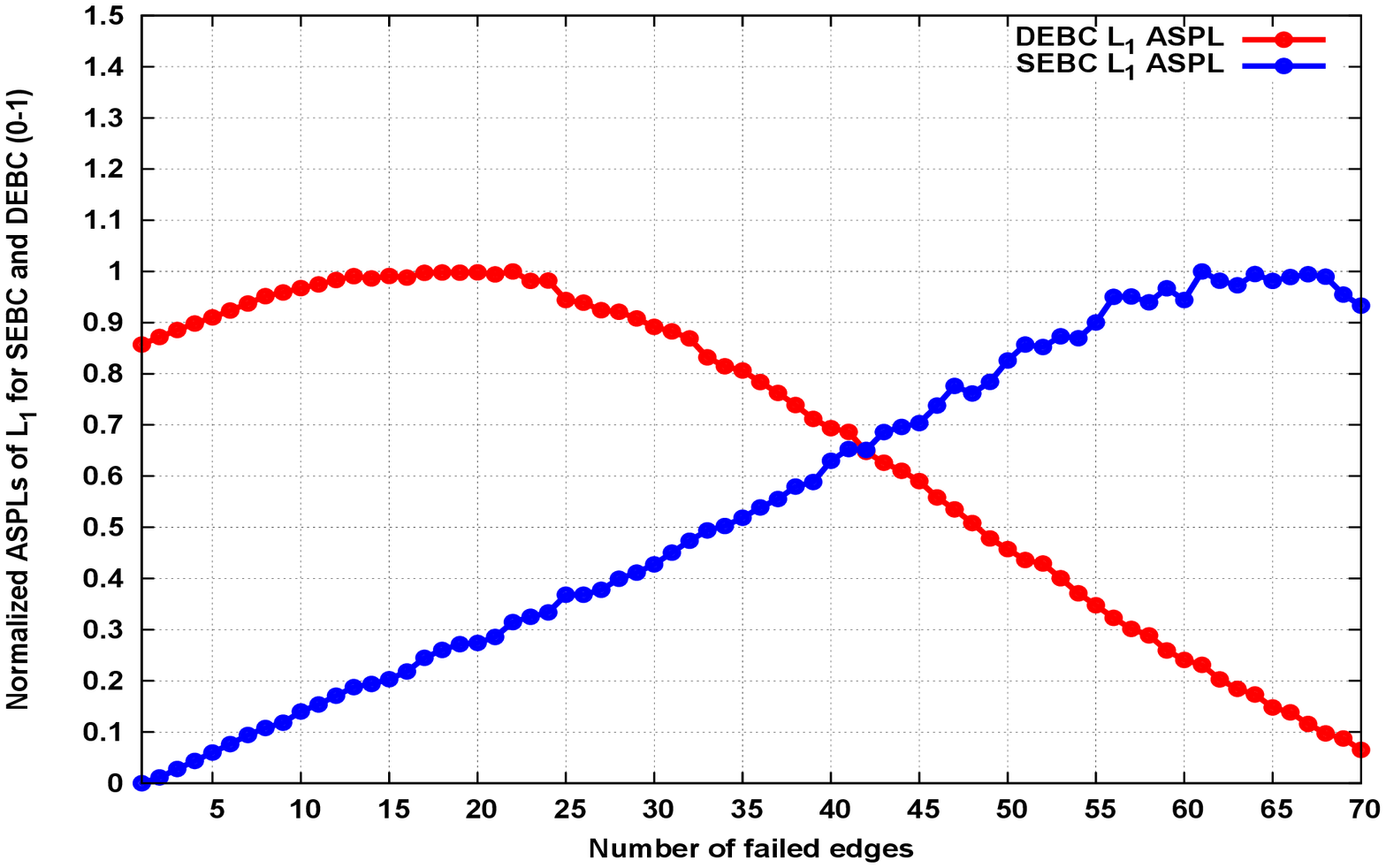}
\caption{DEBC and SEBC Comparison - $L_1$ ASPL vs. number of failed edges}
\label{fig_plot_edge_l1_l2_avg_shortest_path_length_two_scenarios_ink}
\end{figure}

Fig. \ref{fig_plot_edge_l1_l2_avg_shortest_path_length_two_scenarios_ink} shows the $L_1$ ASPLs of the two scenarios of DEBC and SEBC. For DEBC case, $L_1$ ASPL starts with high value and saturates quickly and then starts dropping due to reduction in the graph size. For SEBC, starts with low value and saturates much later and then starts decreasing. Thus, DEBC brings faster changes to the network topology.
\subsubsection{Behaviour of $L_1$ TSPC}
\begin{figure}[ht]
\centering
\includegraphics[width=\columnwidth]{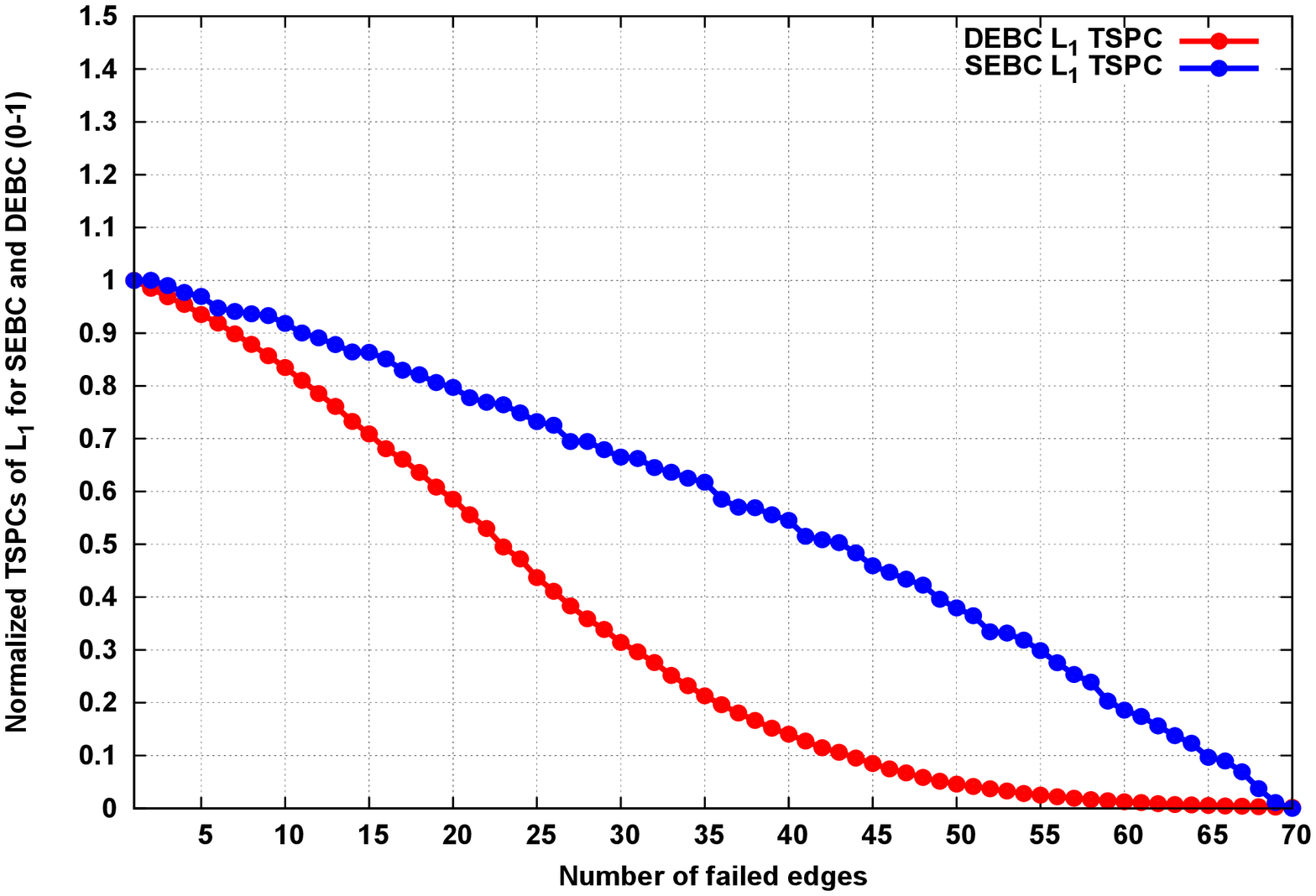}
\caption{DEBC and SEBC Comparison - $L_1$ TSPC vs. number of failed edges}
\label{fig_plot_edge_l1_l2_total_shortest_paths_two_scenarios_ink}
\end{figure}

Fig. \ref{fig_plot_edge_l1_l2_total_shortest_paths_two_scenarios_ink} shows $L_1$ TSPC for the two scenarios. It can be observed that for DEBC the drop in $L_1$ TSPC is much faster compared to SEBC.

TNEs of $L_1$ for the two scenarios is not presented separately because both decreases linearly with number of failed edges unlike the case of node failures where the TNEs need not decrease linearly.
\subsubsection{Behaviour of $L_2$ ASPL}
\begin{figure}[ht]
\centering
\includegraphics[width=\columnwidth]{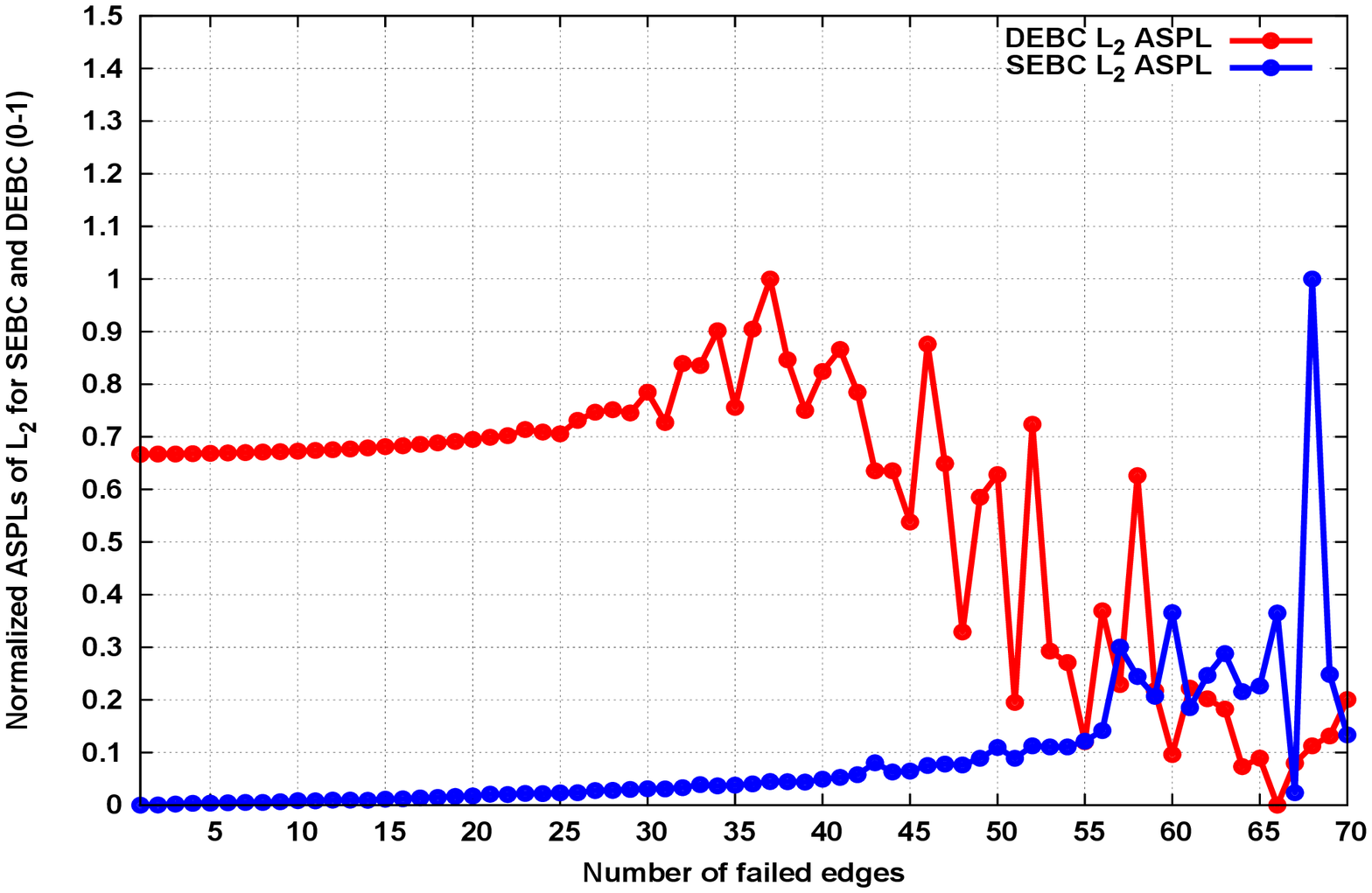}
\caption{DEBC and SEBC Comparison - $L_2$ ASPL vs. number of failed edges}
\label{fig_plot_edge_l3_avg_shortest_path_length_two_scenarios_ink}
\end{figure}

Behaviour of the $L_2$ ASPLs for the two scenarios are shown in Fig. \ref{fig_plot_edge_l3_avg_shortest_path_length_two_scenarios_ink}. For DEBC, $L_2$ ASPL starts at a much higher value compared to SEBC. In both cases, there is a progressive increase. However, the chaotic behaviour starts much early for the DEBC than for SEBC. Also, there is no appreciable change in the values number of failed edges till 20 in both the cases.
\subsubsection{Behaviour of $L_2$ TSPC}
\begin{figure}[ht]
\centering
\includegraphics[width=\columnwidth]{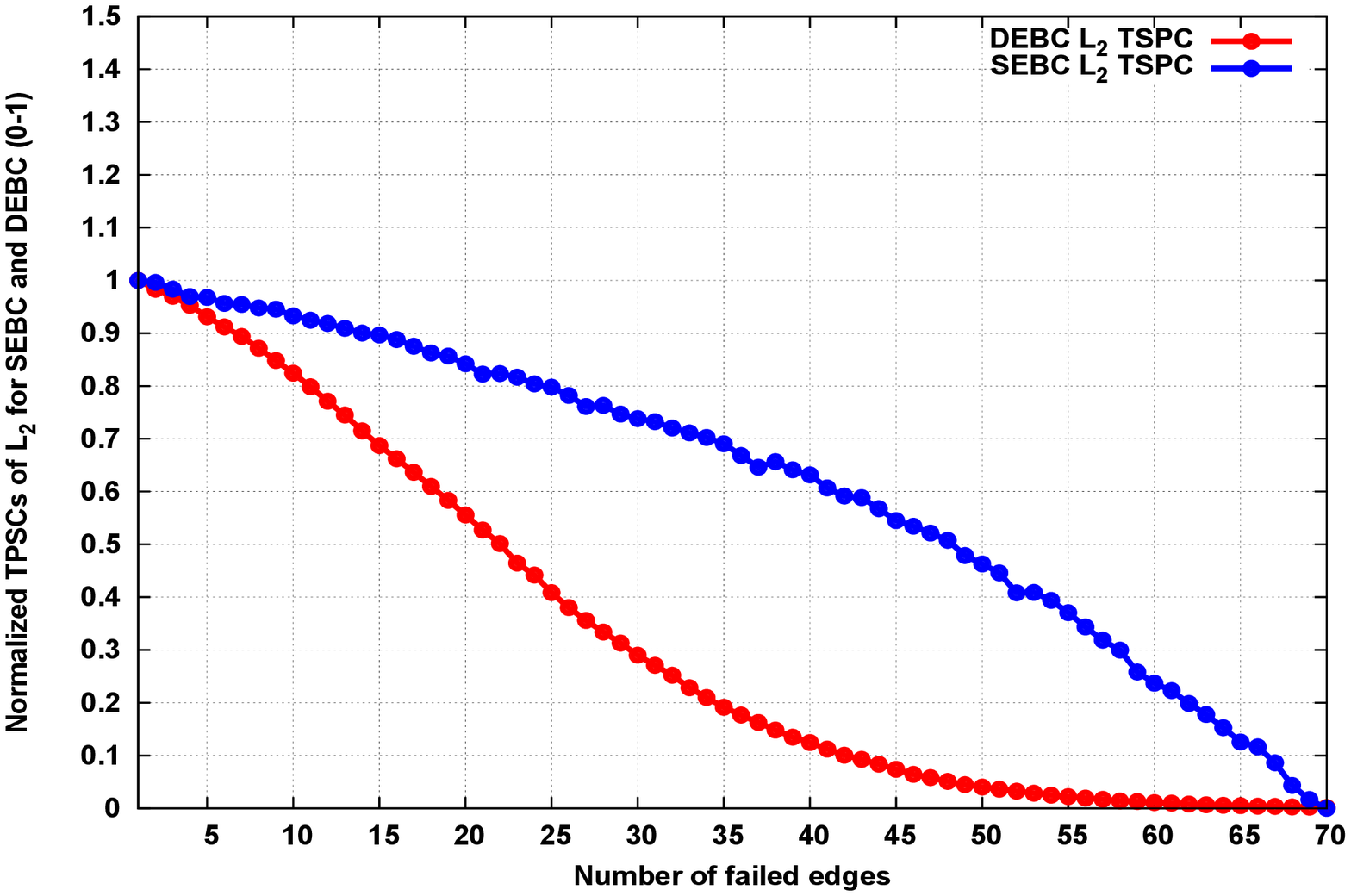}
\caption{DEBC and SEBC Comparison - $L_2$ TSPC vs. number of failed edges}
\label{fig_plot_edge_l3_total_shortest_paths_two_scenarios_ink}
\end{figure}

Almost similar behaviours are observed for $L_2$ TSPCs in Fig. \ref{fig_plot_edge_l3_total_shortest_paths_two_scenarios_ink} as compared to $L_1$ TSPCs.
\subsubsection{Behaviour of $L_2$ TNE}
\begin{figure}[ht]
\centering
\includegraphics[width=\columnwidth]{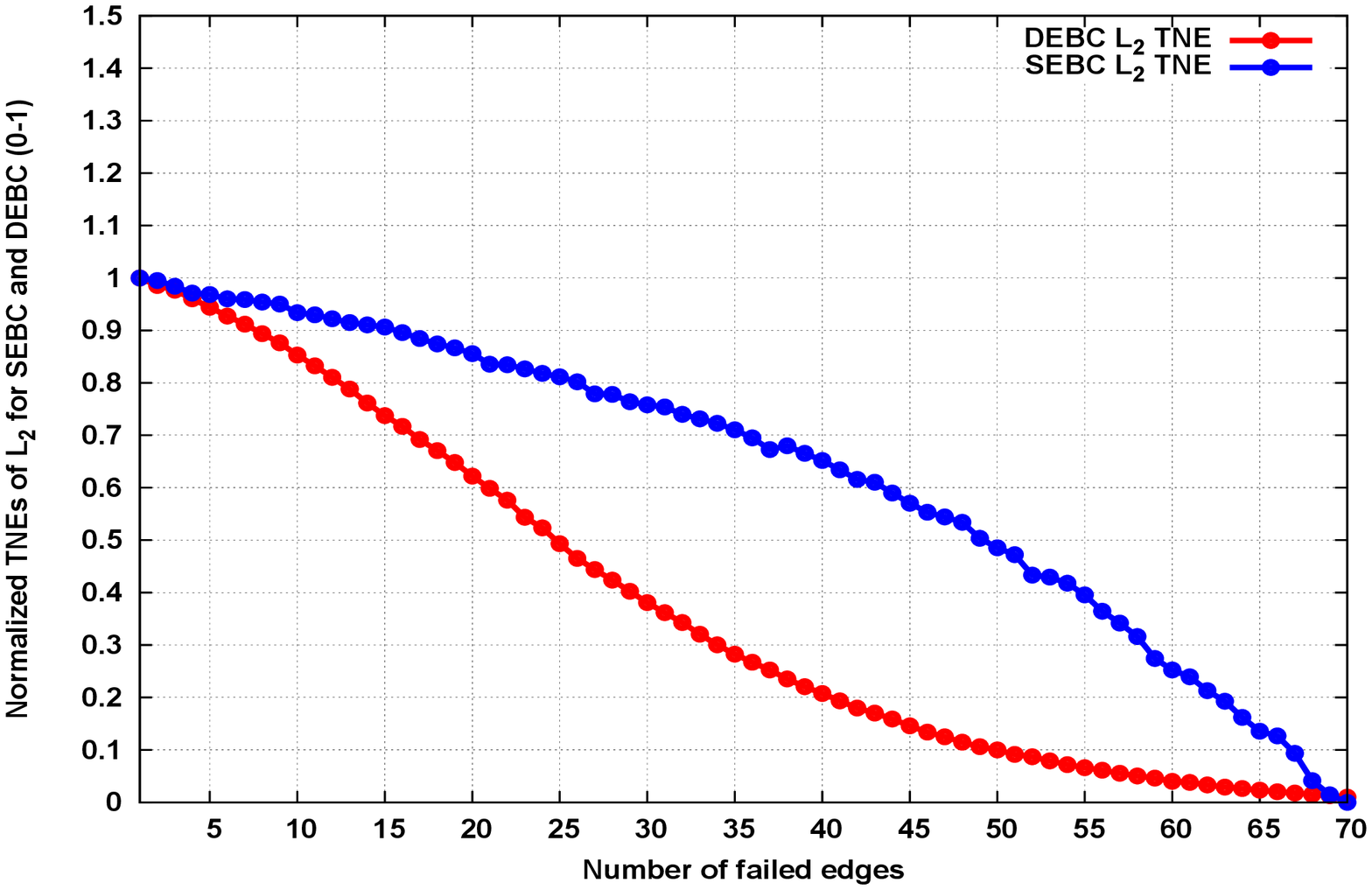}
\caption{DEBC and SEBC Comparison - $L_2$ TNE vs. number of failed edges}
\label{fig_plot_edge_l3_total_edges_two_scenarios_ink}
\end{figure}

Fig. \ref{fig_plot_edge_l3_total_edges_two_scenarios_ink} shows behaviours of $L_2$ TNEs for the two scenarios. They both behave in the similar ways as $L_2$ TSPCs in Fig. \ref{fig_plot_edge_l3_total_shortest_paths_two_scenarios_ink}.
\subsubsection{Behaviour of $L_3$ TNE}
\begin{figure}[ht]
\centering
\includegraphics[width=\columnwidth]{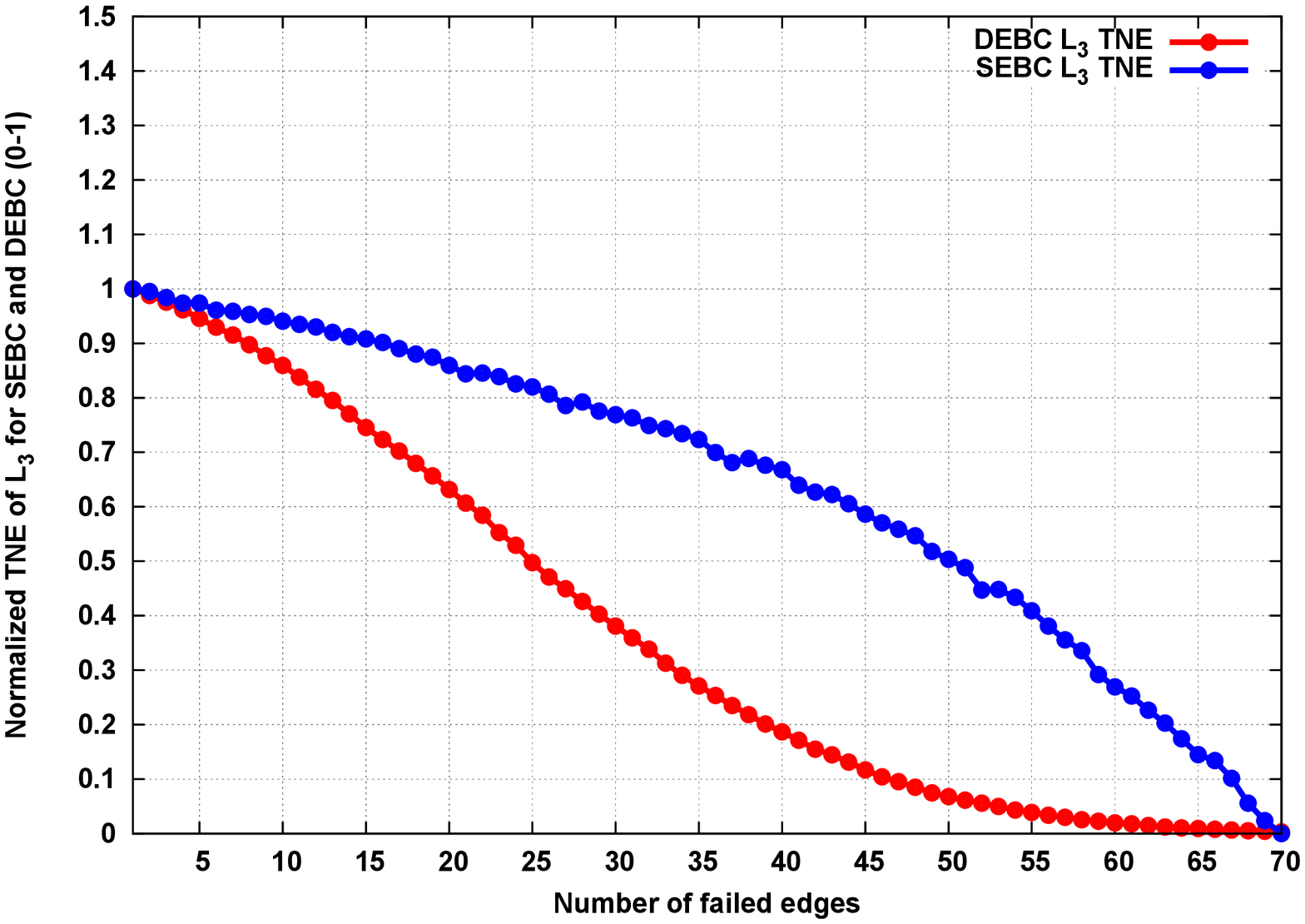}
\caption{DEBC and SEBC Comparison - $L_3$ TNE vs. number of failed edges}
\label{fig_plot_edge_l4_total_edges_two_scenarios_ink}
\end{figure}

Fig. \ref{fig_plot_edge_l4_total_edges_two_scenarios_ink} shows the behaviour of $L_3$ TNEs for the two scenarios. It can be observed the behaviour is similar to $L_2$ TNE.
\subsection{SNBC Behaviour}
In this scenario, each iteration new $L_1$, $L_2$ and $L_3$ graphs are constructed and then the required number of nodes with highest centrality nodes fail in one go using \textbf{Algorithm \ref{algo_static_snbc}}. The 0-1 normalized values of parameters are shown along \emph{y}-axis and \emph{x}-axis shows the number of failed edges for all the plots in the section.
\subsubsection{Behaviour of $L_1$ parameters}
\begin{figure}[ht]
\centering
\includegraphics[width=\columnwidth]{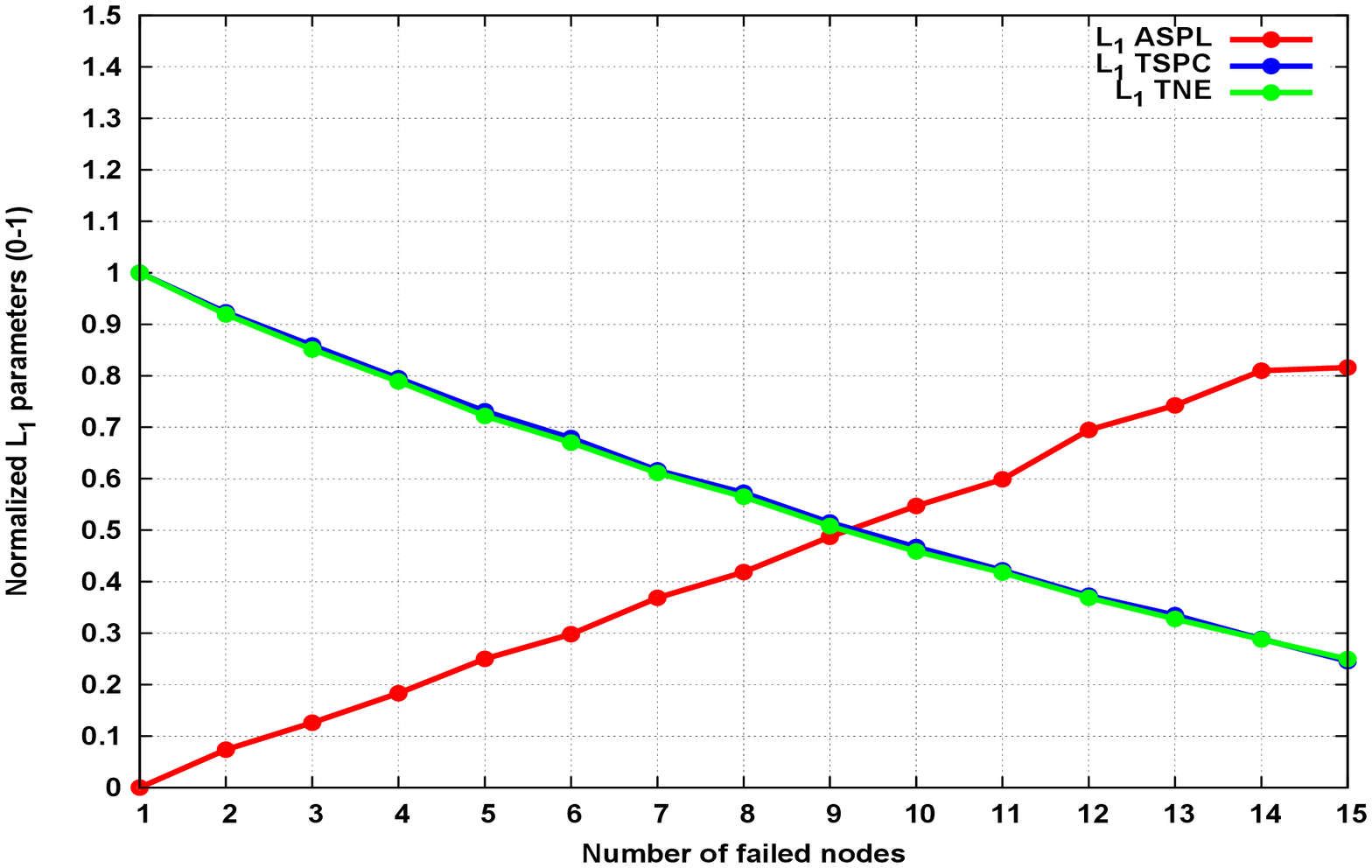}
\caption{SNBC - New $L_1$, $L_2$ and $L_3$ created. Behaviour of $L_1$ ASPL,
TSPC, and TNE vs. number of failed nodes}
\label{fig_plot_node_all_l1_l2_params_scaled_down_l1_l2_change_ink}
\end{figure}

Fig. \ref{fig_plot_node_all_l1_l2_params_scaled_down_l1_l2_change_ink} shows the behaviour of $L_1$ parameters. $L_1$ TNE and TSPC decreases almost linearly at the same rate. $L_1$ ASPL first increases almost linearly till 14 nodes fail, before saturating.
\subsubsection{Behaviour of $L_2$ parameters}
\begin{figure}[ht]
\centering
\includegraphics[width=\columnwidth]{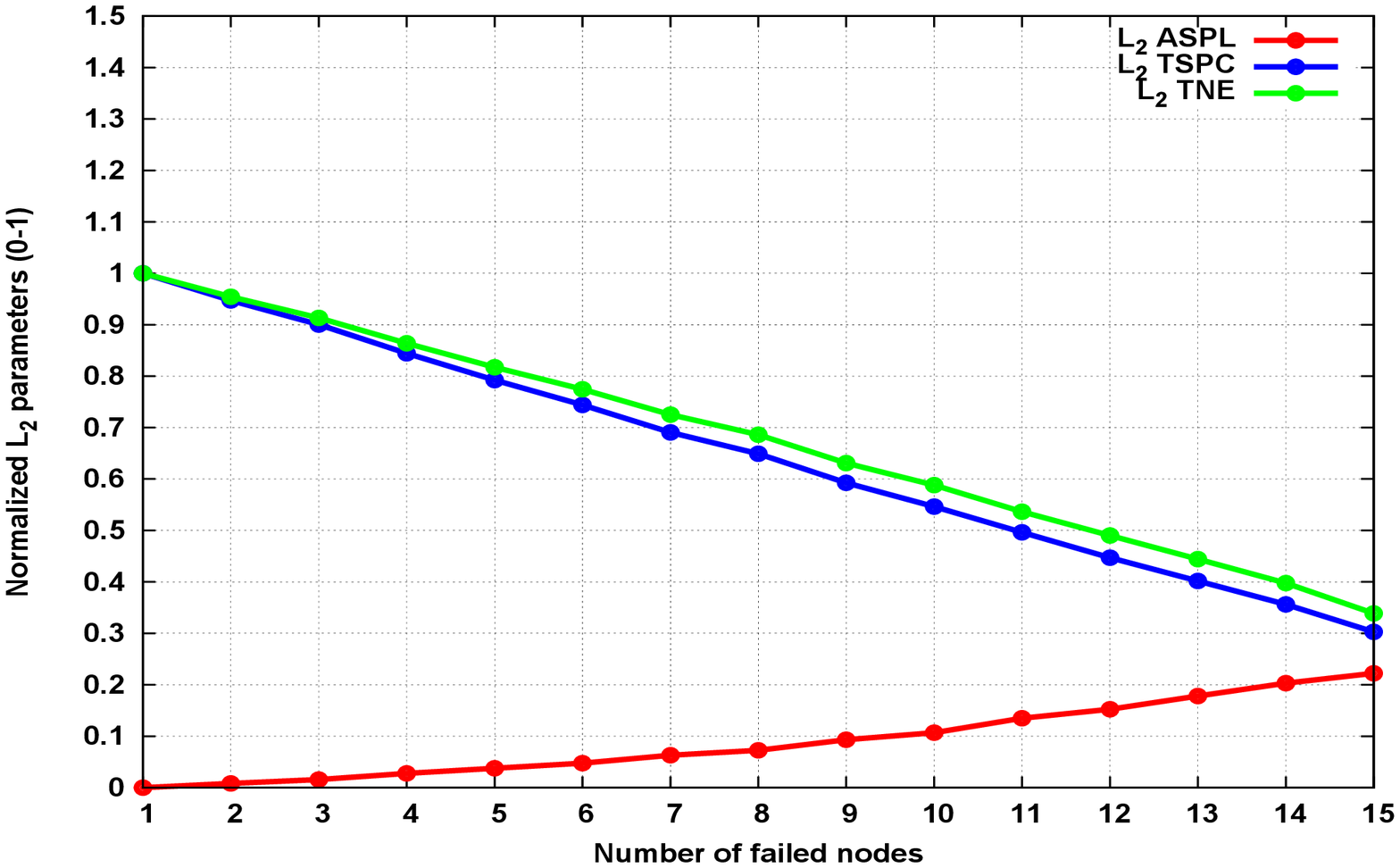}
\caption{SNBC - New $L_1$, $L_2$ and $L_3$ created. Behaviour of $L_2$ ASPL,
TSPC, and TNE vs. number of failed nodes}
\label{fig_plot_node_all_l3_params_scaled_down_l1_l2_change_ink}
\end{figure}

Fig. \ref{fig_plot_node_all_l3_params_scaled_down_l1_l2_change_ink} shows the behaviour of ASPL, TSPC and TNE of $L_2$. $L_2$ ASPL shows a non-linear increase whereas TSPC and TNE show almost linear decrease.  TSPC decreases at a faster rate compared to TNE. No major change in ASPL till high centrality 3 nodes fail.
\subsubsection{Behaviour of $L_3$ parameters}
\begin{figure}[ht]
\centering
\includegraphics[width=\columnwidth]{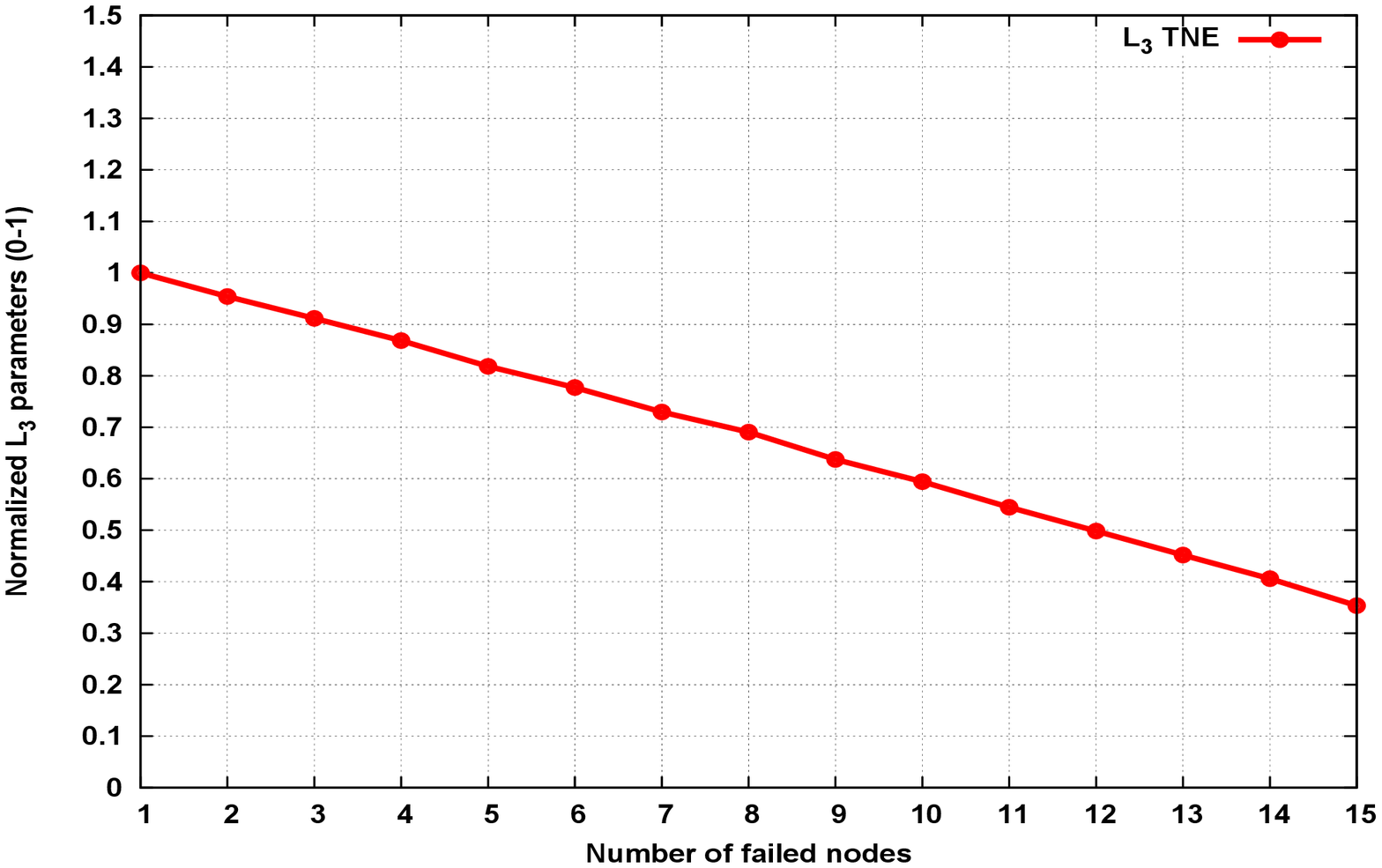}
\caption{SNBC - New $L_1$, $L_2$ and $L_3$ created. Behaviour of $L_3$ TNE vs. number of failed nodes}
\label{fig_plot_node_all_l4_params_scaled_down_l1_l2_change_ink}
\end{figure}

$L_3$ TNE shows a linear decrease of over 60\% as nodes keep failing which is shown in Fig. \ref{fig_plot_node_all_l4_params_scaled_down_l1_l2_change_ink}.
\subsubsection{Comparison of ASPLs of $L_1$ and $L_2$}
\begin{figure}[ht]
\centering
\includegraphics[width=\columnwidth]{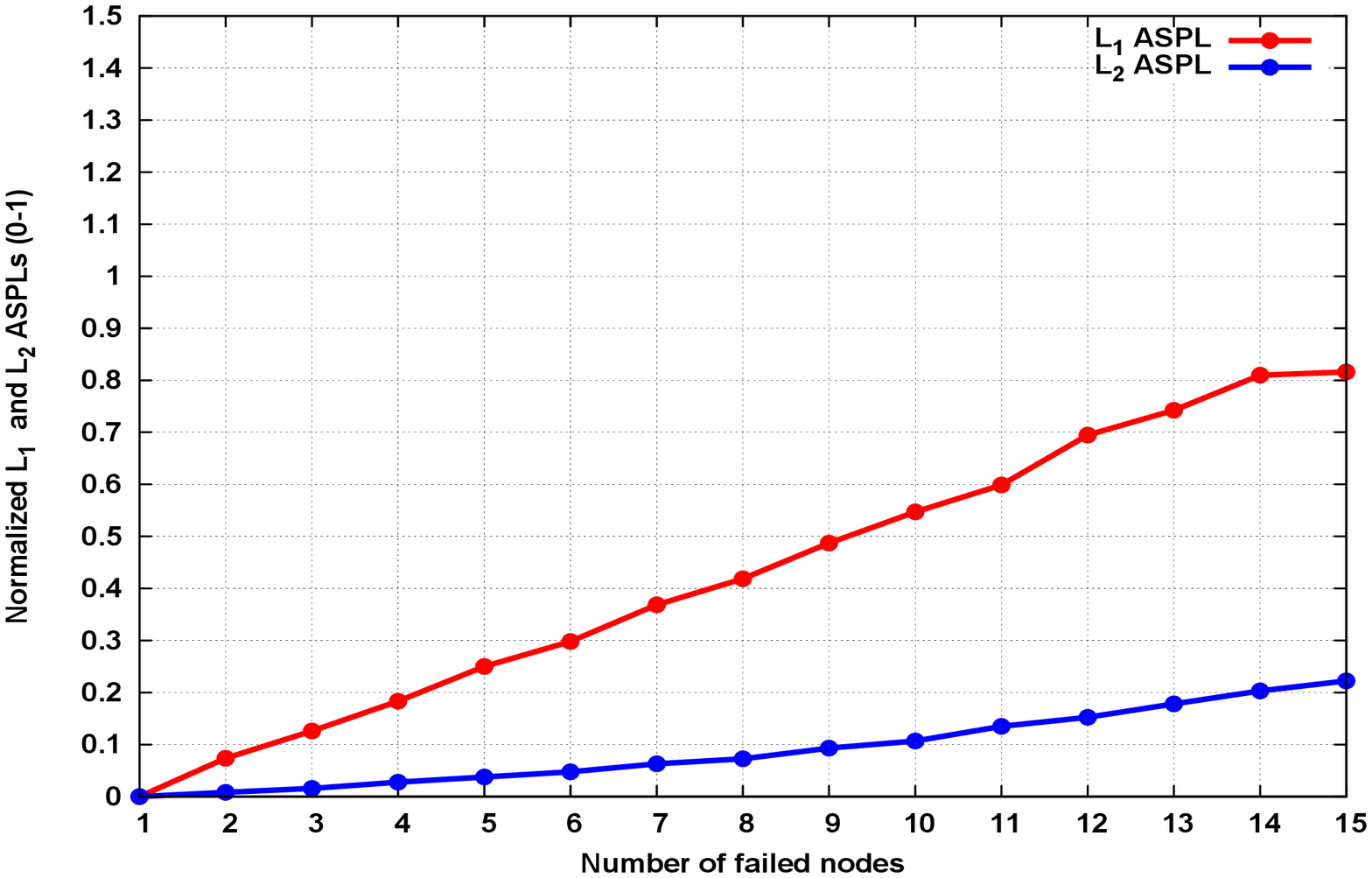}
\caption{SNBC - New $L_1$, $L_2$ and $L_3$ created - Comparison of $L_1$ and $L_2$ ASPLs vs. number of failed nodes}
\label{fig_plot_node_avg_shortest_path_length_all_layers_scaled_down_l1_l2_change_ink}
\end{figure}

Fig. \ref{fig_plot_node_avg_shortest_path_length_all_layers_scaled_down_l1_l2_change_ink} shows the behaviour of ASPLs of $L_1$ and $L_2$. It is observed that $L_1$ ASPL increase faster than that of $L_2$. For almost 80\% increase in $L_1$ ASPL, there is around 20\% rise in $L_2$ ASPL. Also $L_2$ ASPL can withstand small amount of node failures (around 3) without any major degradation.
\subsubsection{Comparison of TSPCs of $L_1$ and $L_2$}
\begin{figure}[ht]
\centering
\includegraphics[width=\columnwidth]{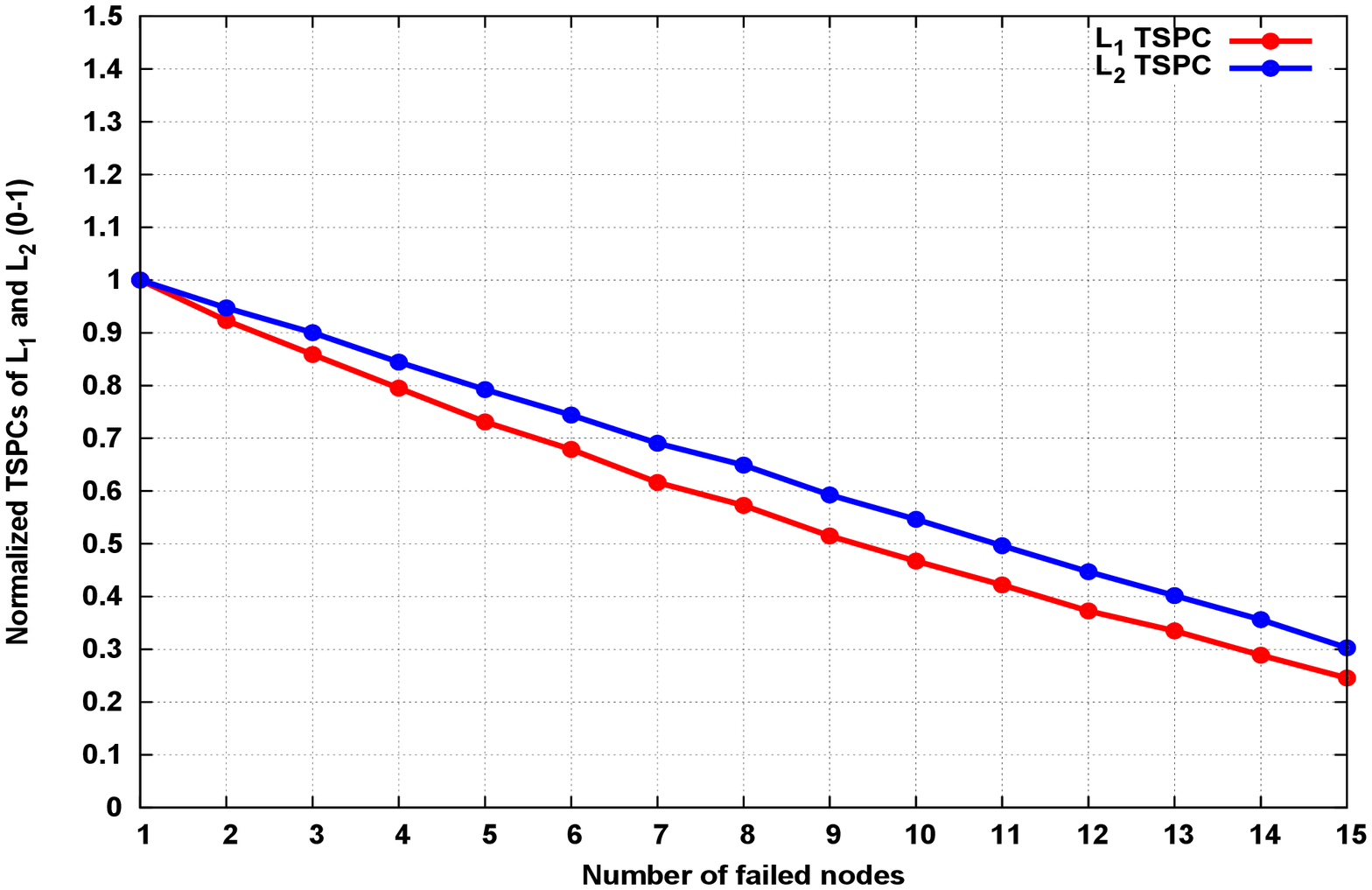}
\caption{SNBC - New $L_1$, $L_2$ and $L_3$ created - Comparison of $L_1$ and $L_2$ TSPCs vs. number of failed nodes}
\label{fig_plot_node_total_shortest_paths_all_layers_scaled_down_l1_l2_change_ink}
\end{figure}

It can be observed that in Fig. \ref{fig_plot_node_total_shortest_paths_all_layers_scaled_down_l1_l2_change_ink} that $L_1$ TSPC decreases at a faster rate at maximum of $\approx$ 10\% compared to that of $L_2$. However, beyond 8 node failures the difference is more or less constant.
\subsubsection{Comparison of TNEs of $L_1$, $L_2$ and $L_3$}
\begin{figure}[ht]
\centering
\includegraphics[width=\columnwidth]{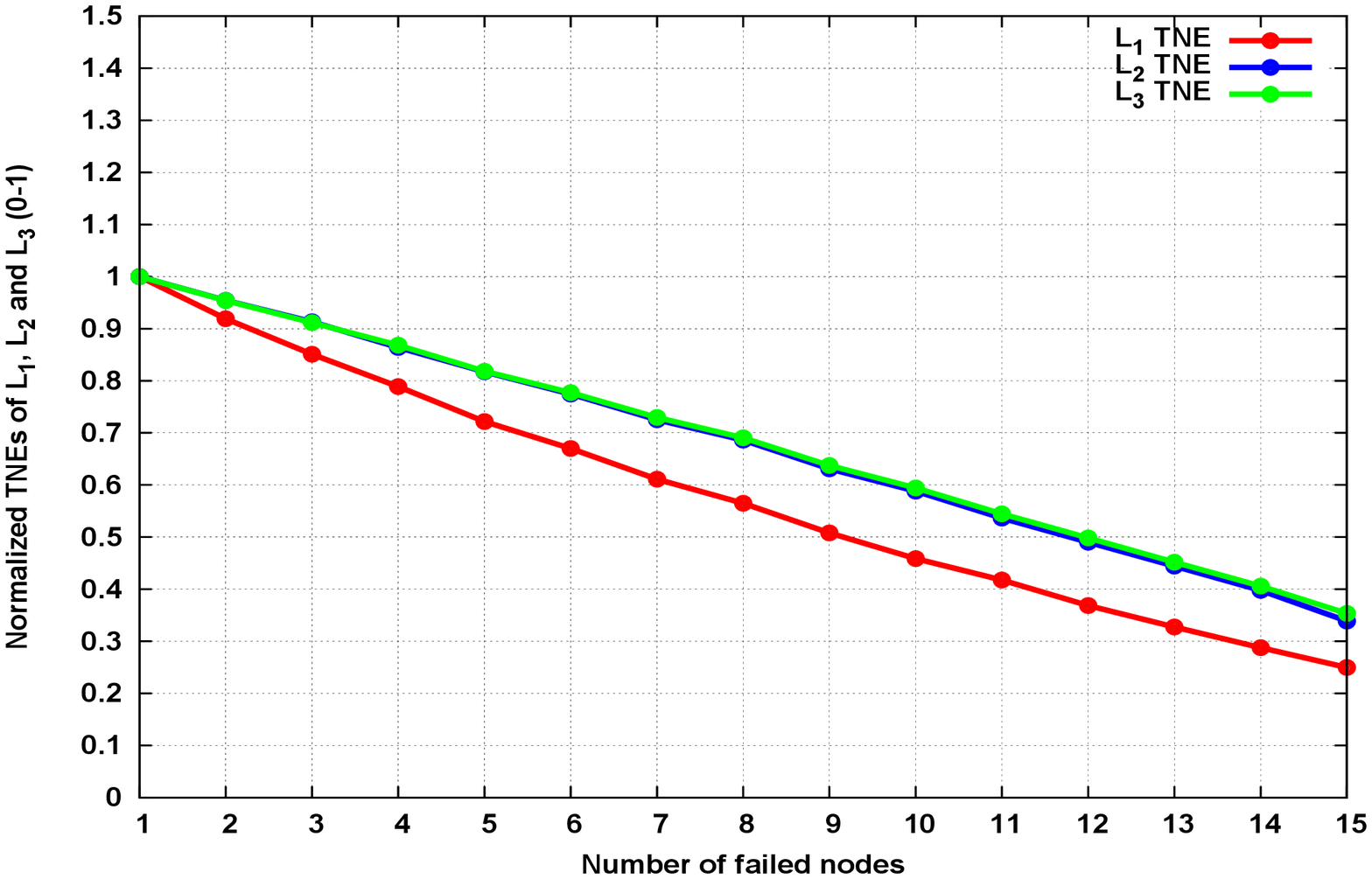}
\caption{SNBC - New $L_1$, $L_2$ and $L_3$ created - Comparison of TNEs of $L_1$, $L_2$ and $L_3$ vs. number of failed nodes}
\label{fig_plot_node_total_edges_all_layers_scaled_down_l1_l2_change_ink}
\end{figure}

Fig. \ref{fig_plot_node_total_edges_all_layers_scaled_down_l1_l2_change_ink} shows the TNEs of $L_1$, $L_3$ and $L_2$. All of them are linear more or less. $L_1$ TNE shows faster decrease by amount maximum of 75\%. TNEs of $L_2$ and $L_3$ both decrease at same rate of 65\% as nodes fail.
\subsection{DNBC Behaviour}
This section describes the emergent behaviour of different parameters when the $L_1$ graph topology is not regenerated in each iteration and only the node with highest centrality fails progressively as described in \textbf{Algorithm \ref{algo_dynamic_dnbc}}. For the figures the relevant parameter is along \emph{y}-axis vs. number of failed nodes along \emph{x}-axis.
\subsubsection{Behaviour of $L_1$ parameters}
\begin{figure}[ht]
\centering
\includegraphics[width=\columnwidth]{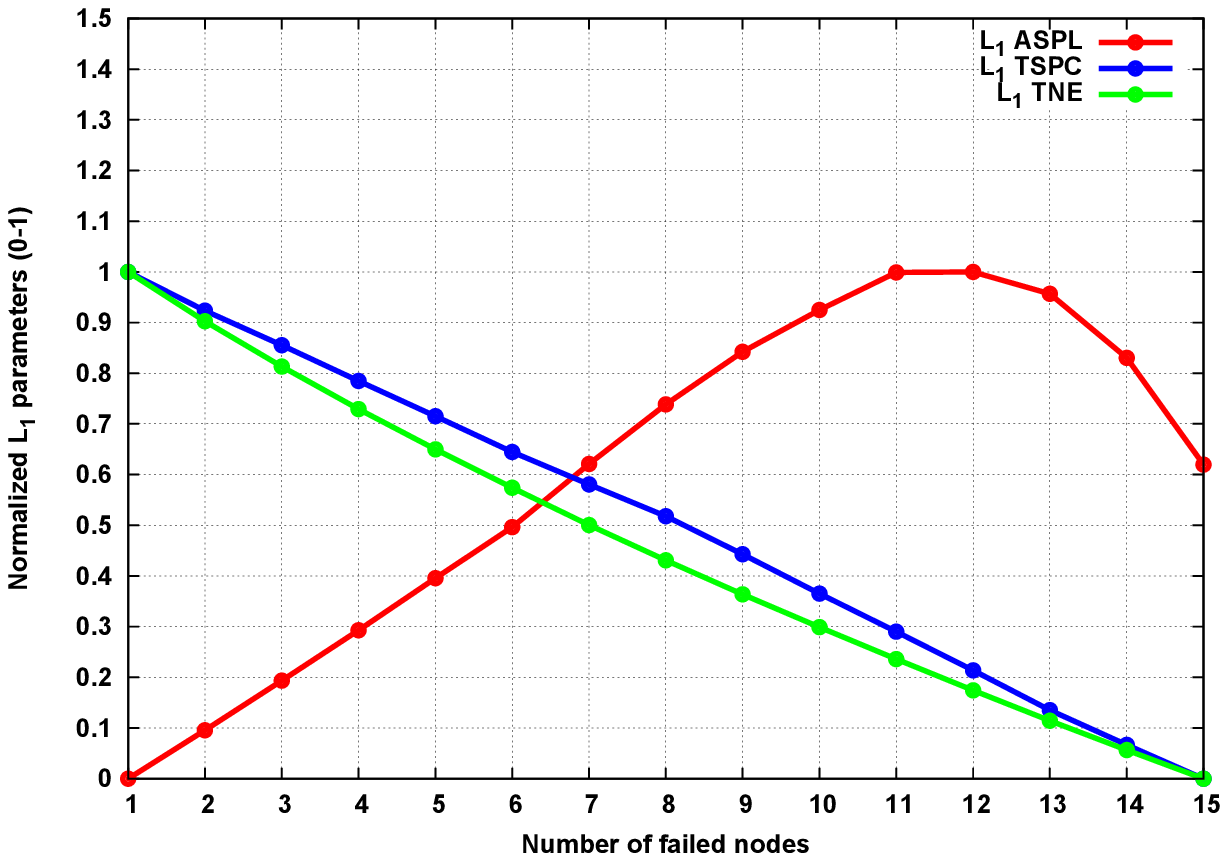}
\caption{DNBC - $L_1$ graph created only once. Behaviour of $L_1$ ASPL, TSPC, and TNE vs. number of failed nodes}
\label{fig_plot_node_all_l1_l2_params_scaled_down_l1_l2_fixed_ink}
\end{figure}

Fig. \ref{fig_plot_node_all_l1_l2_params_scaled_down_l1_l2_fixed_ink} shows the behaviour of emergent $L_1$ parameters. $L_1$ TNE and $L_1$ TSPC decreases slightly  non-linearly versus number of failed nodes. $L_1$ ASPL first increases almost linearly till 11 nodes fail, before saturating for couple of nodes and then decreases non-linearly due to reduction in the graph size (less number of nodes and edges).
\subsubsection{Behaviour of $L_2$ parameters}
\begin{figure}[ht]
\centering
\includegraphics[width=\columnwidth]{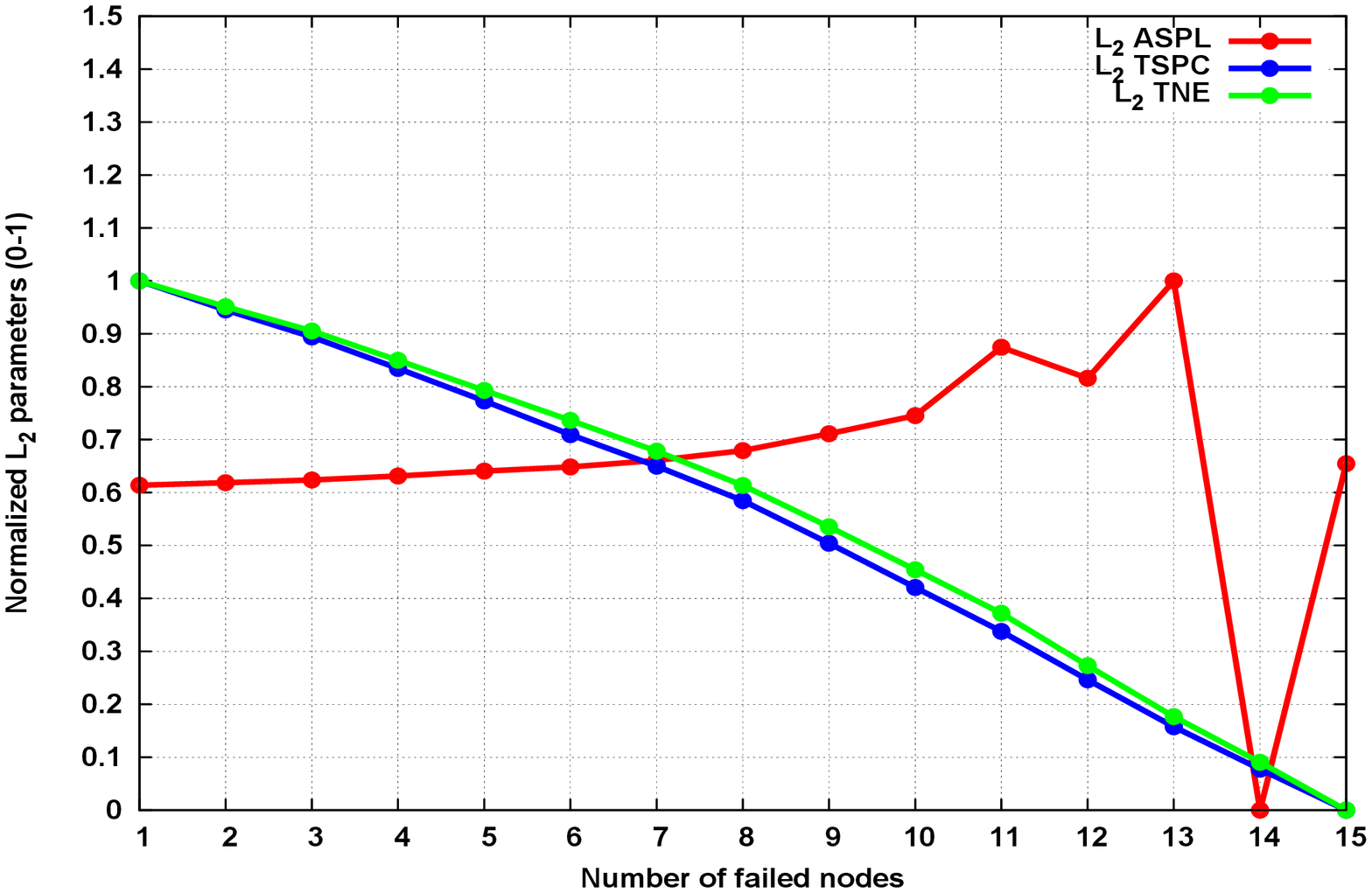}
\caption{DNBC - $L_1$ graph created only once. Behaviour of $L_2$ ASPL, TSPC, and TNE vs. number of failed nodes}
\label{fig_plot_node_all_l3_params_scaled_down_l1_l2_fixed_ink}
\end{figure}

Fig. \ref{fig_plot_node_all_l3_params_scaled_down_l1_l2_fixed_ink} shows the behaviour of $L_2$ parameters. $L_2$ TSPC and TNE decreases non-linearly almost in the same fashion. However, $L_2$ ASPL increases progressive as nodes continue to fail and starts showing chaotic behaviour after 11 nodes have failed which is roughly 10\% of the total number of nodes. This behaviour is consistent with edge failures in the section below.
\subsubsection{Behaviour of $L_3$ parameters}
\begin{figure}[ht]
\centering
\includegraphics[width=\columnwidth]{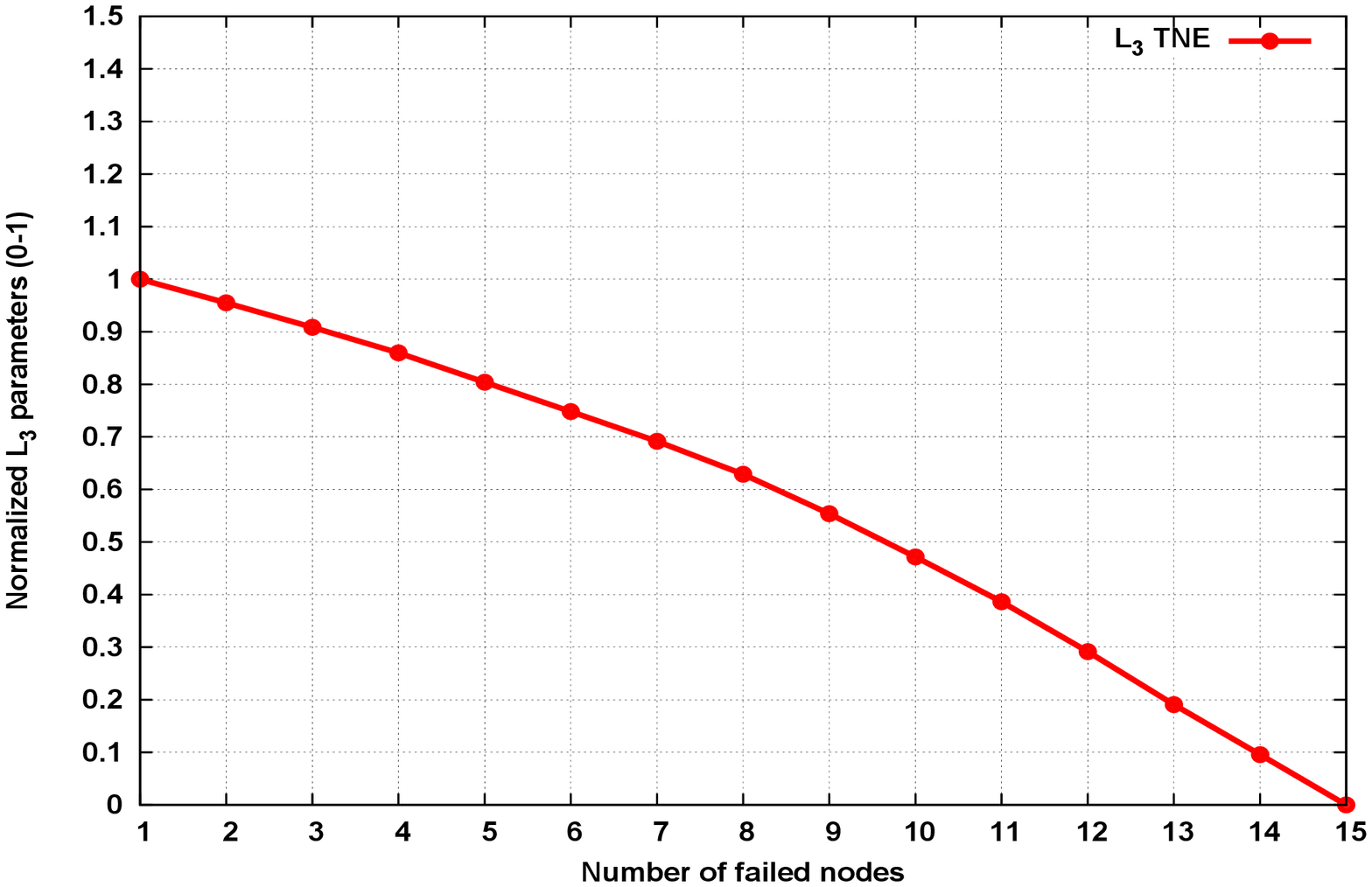}
\caption{DNBC - $L_1$ graph created only once. Behaviour of $L_3$ TNE vs. number of failed nodes}
\label{fig_plot_node_all_l4_params_scaled_down_l1_l2_fixed_ink}
\end{figure}

The behaviour of $L_3$ TNEs is shown in Fig. \ref{fig_plot_node_all_l4_params_scaled_down_l1_l2_fixed_ink}. It can be observed that it decreases non-linearly from the highest to the lowest value with a sharper decline as more nodes fail.
\subsubsection{Comparison of ASPLs of $L_1$ and $L_2$}
\begin{figure}[ht]
\centering
\includegraphics[width=\columnwidth]{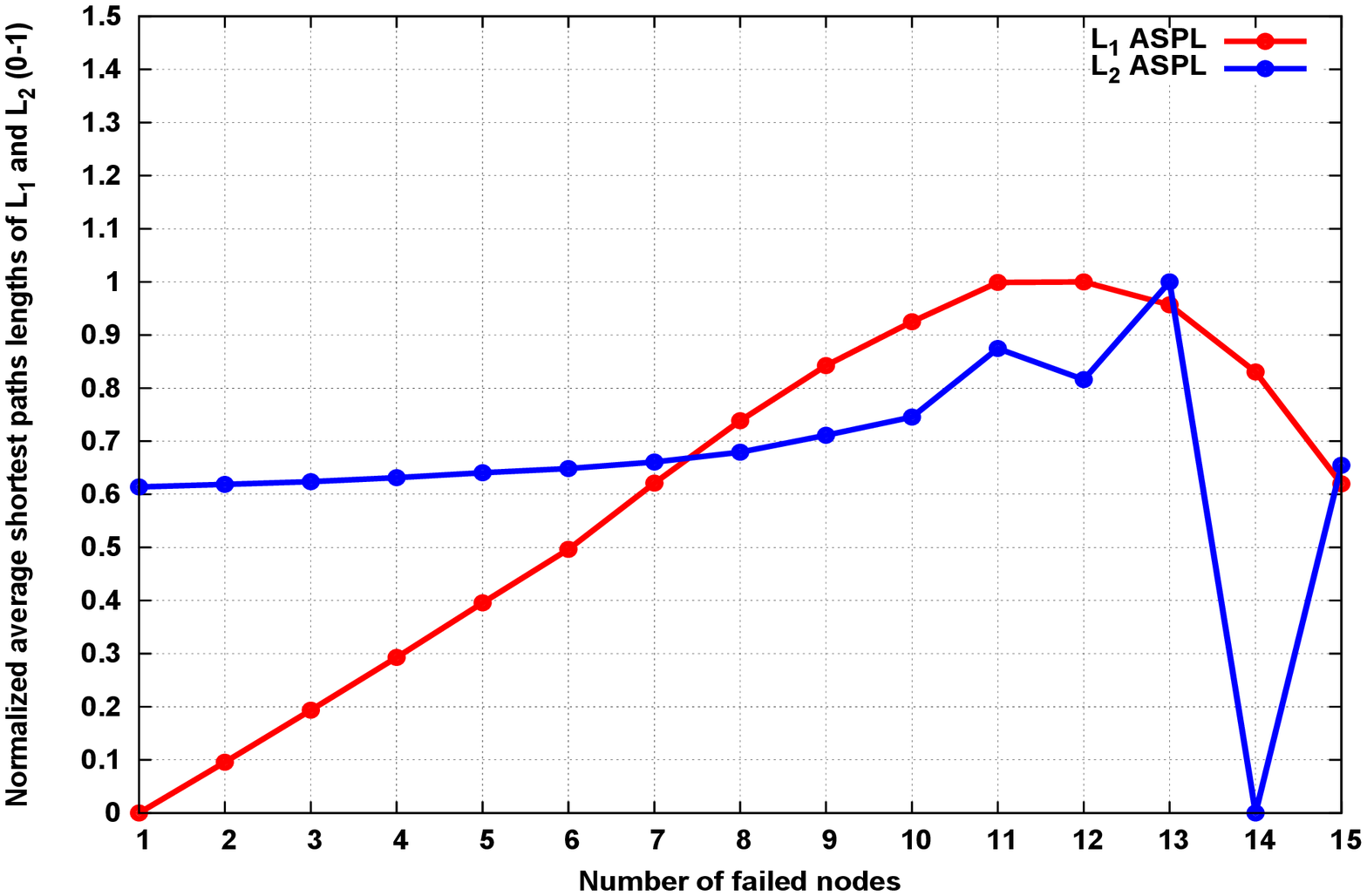}
\caption{DNBC - $L_1$ graph created only once. Comparison of ASPLs of $L_1$ and $L_2$ vs. number of failed nodes}
\label{fig_plot_node_avg_shortest_path_length_all_layers_scaled_down_l1_l2_fixed_ink}
\end{figure}

Fig. \ref{fig_plot_node_avg_shortest_path_length_all_layers_scaled_down_l1_l2_fixed_ink} compares the ASPL of $L_1$ and $L_2$. It can be observed that even if 4 nodes with high centrality fail the $L_2$ ASPL does not increase drastically showing some amount of resilience. However, with higher node failures the $L_2$ ASPL increases progressively before showing a chaotic behaviour whereas $L_1$ ASPL shows a more definite pattern.
\subsubsection{Comparison of TSPCs of $L_1$ and $L_2$ }
\begin{figure}[ht]
\centering
\includegraphics[width=\columnwidth]{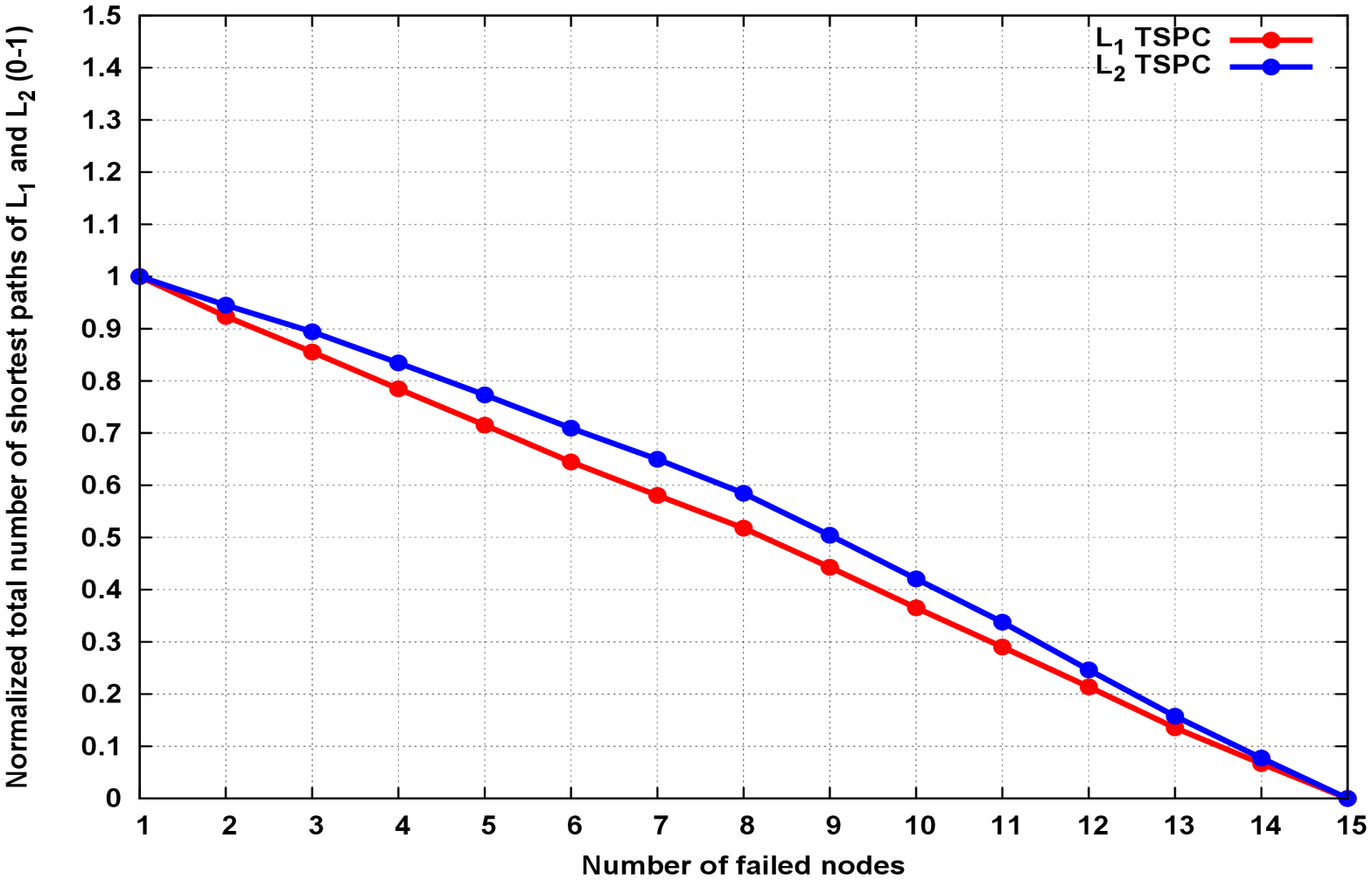}
\caption{DNBC - $L_1$ graph created only once. Comparison of TSPC of $L_1$ and $L_2$ vs. number of failed nodes}
\label{fig_plot_node_total_shortest_paths_all_layers_scaled_down_l1_l2_fixed_ink}
\end{figure}

TSPCs of $L_1$ and $L_2$ are compared in Fig. \ref{fig_plot_node_total_shortest_paths_all_layers_scaled_down_l1_l2_fixed_ink}. Though, both of them decrease non-linearly with node failures, $L_1$ TSPC decreases at a marginally faster rate than that of $L_2$.
\subsubsection{Comparison of TNEs of $L_1$, $L_2$ and $L_3$}
\begin{figure}[ht]
\centering
\includegraphics[width=\columnwidth]{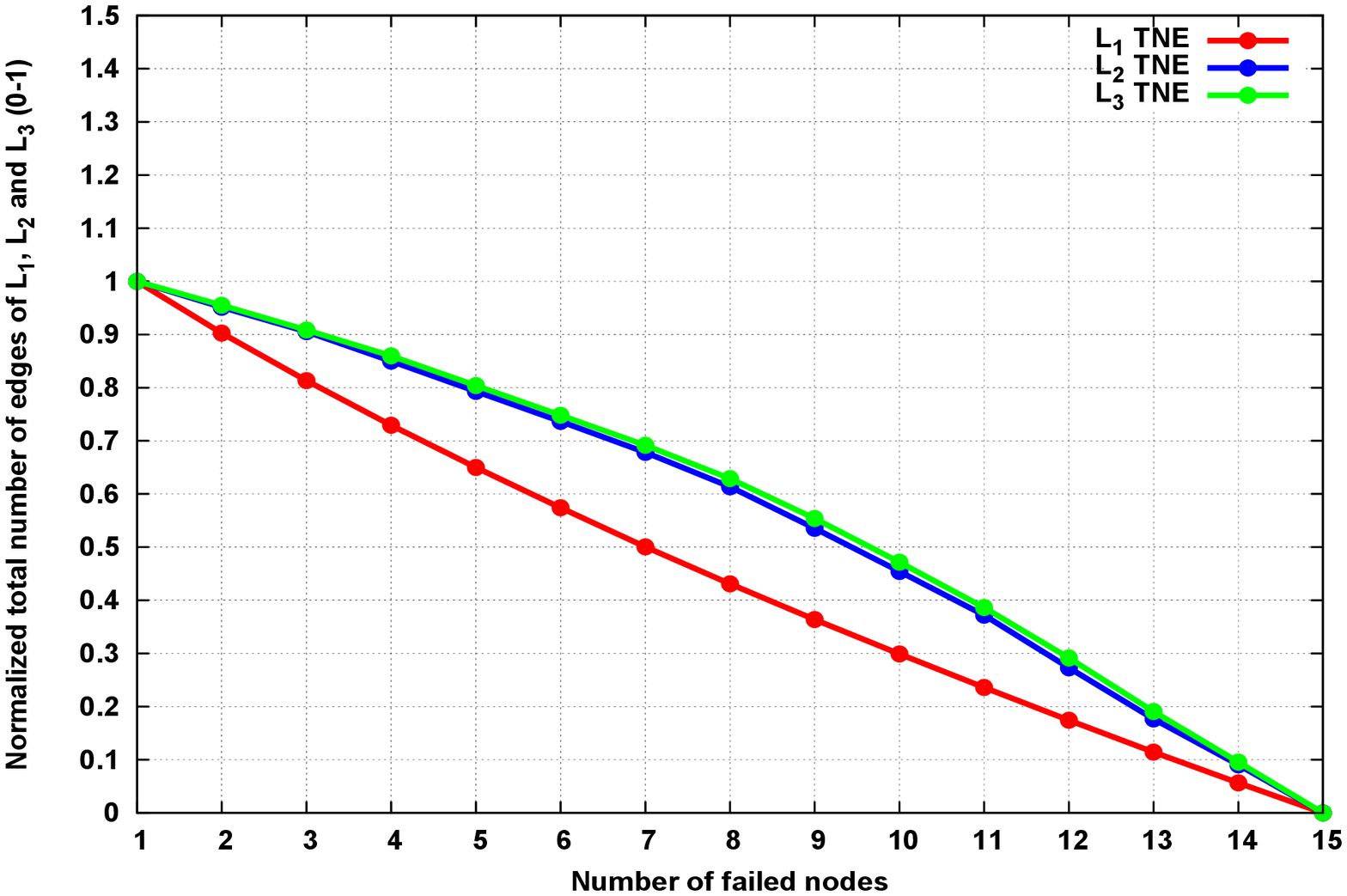}
\caption{DNBC - $L_1$ graph created only once. Comparison of TNE of $L_1$, $L_2$ and $L_3$ vs. number of failed nodes}
\label{fig_plot_node_total_edges_all_layers_scaled_down_l1_l2_fixed_ink}
\end{figure}

Behaviour of TNEs of $L_1$, $L_2$ and $L_3$ is shown in Fig. \ref{fig_plot_node_total_edges_all_layers_scaled_down_l1_l2_fixed_ink}. TNEs of $L_1$ decrease marginally non-linearly, with $L_2$ and $L_3$ decreases non-linearly almost at the same rate with $L_2$ TNE showing a slightly faster decline.

\subsection{Comparison of SNBC and DNBC }
This section compares the parameters of the two scenarios of SNBC and DNBC against same number of failed nodes.
\subsubsection{Behaviour of $L_1$ ASPL}
\begin{figure}[ht]
\centering
\includegraphics[width=\columnwidth]{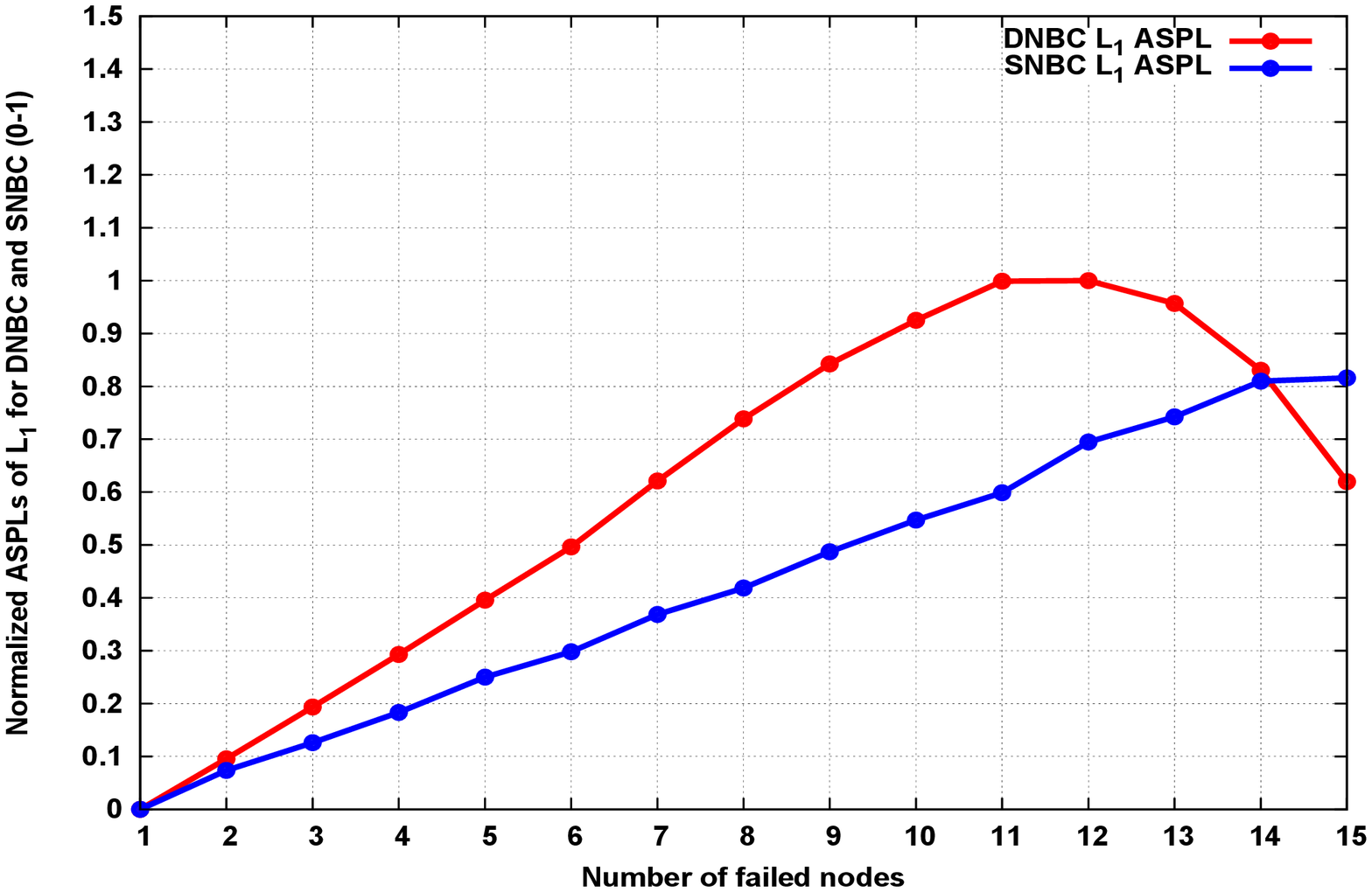}
\caption{SNBC and DNBC Comparison - $L_1$ ASPL vs. number of failed nodes}
\label{fig_plot_node_l1_l2_avg_shortest_path_length_two_scenarios_ink}
\end{figure}

Fig. \ref{fig_plot_node_l1_l2_avg_shortest_path_length_two_scenarios_ink} compares the behaviour of ASPL of $L_1$. It can be observed that for the DNBC the increase in ASPL is much steeper (40\%) and faster (when 11 nodes failed) compared to SNBC.
\subsubsection{Behaviour of $L_1$ TSPC}
\begin{figure}[ht]
\centering
\includegraphics[width=\columnwidth]{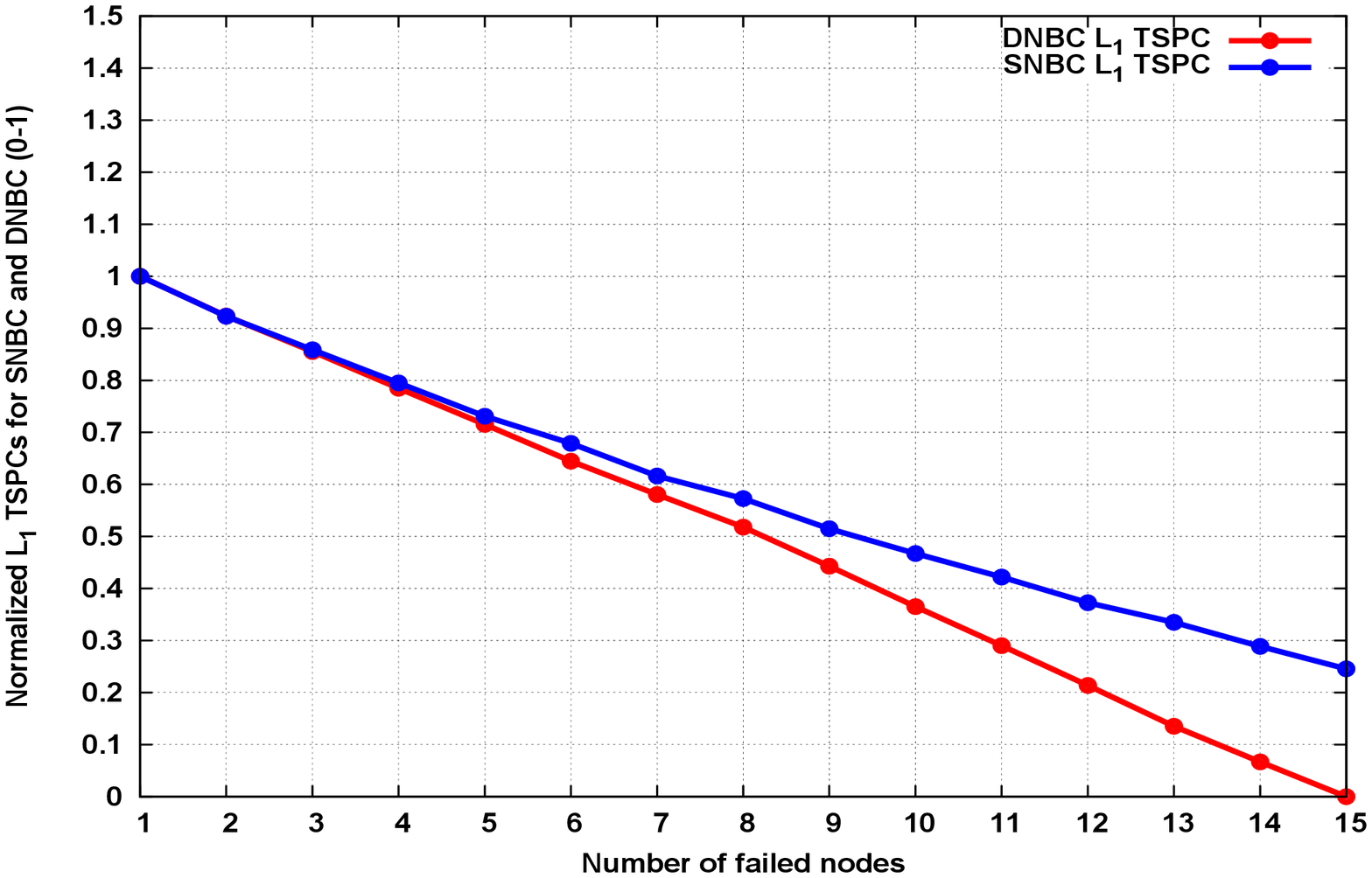}
\caption{SNBC and DNBC Comparison - $L_1$ TSPC vs. number of failed nodes}
\label{fig_plot_node_l1_l2_total_shortest_paths_two_scenarios_ink}
\end{figure}

Fig. \ref{fig_plot_node_l1_l2_total_shortest_paths_two_scenarios_ink} compares the $L_1$ TSPCs for the two scenarios. Both decreases at the same rate initially till 4 nodes fail. Thereafter, $L_1$ TSPC for DNBC scenario decreases at a much faster rate with difference growing above 20\% as more nodes fail till the end of the plot.
\subsubsection{Behaviour of $L_1$ TNE}
\begin{figure}[ht]
\centering
\includegraphics[width=\columnwidth]{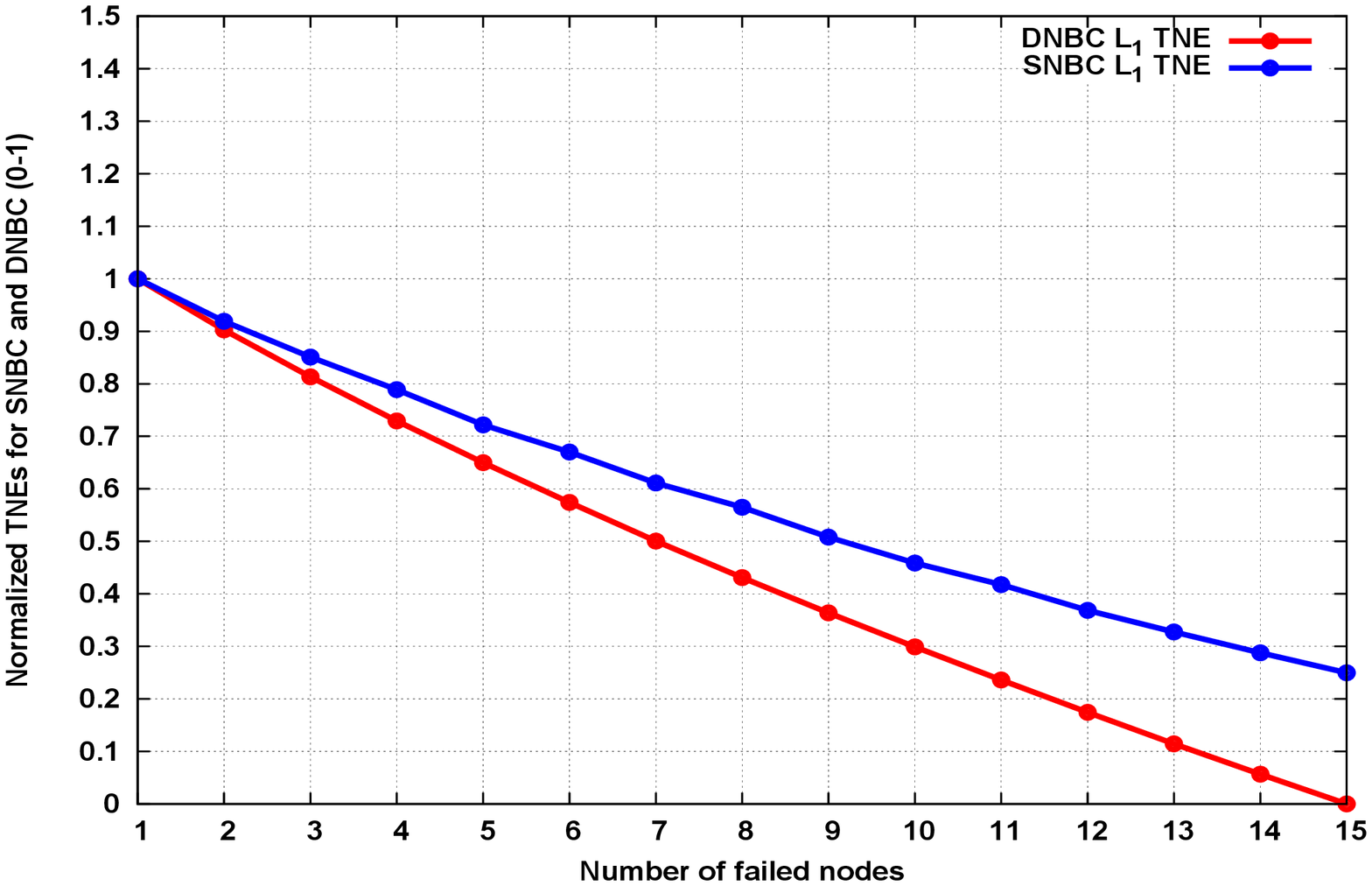}
\caption{SNBC and DNBC Comparison - $L_1$ TNE vs. number of failed nodes}
\label{fig_plot_node_l1_l2_total_edges_all_layers_two_scenarios_ink}
\end{figure}

The TNEs of DNBC decreases at a faster rate compared to that of SNBC case with further widening of the gap to about 25\% as seen in Fig. \ref{fig_plot_node_l1_l2_total_edges_all_layers_two_scenarios_ink}.
\subsubsection{Behaviour of $L_2$ ASPL}
\begin{figure}[ht]
\centering
\includegraphics[width=\columnwidth]{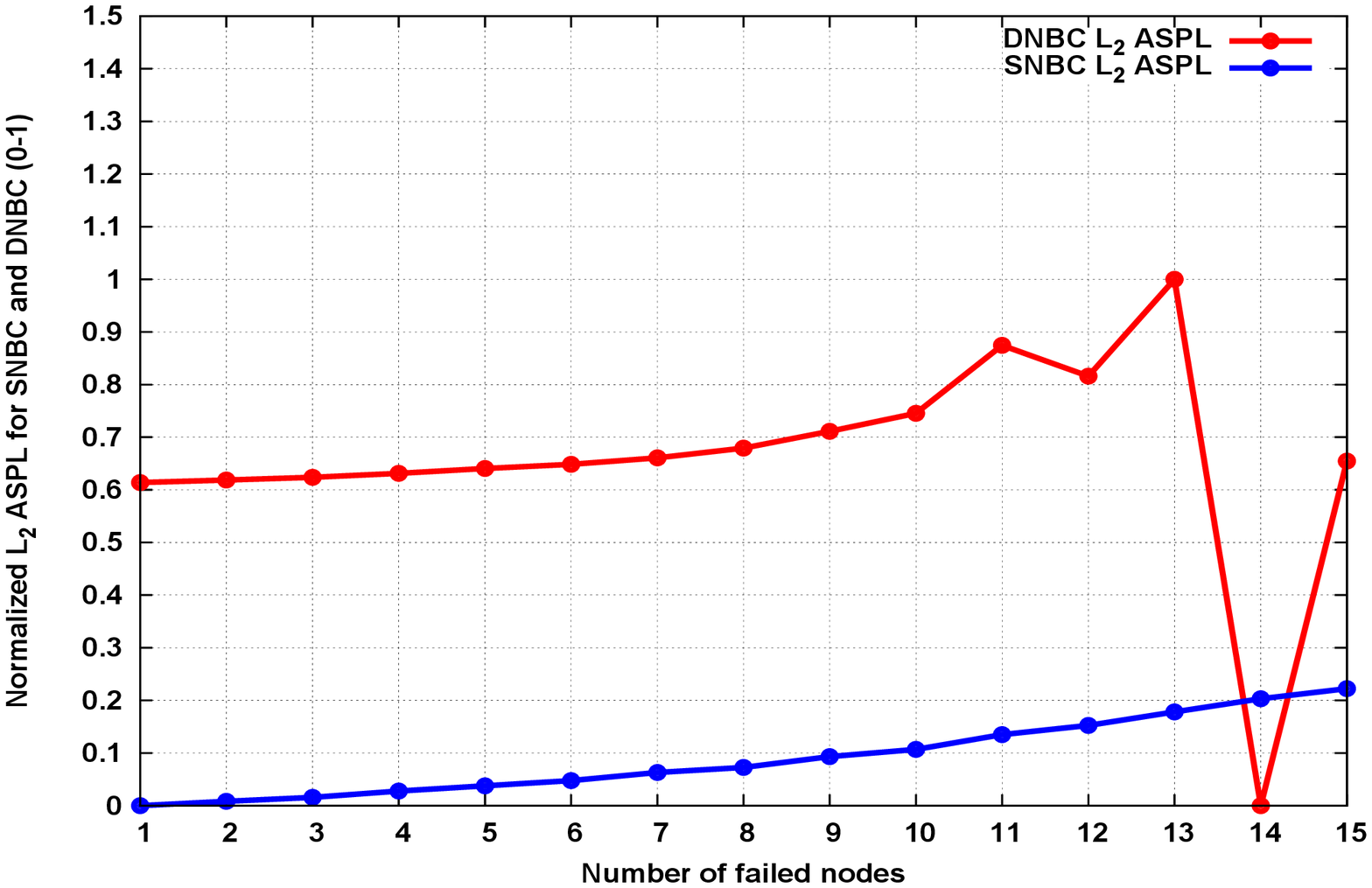}
\caption{SNBC and DNBC Comparison - $L_2$ ASPL vs. number of failed nodes}
\label{fig_plot_node_l3_avg_shortest_path_length_two_scenarios_ink}
\end{figure}

Fig. \ref{fig_plot_node_l3_avg_shortest_path_length_two_scenarios_ink} compares $L_2$ ASPL for the two scenarios. For the DNBC scenario, the ASPL is much higher and shows chaotic behaviour after 11 node failures whereas for SNBC same increases at much less rate and probably the chaotic behaviour will be observed subsequently when more nodes fail beyond 15. 
\subsubsection{Behaviour of $L_2$ TSPC}
\begin{figure}[ht]
\centering
\includegraphics[width=\columnwidth]{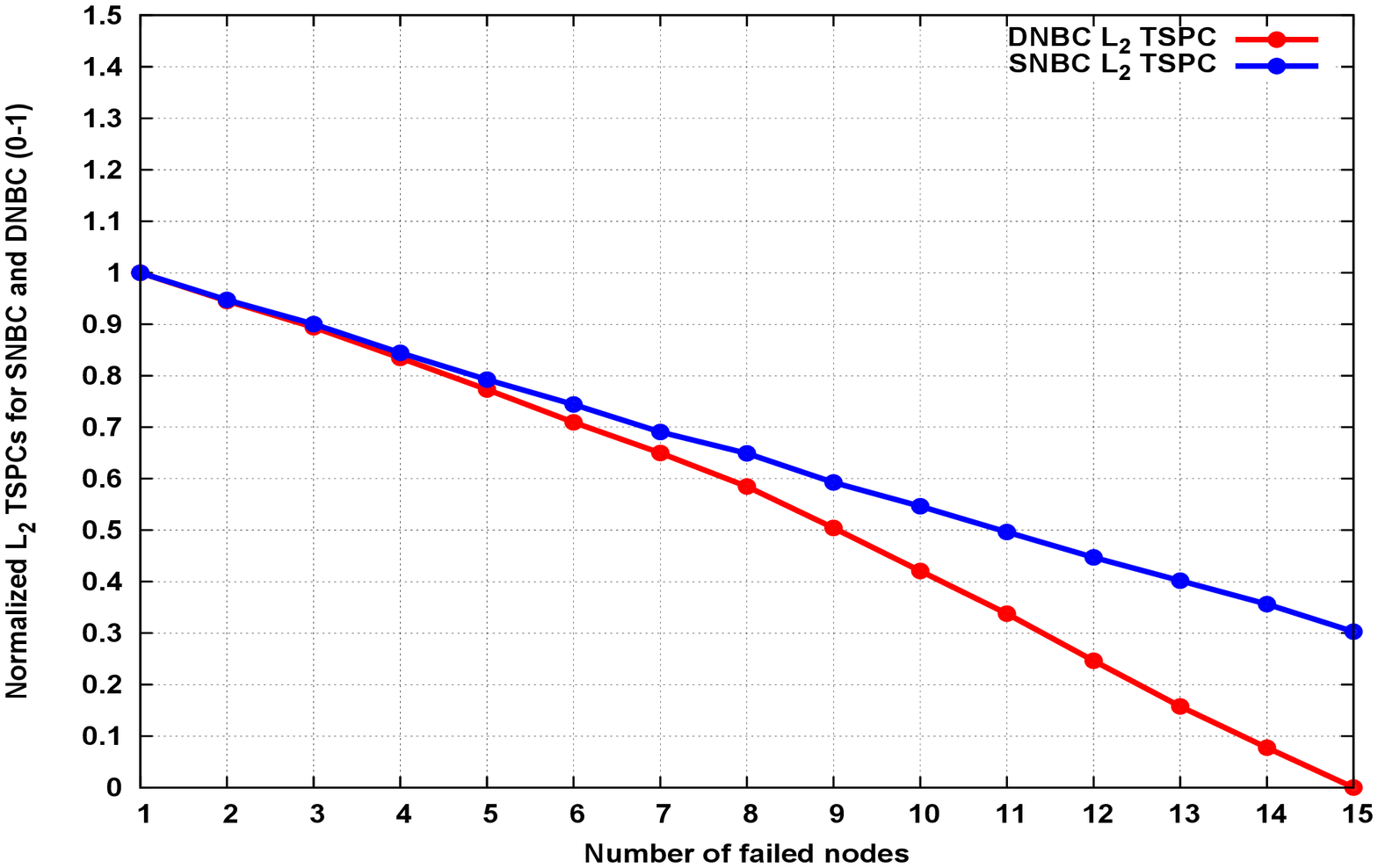}
\caption{SNBC and DNBC Comparison - $L_2$ TSPC vs. number of failed nodes}
\label{fig_plot_node_l3_total_shortest_paths_two_scenarios_ink}
\end{figure}

Fig. \ref{fig_plot_node_l3_total_shortest_paths_two_scenarios_ink} compares the $L_2$ TSPCs for the two scenarios. Both decreases at the same rate initially till 4 nodes fail. Thereafter, $L_2$ TSPC for DNBC decreases at a much faster rate with difference growing above 25\% as more nodes fail till the end of the plot.
\subsubsection{Behaviour of $L_2$ TNE}
\begin{figure}[ht]
\centering
\includegraphics[width=\columnwidth]{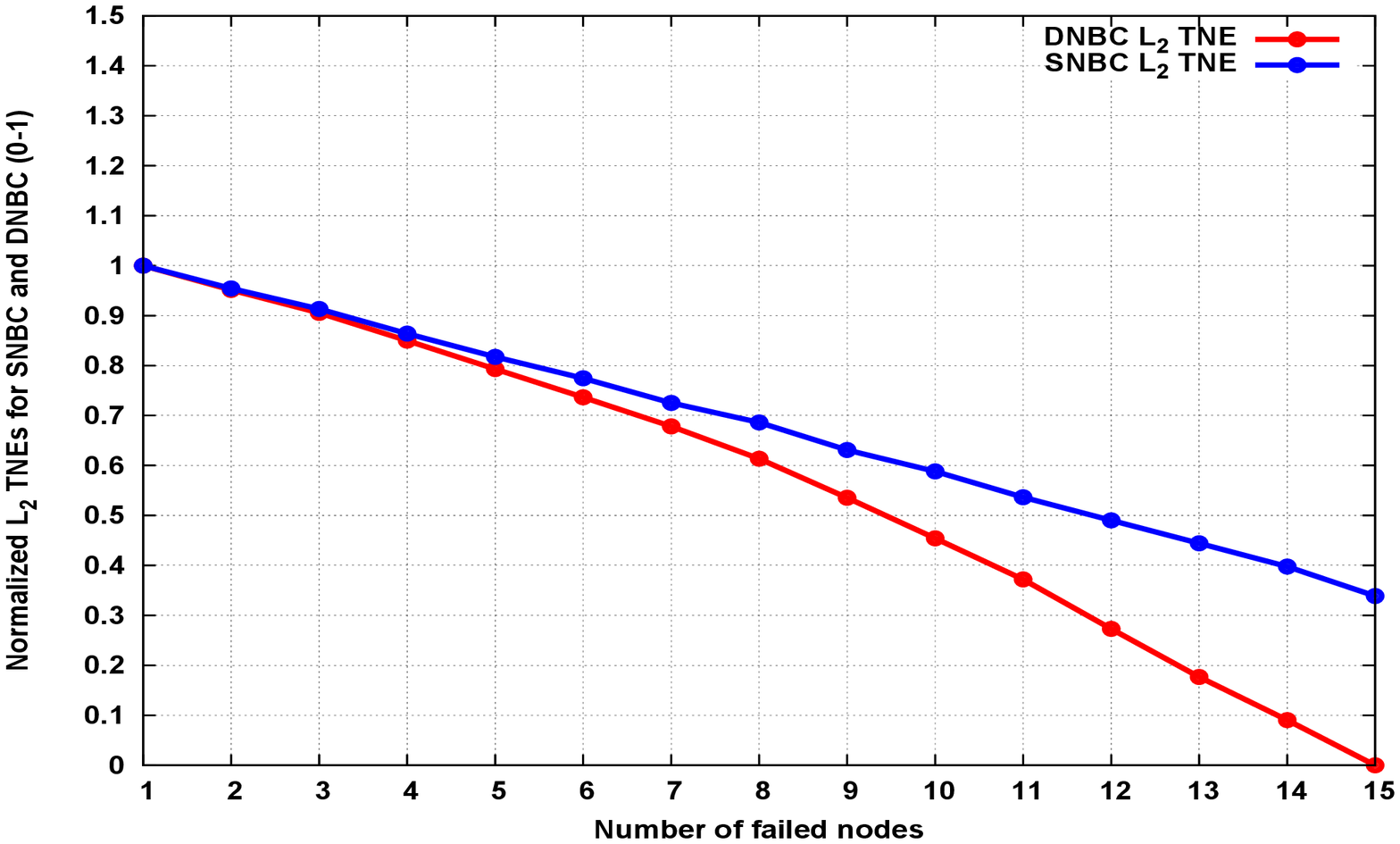}
\caption{SNBC and DNBC Comparison - $L_2$ TNEs vs. number of failed nodes}
\label{fig_plot_node_l3_total_edges_two_scenarios_ink}
\end{figure}

Fig. \ref{fig_plot_node_l3_total_edges_two_scenarios_ink} compares the $L_2$ TNEs for the two scenarios. Both decreases at the same rate initially till 4 nodes fail. Thereafter, $L_2$ TNE for DNBC decreases at a much faster rate with difference growing above 35\% as more nodes fail till the end of the plot.
\subsubsection{Behaviour of $L_3$ TNE}
\begin{figure}[ht]
\centering
\includegraphics[width=\columnwidth]{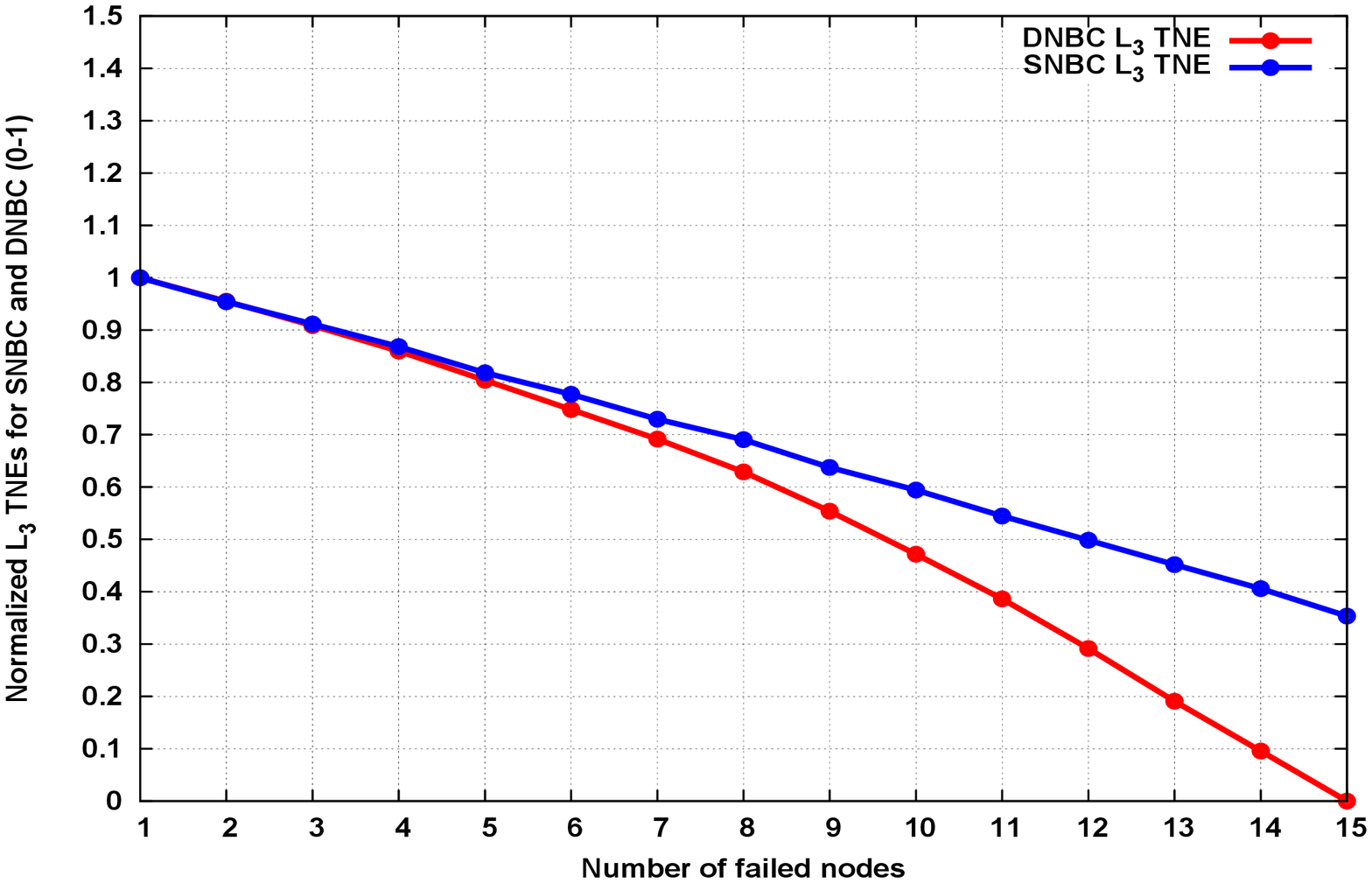}
\caption{SNBC and DNBC Comparison - $L_3$ TNEs vs. number of failed nodes}
\label{fig_plot_node_l4_total_edges_two_scenarios_ink}
\end{figure}

Fig. \ref{fig_plot_node_l4_total_edges_two_scenarios_ink} compares the $L_3$ TNEs for the two scenarios. Behaviour is more or less same as that of $L_2$ TNE as shown in Fig. \ref{fig_plot_node_l3_total_edges_two_scenarios_ink}.

\subsection{Discussion}
The above detailed results throws up following broad observations. Firstly, node failures have higher degrading consequences on the MLCN compared to edge failures. Secondly, dynamic centrality based failures both for nodes and edges have higher consequences than static centrality based failures. Thirdly, all but one parameters of the three layers shows predictable degradation. Only $L_2$ ASPL starts showing chaotic behaviour after certain number of edge and node failures.

\section{Conclusion and Future Work}\label{section_conclusion}
Failures in MLCN can cause disruptions in their functionalities. Research works have primarily focussed on different kinds of failures, such as, cascades, their reasons and ways to avoid them. This paper considers failures in a specific type of MLCN where higher layer uses the services of lower layers and failures occur only in the lowest layer. A three layer MLCN is constructed with the same set of nodes having different characteristics with the bottommost layer as ER random graph shortest path hop count among the nodes is gaussian, the middle layer is ER graph with higher number of edges from the previous, and the top most layer is scale free graph with even higher number of edges. Only the edges of bottommost layer fail and these leads to upper layer edge failures when there is no shortest path in the underlying layer between the nodes. However, node failures will affect all the layers which has edges incident on the failed node. Failures occur with decreasing order of edge or node centralities centralities in static batch mode and when the centralities change dynamically due to progressive failures. Emergent pattern of three key parameters, namely, ASPL, TSPC and TNE for all the three layers after node or edge failures are studied through simulations. One of the key results show that all  but one parameters follow a definite degrading pattern, surprising, ASPL for $L_2$ starts showing a chaotic behavior beyond a certain point for all types of failures.

Future work would concentrate on studying failures by changing the properties of each layer beyond the present configuration.

\section*{Acknowledgment}
This research project is funded by Tejas Networks, Bangalore, India.

\bibliographystyle{IEEEtran}
\bibliography{IEEEabrv,net_failure_pred}

\begin{thebibliography}{10}
\providecommand{\url}[1]{#1}
\csname url@samestyle\endcsname
\providecommand{\newblock}{\relax}
\providecommand{\bibinfo}[2]{#2}
\providecommand{\BIBentrySTDinterwordspacing}{\spaceskip=0pt\relax}
\providecommand{\BIBentryALTinterwordstretchfactor}{4}
\providecommand{\BIBentryALTinterwordspacing}{\spaceskip=\fontdimen2\font plus
\BIBentryALTinterwordstretchfactor\fontdimen3\font minus
  \fontdimen4\font\relax}
\providecommand{\BIBforeignlanguage}[2]{{%
\expandafter\ifx\csname l@#1\endcsname\relax
\typeout{** WARNING: IEEEtran.bst: No hyphenation pattern has been}%
\typeout{** loaded for the language `#1'. Using the pattern for}%
\typeout{** the default language instead.}%
\else
\language=\csname l@#1\endcsname
\fi
#2}}
\providecommand{\BIBdecl}{\relax}
\BIBdecl

\bibitem{cite_book_complex_networks_principles_methods_apps}
V.~Latora, V.~Nicosia, and G.~Russo, \emph{Complex Networks: Principles,
  Methods and Applications}, 1st~ed.\hskip 1em plus 0.5em minus 0.4em\relax
  USA: Cambridge University Press, 2017.

\bibitem{cite_complex_networks_apps_survey}
\BIBentryALTinterwordspacing
L.~da~Fontoura~Costa, O.~N.~O. Jr., G.~Travieso, F.~A. Rodrigues, P.~R.~V.
  Boas, L.~Antiqueira, M.~P. Viana, and L.~E.~C. Rocha, ``Analyzing and
  modeling real-world phenomena with complex networks: a survey of
  applications,'' \emph{Advances in Physics}, vol.~60, no.~3, pp. 329--412,
  2011. [Online]. Available: \url{https://doi.org/10.1080/00018732.2011.572452}
\BIBentrySTDinterwordspacing

\bibitem{cite_multilayer_complex_networks_structure_and_dynamics}
S.~Boccaletti, G.~Bianconi, R.~Criado, C.~I. Del~Genio, J.~G{\'o}mez-Gardenes,
  M.~Romance, I.~Sendina-Nadal, Z.~Wang, and M.~Zanin, ``The structure and
  dynamics of multilayer networks,'' \emph{Physics reports}, vol. 544, no.~1,
  pp. 1--122, 2014.

\bibitem{cite_cn_cascade_failures_detailed_study}
\BIBentryALTinterwordspacing
D.~J. Watts, ``A simple model of global cascades on random networks,''
  \emph{Proceedings of the National Academy of Sciences}, vol.~99, no.~9, pp.
  5766--5771, 2002. [Online]. Available:
  \url{https://www.pnas.org/doi/abs/10.1073/pnas.082090499}
\BIBentrySTDinterwordspacing

\bibitem{cite_cn_cascade_attacks}
A.~E. Motter and Y.-C. Lai, ``Cascade-based attacks on complex networks,''
  \emph{Physical Review E}, vol.~66, no.~6, p. 065102, 2002.

\bibitem{cite_cn_cascade_failures_in_er_and_scale_free_nets_load_dist}
\BIBentryALTinterwordspacing
N.~Wang, Z.-Y. Jin, and J.~Zhao, ``Cascading failures of overload behaviors on
  interdependent networks,'' \emph{Physica A: Statistical Mechanics and its
  Applications}, vol. 574, p. 125989, 2021. [Online]. Available:
  \url{https://www.sciencedirect.com/science/article/pii/S0378437121002612}
\BIBentrySTDinterwordspacing

\bibitem{cite_cn_dynamic_topological_models_to_predict_system_failure_from_cascade}
\BIBentryALTinterwordspacing
D.~Duan, C.~Lv, S.~Si, Z.~Wang, D.~Li, J.~Gao, S.~Havlin, H.~E. Stanley, and
  S.~Boccaletti, ``Universal behavior of cascading failures in interdependent
  networks,'' \emph{Proceedings of the National Academy of Sciences}, vol. 116,
  no.~45, pp. 22\,452--22\,457, 2019. [Online]. Available:
  \url{https://www.pnas.org/doi/abs/10.1073/pnas.1904421116}
\BIBentrySTDinterwordspacing

\bibitem{cite_cn_multiplex_network_cascade_failures_emergent_coupling}
\BIBentryALTinterwordspacing
K.-M. Lee and K.-I. Goh, ``Strength of weak layers in cascading failures on
  multiplex networks: case of the international trade network,''
  \emph{Scientific Reports}, vol.~6, no.~1, p. 26346, May 2016. [Online].
  Available: \url{https://doi.org/10.1038/srep26346}
\BIBentrySTDinterwordspacing

\bibitem{cite_cn_scale_free_interdependen_network_cascade_prevent_with_different_properties}
\BIBentryALTinterwordspacing
M.~Turalska, K.~Burghardt, M.~Rohden, A.~Swami, and R.~M. D'Souza, ``Cascading
  failures in scale-free interdependent networks,'' \emph{Phys. Rev. E},
  vol.~99, p. 032308, Mar 2019. [Online]. Available:
  \url{https://link.aps.org/doi/10.1103/PhysRevE.99.032308}
\BIBentrySTDinterwordspacing

\bibitem{cite_cn_mitigate_cascade_failures_protect_critical_nodes_and_local_structure}
\BIBentryALTinterwordspacing
A.~Smolyak, O.~Levy, I.~Vodenska, S.~Buldyrev, and S.~Havlin, ``Mitigation of
  cascading failures in complex networks,'' \emph{Scientific Reports}, vol.~10,
  no.~1, p. 16124, Sep 2020. [Online]. Available:
  \url{https://doi.org/10.1038/s41598-020-72771-4}
\BIBentrySTDinterwordspacing

\bibitem{cite_cn_cascade_failures_avoided_with_reinforced_nodes}
\BIBentryALTinterwordspacing
X.~Yuan, Y.~Hu, H.~E. Stanley, and S.~Havlin, ``Eradicating catastrophic
  collapse in interdependent networks via reinforced nodes,'' \emph{Proceedings
  of the National Academy of Sciences}, vol. 114, no.~13, pp. 3311--3315, 2017.
  [Online]. Available:
  \url{https://www.pnas.org/doi/abs/10.1073/pnas.1621369114}
\BIBentrySTDinterwordspacing

\bibitem{cite_cn_self_heal_reconstruct_after_failure_using_local_info}
L.~K. Gallos and N.~H. Fefferman, ``Simple and efficient self-healing strategy
  for damaged complex networks,'' \emph{Physical Review E}, vol.~92, no.~5, p.
  052806, 2015.

\bibitem{cite_cn_percolulation_of_failures_in_ER_scale_free_random_regular_nets}
\BIBentryALTinterwordspacing
S.~Shao, X.~Huang, H.~E. Stanley, and S.~Havlin, ``Percolation of localized
  attack on complex networks,'' \emph{New Journal of Physics}, vol.~17, no.~2,
  p. 023049, feb 2015. [Online]. Available:
  \url{https://doi.org/10.1088/1367-2630/17/2/023049}
\BIBentrySTDinterwordspacing

\bibitem{cite_cn_knowledge_from_random_attacks_lead_to_targetted_attack}
\BIBentryALTinterwordspacing
S.~Wandelt, W.~Lin, X.~Sun, and M.~Zanin, ``From random failures to targeted
  attacks in network dismantling,'' \emph{Reliability Engineering \& System
  Safety}, vol. 218, p. 108146, 2022. [Online]. Available:
  \url{https://www.sciencedirect.com/science/article/pii/S0951832021006335}
\BIBentrySTDinterwordspacing

\bibitem{cite_cn_assortativity_decreases_robustness}
\BIBentryALTinterwordspacing
D.~Zhou, H.~E. Stanley, G.~D'Agostino, and A.~Scala, ``Assortativity decreases
  the robustness of interdependent networks,'' \emph{Phys. Rev. E}, vol.~86, p.
  066103, Dec 2012. [Online]. Available:
  \url{https://link.aps.org/doi/10.1103/PhysRevE.86.066103}
\BIBentrySTDinterwordspacing

\bibitem{cite_cn_assymmetry_lead_to_perculation_transition_and_node_and_its_peer_position}
\BIBentryALTinterwordspacing
R.-R. Liu, C.-X. Jia, and Y.-C. Lai, ``Asymmetry in interdependence makes a
  multilayer system more robust against cascading failures,'' \emph{Phys. Rev.
  E}, vol. 100, p. 052306, Nov 2019. [Online]. Available:
  \url{https://link.aps.org/doi/10.1103/PhysRevE.100.052306}
\BIBentrySTDinterwordspacing

\bibitem{cite_cn_evolutionary_algo_for_cascade_failure_resilience}
\BIBentryALTinterwordspacing
J.~Ash and D.~Newth, ``Optimizing complex networks for resilience against
  cascading failure,'' \emph{Physica A: Statistical Mechanics and its
  Applications}, vol. 380, pp. 673--683, 2007. [Online]. Available:
  \url{https://www.sciencedirect.com/science/article/pii/S0378437107002543}
\BIBentrySTDinterwordspacing

\bibitem{cite_cn_triple_points_help_restore_failures}
\BIBentryALTinterwordspacing
A.~Majdandzic, L.~A. Braunstein, C.~Curme, I.~Vodenska, S.~Levy-Carciente,
  H.~Eugene~Stanley, and S.~Havlin, ``Multiple tipping points and optimal
  repairing in interacting networks,'' \emph{Nature Communications}, vol.~7,
  no.~1, p. 10850, Mar 2016. [Online]. Available:
  \url{https://doi.org/10.1038/ncomms10850}
\BIBentrySTDinterwordspacing

\bibitem{cite_cn_sparse_coupling_enhanced_coupling_prob_for_robustness_against_cascade}
\BIBentryALTinterwordspacing
F.~Tan, Y.~Xia, W.~Zhang, and X.~Jin, ``Cascading failures of loads in
  interconnected networks under intentional attack,'' \emph{{EPL} (Europhysics
  Letters)}, vol. 102, no.~2, p. 28009, apr 2013. [Online]. Available:
  \url{https://doi.org/10.1209/0295-5075/102/28009}
\BIBentrySTDinterwordspacing

\bibitem{cite_cn_cascade_failures_avoid_with_split_and_combine_layers}
\BIBentryALTinterwordspacing
C.~D. Brummitt, K.-M. Lee, and K.-I. Goh, ``Multiplexity-facilitated cascades
  in networks,'' \emph{Phys. Rev. E}, vol.~85, p. 045102, Apr 2012. [Online].
  Available: \url{https://link.aps.org/doi/10.1103/PhysRevE.85.045102}
\BIBentrySTDinterwordspacing

\bibitem{cite_cn_node_cascade_node_community_cascade}
L.~Ma, X.~Zhang, J.~Li, Q.~Lin, M.~Gong, C.~A.~C. Coello, and A.~K. Nandi,
  ``Enhancing robustness and resilience of multiplex networks against
  node-community cascading failures,'' \emph{IEEE Transactions on Systems, Man,
  and Cybernetics: Systems}, vol.~52, no.~6, pp. 3808--3821, 2022.

\bibitem{cite_cn_recovery_coupling_from_recoveries_of_power_grid_failures}
\BIBentryALTinterwordspacing
M.~M. Danziger and A.-L. Barab{\~A}{\textexclamdown}si, ``Recovery coupling in
  multilayer networks,'' \emph{Nature Communications}, vol.~13, no.~1, p. 955,
  Feb 2022. [Online]. Available:
  \url{https://doi.org/10.1038/s41467-022-28379-5}
\BIBentrySTDinterwordspacing

\bibitem{cite_cn_optimal_subnetwork_interaction_for_resilience}
\BIBentryALTinterwordspacing
G.~Dong, F.~Wang, L.~M. Shekhtman, M.~M. Danziger, J.~Fan, R.~Du, J.~Liu,
  L.~Tian, H.~E. Stanley, and S.~Havlin, ``Optimal resilience of modular
  interacting networks,'' \emph{Proceedings of the National Academy of
  Sciences}, vol. 118, no.~22, p. e1922831118, 2021. [Online]. Available:
  \url{https://www.pnas.org/doi/abs/10.1073/pnas.1922831118}
\BIBentrySTDinterwordspacing

\bibitem{cite_cn_degree_heterogeneity_increase_chances_of_failures}
\BIBentryALTinterwordspacing
S.~Sun, Y.~Wu, Y.~Ma, L.~Wang, Z.~Gao, and C.~Xia, ``Impact of degree
  heterogeneity on attack vulnerability of interdependent networks,''
  \emph{Scientific Reports}, vol.~6, no.~1, p. 32983, Sep 2016. [Online].
  Available: \url{https://doi.org/10.1038/srep32983}
\BIBentrySTDinterwordspacing

\bibitem{cite_centrality_definitions}
\BIBentryALTinterwordspacing
U.~Brandes, ``On variants of shortest-path betweenness centrality and their
  generic computation,'' \emph{Social Networks}, vol.~30, no.~2, pp. 136--145,
  2008. [Online]. Available:
  \url{https://www.sciencedirect.com/science/article/pii/S0378873307000731}
\BIBentrySTDinterwordspacing

\end{thebibliography}

\end{document}